\documentclass[prc,floatfix,groupedaddress,nofootinbib,showpacs,preprintnumbers,
amsmath,amssymb,amsfonts,superscriptaddress,widetable] {revtex4-1}
\usepackage{graphicx}
\usepackage{dcolumn}
\usepackage{mathrsfs}
\usepackage{bm}
\usepackage{float}
\usepackage{soul}
\usepackage[usenames]{color}

\newcommand{\FSUt}{FSU\hspace{0.25pt}2}
\newcommand{\rhoz}{\rho_{\raisebox{-0.75pt}{\tiny 0}}}
\newcommand{\epsz}{\varepsilon_{\raisebox{-0.75pt}{\tiny 0}}}
\newcommand{\chisq}{{\chi}^{\raisebox{-1pt}{$\scriptstyle{2}$}}}

\newcommand{\psub}[1]{{\bf p}_{\raisebox{-2pt}{$\scriptstyle#1$}}}
\newcommand{\sigmasub}[1]{{\sigma}_{\raisebox{-2pt}{$\scriptscriptstyle\hspace{-1pt}\!\!#1$}}}
\newcommand{\mc}[1]{\mathcal{#1}}

\begin{document}

\title{Building relativistic mean field models for finite nuclei and neutron stars}
\author{Wei-Chia Chen}
\email{wc09c@my.fsu.edu} 
\affiliation{Department of Physics, Florida State University, Tallahassee, FL 32306} 
\author{J. Piekarewicz}
\email{jpiekarewicz@fsu.edu}
\affiliation{Department of Physics, Florida State University, Tallahassee, FL 32306}
\date{\today}
\begin{abstract}
\begin{description}
\item[Background] Theoretical approaches based on density functional theory 
provide the only tractable method to incorporate the wide range of densities and
isospin asymmetries required to describe finite nuclei, infinite nuclear matter, 
and neutron stars. 
\item[Purpose] A relativistic energy density functional (EDF) is developed to address
the complexity of such diverse nuclear systems. Moreover, a statistical perspective 
is adopted to describe the information content of various physical observables.
\item[Methods] We implement the model optimization by minimizing a suitably
constructed $\chisq$ objective function using various properties of finite nuclei 
and neutron stars. The minimization is then supplemented by a covariance 
analysis that includes both uncertainty estimates and correlation coefficients.
\item[Results] A new model, ``\emph{FSUGold\,2}'', is created that can well reproduce 
the ground-state properties of finite nuclei, their monopole response, and that 
accounts for the maximum neutron star mass observed up to date. In particular,
the model predicts both a stiff symmetry energy and a soft equation of state for 
symmetric nuclear matter---suggesting a fairly large neutron-skin thickness in 
${}^{208}$Pb and a moderate value of the nuclear incompressibility.  
\item[Conclusions] We conclude that without any meaningful constraint on the
isovector sector, relativistic EDFs will continue to predict significantly large
neutron skins. However, the calibration scheme adopted here is flexible enough 
to create models with different assumptions on various observables. Such a 
scheme---properly supplemented by a covariance analysis---provides a powerful 
tool to identify the critical measurements required to place meaningful constraints 
on theoretical models.
\end{description}
\end{abstract}

\maketitle

\section{Introduction}
\label{intro}

Finite nuclei, infinite nuclear matter, and neutron stars are complex,
many-body systems governed largely by the strong nuclear force. 
Although Quantum Chromodynamics (QCD) is the fundamental theory  
of the strong interaction, enormous challenges have prevented us from
solving the theory in the non-perturbative regime of relevance to
nuclear systems. To date, these complex systems can be investigated 
only in the framework of an effective theory with appropriate degrees
of freedom. Among the effective approaches, the one based on 
\emph{density functional theory} (DFT) is most promising, as it is 
the only microscopic approach that may be applied to the entire 
nuclear landscape and to neutron stars. In the past decades
numerous \emph{energy density functionals} (EDFs) have been proposed
which can be grouped into two main branches: non-relativistic 
and relativistic. Skyrme-type functionals are the most 
popular ones within the non-relativistic domain, where nucleons 
interact via density-dependent effective potentials. Using such a 
framework, the \emph{Universal Nuclear Energy Density Functional} 
(UNEDF) Collaboration\,\cite{UNEDF} aims to achieve a comprehensive
understanding of finite nuclei and the reactions involving 
them\,\cite{Kortelainen:2010hv, Kortelainen:2012,Kortelainen:2013}. 
On the other end, relativistic mean field (RMF) models, based on a 
quantum field theory having nucleons interacting via the exchange
of various mesons, have been successfully used since the 1970's 
and provide a covariant description of both infinite nuclear matter and
finite nuclei\,\cite{Walecka:1974qa,Serot:1984ey,Horowitz:1981xw,  
Lalazissis:1996rd,Lalazissis:1999,Todd-Rutel:2005fa}. 

In the traditional spirit of effective theories, both non-relativistic and 
relativistic EDFs are calibrated from nuclear experimental data that
is obtained under normal laboratory conditions, namely, at or slightly
below nuclear saturation density and with small to moderate isospin 
asymmetries. The lack of experimental data at both higher densities 
and with extreme isospin asymmetries leads to a large spread in the 
predictions of the models---even when they may all be calibrated to 
the same experimental data. Consequently, fundamental nuclear
properties, such as the neutron density of medium-to-heavy 
nuclei\,\cite{Brown:2000,Furnstahl:2001un,Centelles:2008vu,
RocaMaza:2011pm}, proton and neutron drip 
lines\,\cite{Erler:2012, Afanasjev:2013}, and a variety of 
neutron star properties\,\cite{Lattimer:2004pg,
Demorest:2010bx,Antoniadis:2013pzd} remain largely 
undetermined.

It has been a common practice for a long time to supplement
experimental results with uncertainty estimates. Indeed, no 
experimental measurement could ever be published without 
properly estimated ``error bars". Often, the most difficult part of 
an experiment is a reliable quantification of systematic errors and 
improving the precision of the measurement consists of painstaking 
efforts at reducing the sources of such uncertainties. On the 
contrary, theoretical predictions merely involve reporting a ``central 
value'' without any information on the uncertainties inherent in the 
formulation or the calculation. Thus, to determine whether a theory 
is successful or not, the only required criterion is to reproduce the 
experimental data. Although this approach has certain value---especially 
if the examined model reproduces a vast amount of experimental 
data---such a criterion is often neither helpful nor meaningful. And the 
situation becomes even worse if the predictions of an effective theory 
are extrapolated into unknown regions, such as the boundaries of the
nuclear landscape and the interior of neutron stars. How can a model 
provide experimental or observational guidance without supplementing
its predictions with theoretical errors? In recent years, \emph{``the 
importance of including uncertainty estimates in papers involving 
theoretical calculations of physical quantities''} has been 
underscored\,\cite{PhysRevA.83.040001}. This is particularly critical when 
theoretical models are used to extrapolate experimental data to uncharted 
regions of the observable landscape. Thus, theoretical uncertainty estimates 
are critical in assessing the reliability of the extrapolations. Moreover, if 
these theoretical errors are large, then one can perform a correlation 
analysis to uncover observables that can help reduce the size of the 
uncertainties. Several manuscripts highlighting the role of information 
and statistics in nuclear physics have been published 
recently\,\cite{Reinhard:2010wz,Fattoyev:2011ns,Fattoyev:2012rm,
Reinhard:2012vw,Reinhard:2013fpa,Dobaczewski:2014jga,
Piekarewicz:2014kza}. Moreover, at the time of this writing, a 
focus issue devoted to \emph{``Enhancing the interaction between 
nuclear experiment and theory through information and statistics''}
was under development.

In this work we develop a modeling scheme within the framework of the
RMF theory that consists of both the optimization of a theoretical model 
and the follow-up covariance analysis. However, unlike the UNEDF 
Collaboration, our goals are rather modest as we do not attempt to study 
all the facets of finite nuclei. Instead, we limit ourselves to a treatment 
of the ground-state properties of magic (or semi-magic) finite nuclei, 
centroid energies of monopole resonances, and properties of neutron 
stars. We would like to emphasize that all the data that we use in the 
optimization of the relativistic EDF consists of real physical observables 
without any reliance on bulk properties of infinite nuclear matter. This is 
now possible due to the remarkable advances in land- and space-based
telescopes that have started to place meaningful constraints on the 
high-density component of the equation of state. In particular, observations 
made with the Green Bank Telescope have provided 
highly precise measurements of two massive (of about 2\,$M_{\odot}$) 
neutron stars\,\cite{Demorest:2010bx,Antoniadis:2013pzd}. Further, an 
enormous effort is also being devoted to the extraction of stellar radii 
from x-ray observations\,\cite{Ozel:2010fw,Steiner:2010fz,
Suleimanov:2010th,Guillot:2013wu}. Such astronomical observations 
will be instrumental in constraining the nuclear EDF in regions inaccessible 
to laboratory experiments.

Not having to rely on the bulk properties of nuclear matter in the calibration 
procedure implies that these properties now become genuine model 
predictions---with associated theoretical errors---that may be compared 
against results from {\it ab initio} calculations or other microscopic
approaches\,\cite{Schwenk:2005ka,Gezerlis:2009iw,Vidana:2009is,
Hebeler:2010jx,Hebeler:2013nza}. Although not directly measurable,
a determination of the bulk properties of infinite nuclear matter provides
valuable constraints on the equation of state (EOS) of dense neutron-rich
matter. Moreover, some of these critical parameters are known to be
strongly correlated to observables that may be directly measured. 
This fact provides a powerful bridge between observation, experiment, 
and theory. However, until very recently most of these correlations 
were inferred by comparing a large set of EDFs; see 
Ref.\,\cite{RocaMaza:2011pm} for a particularly illustrative example.
Although such an analysis provides critical insights into the 
\emph{systematic} errors associated with the biases and limitations 
of each model, it is essential that it be supplemented with a proper 
\emph{statistical} analysis. Indeed, such a covariance analysis represents 
the least biased and most reliable approach to uncover correlations among 
physical observables\,\cite{Reinhard:2010wz, Fattoyev:2011ns,Fattoyev:2012rm,
Reinhard:2012vw,Reinhard:2013fpa,Dobaczewski:2014jga,Piekarewicz:2014kza}. 

The paper has been organized as follows. Following this introduction, 
we outline the theoretical framework in Sec.\,\ref{Formalism}. We follow
closely the approach developed in Ref.\,\cite{Piekarewicz:2014kza} that
starts from a gaussian approximation to a suitably defined likelihood 
function. To demonstrate the power of the approach, we construct in
Sec.\,\ref{Results} a brand new functional (\emph{FSUGold\,2}) that is 
calibrated from the ground-state properties of finite nuclei, their isoscalar 
monopole response, and a maximum neutron star mass. Finally, we 
conclude with a summary and outlook in Sec.\,\ref{Conclusions}.

\section{Theoretical Framework}
\label{Formalism}

In this section we outline the theoretical framework required to 
accurately calibrate an energy density functional. The section 
itself is divided into three components. First, we introduce the
RMF model that will be used to compute all required nuclear
properties---from finite nuclei to neutron stars. Second, we 
develop, to our knowledge for the first time in the RMF
context, a transformation that links the model parameters 
to ``pseudo data'' given in the form of bulk properties of infinite
nuclear matter. Such a transformation enables us to implement
the optimization in the space of pseudo data, resulting in a 
more intuitive and more efficient approach. Finally, in the third 
and last subsection we describe details of the optimization 
procedure followed by a covariance analysis that is used to 
estimate both theoretical uncertainties and correlations among 
observables.

\subsection{Relativistic Mean Field Theory}

In the framework of the RMF theory, the basic degrees of freedom 
include nucleons (protons and neutrons), three ``mesons'', and the 
photon. The nucleons are the constituents of the nuclear many-body
system, which interact via the transfer of the force carriers, with the
various mesons conveying the strong force between the nucleons and
the photons mediating the additional electromagnetic force between 
the protons. The interactions among the particles can be depicted by 
an effective Lagrangian density of the following 
form\,\cite{Walecka:1974qa,Serot:1984ey, Mueller:1996pm, Serot:1997xg, 
Horowitz:2000xj}:
\begin{eqnarray}
{\mathscr L}_{\rm int} &=&
\bar\psi \left[g_{\rm s}\phi   \!-\! 
         \left(g_{\rm v}V_\mu  \!+\!
    \frac{g_{\rho}}{2}{\mbox{\boldmath $\tau$}}\cdot{\bf b}_{\mu} 
                               \!+\!    
    \frac{e}{2}(1\!+\!\tau_{3})A_{\mu}\right)\gamma^{\mu}
         \right]\psi \nonumber \\
                   &-& 
    \frac{\kappa}{3!} (g_{\rm s}\phi)^3 \!-\!
    \frac{\lambda}{4!}(g_{\rm s}\phi)^4 \!+\!
    \frac{\zeta}{4!}   g_{\rm v}^4(V_{\mu}V^\mu)^2 +
   \Lambda_{\rm v}\Big(g_{\rho}^{2}\,{\bf b}_{\mu}\cdot{\bf b}^{\mu}\Big)
                           \Big(g_{\rm v}^{2}V_{\nu}V^{\nu}\Big)\;,
 \label{LDensity}
\end{eqnarray}
where $\psi$ is the isodoublet nucleon field, $A_{\mu}$ is the photon
field, and $\phi$, $V_{\mu}$, and ${\bf b}_{\mu}$ represent the
isoscalar-scalar $\sigma$-, isoscalar-vector $\omega$-, and
isovector-vector $\rho$-meson field, respectively. The first line of
the above equation contains the conventional Yukawa couplings between
the nucleons and the mesons, while the second line includes some
nonlinear self and mixed interactions between the mesons.  In the
spirit of an effective field theory, one should incorporate all
possible meson interactions that are allowed by symmetry
considerations to a given order in a power-counting scheme. Moreover,
once the dimensionful meson fields have been properly scaled using
strong-interaction mass scales, the remaining dimensionless
coefficients of the effective Lagrangian should all be ``natural'',
namely, of order one (i.e., neither too small nor too
large)\,\cite{Furnstahl:1996wv,Furnstahl:1996zm,
Rusnak:1997dj,Furnstahl:1997hq,Kortelainen:2010dt}. However,
given the limited experimental database of nuclear observables, 
certain empirical coefficients---or linear combinations of 
them---may remain poorly constrained after the optimization
procedure. This results in ``unnatural''  coefficients that deviate 
significantly from unity. Therefore, in an effort to avoid this problem 
only a subset of nonlinear meson interactions is retained in the 
formalism. For instance, in the interaction Lagrangian density depicted
in Eq.\,(\ref{LDensity}), one only keeps the four meson interactions 
denoted by the coefficients: $\kappa$, $\lambda$, $\zeta$, and 
$\Lambda_{\rm v}$. In particular, these terms are found to have a
clear physical connection to various properties of the nuclear equation of 
state. Two of the isoscalar parameters, $\kappa$ and $\lambda$, 
were introduced by Boguta and Bodmer\,\cite{Boguta:1977xi} 
to reduce the nuclear incompressibility coefficient of symmetric nuclear
matter from an unreasonably large value in the original Walecka 
model\,\cite{Walecka:1974qa,Serot:1984ey} to one that can be made
consistent with measurements of giant monopole resonances in finite
nuclei. In turn, $\zeta$ may be used to efficiently tune the maximum 
neutron star mass without sacrificing the agreement with other well 
reproduced observables\,\cite{Mueller:1996pm}. Finally, 
$\Lambda_{\rm v}$ is highly sensitive to the density dependence of 
symmetry energy---and in particular to its slope at saturation 
density---which has important implications in the structure and
dynamics of neutron stars\,\cite{Horowitz:2000xj,Horowitz:2001ya,
Carriere:2002bx,Horowitz:2004yf}. 

With the Lagrangian density given in Eq.\,(\ref{LDensity}), one can 
derive the equation of motion for each of the constituent particles 
in the mean-field limit\,\cite{Todd:2003xs}. The nucleons satisfy
a Dirac equation in the presence of mean-field potentials having
Lorentz scalar and vector character. In turn, the various mesons 
satisfy nonlinear and inhomogeneous Klein-Gordon equations with 
the various nuclear densities acting as source terms. Lastly, the
photon obeys the Poisson equation with the proton density being
the relavant source term. Given that the nuclear densities act as sources 
for the meson fields and, in turn, the meson fields determine the 
mean-field potentials for the nucleons, the set of equations must 
be solved self-consistently. Once solved, these equations determine
the  ground-state properties of the nucleus of interest---such as its
total binding energy, single-nucleon energies and wave functions, 
distribution of meson fields, and density profiles.

The solution of the mean-field equations is simplified significantly
in the case of infinite nuclear matter, which we assume to be spatially 
uniform. For this uniform case, the meson fields are uniform (i.e., 
constant throughout space) and the nucleon orbitals are plane-wave 
Dirac spinors with medium-modified effective masses and energies. 
By forming the energy-momentum tensor in the mean-field 
approximation\,\cite{Serot:1984ey}, one can readily infer (in the 
rest frame of the fluid) the energy density and pressure of the system as a 
function of the conserved baryon density $\rho\!=\rho_{n}\!+\!\rho_{p}$ 
and the neutron-proton asymmetry
$\alpha\!\equiv\!(\rho_{n}\!-\!\rho_{p})/(\rho_{n}\!+\!\rho_{p})$. 
In particular, the energy per nucleon of the system may be expanded 
in even powers of $\alpha$. That is, 
\begin{equation}
  \frac{E}{A}(\rho,\alpha) -\!M \equiv {\cal E}(\rho,\alpha)
                          = {\cal E}_{\rm SNM}(\rho)
                          + \alpha^{2}{\cal S}(\rho)  
                          + {\cal O}(\alpha^{4}) \,,
 \label{EOS}
\end {equation}
where ${\cal E}_{\rm SNM}(\rho)\!=\!{\cal E}(\rho,\alpha\!\equiv\!0)$
is the energy per nucleon of symmetric nuclear matter (SNM) and the 
symmetry energy ${\cal S}(\rho)$ represents the first-order correction to 
the symmetric limit. Note that no odd powers of $\alpha$ appear as the
nuclear force is assumed to be isospin symmetric and electromagnetic
effects have been ``turned off''. Also note that, although model
dependent, to a very good approximation the symmetry energy has
a very intuitive interpretation: it represents the energy cost required
to convert symmetric nuclear matter into pure neutron matter (PNM):
\begin{equation}
 {\cal S}(\rho)\!\approx\!{\cal E}(\rho,\alpha\!=\!1) \!-\! 
 {\cal E}(\rho,\alpha\!=\!0) \;.
 \label{SymmE}
\end {equation}

It is also customary to characterize the behavior of both symmetric nuclear 
matter and the symmetry energy in terms of a few bulk parameters. To do
so, we perform a Taylor series expansion around nuclear matter saturation
density $\rhoz$. That is\,\cite{Piekarewicz:2008nh},
\begin{subequations}
\begin{align}
 & {\cal E}_{\rm SNM}(\rho) = \epsz + \frac{1}{2}Kx^{2}+\ldots ,\label{EandSa}\\
 & {\cal S}(\rho) = J + Lx + \frac{1}{2}K_{\rm sym}x^{2}+\ldots ,\label{EandSb}
\end{align} 
\label{EandS}
\end{subequations}
\!\!\!where $x\!=\!(\rho-\rhoz)\!/3\rhoz$ is a dimensionless parameter that
quantifies the deviations of the density from its value at saturation. Here
$\epsz$ and $K$ represent the energy per nucleon and the 
incompressibility coefficient of SNM; $J$ and 
$K_{\rm sym}$ are the corresponding quantities for the symmetry energy. 
However, unlike symmetric nuclear matter whose pressure vanishes at
$\rhoz$, the slope of the symmetry energy $L$ does not vanish at
saturation density. Indeed, assuming the validity of Eq.\,(\ref{SymmE}),
$L$ is directly proportional to the pressure of PNM ($P_{0}$) at saturation 
density, namely,
\begin{equation}
   P_{0} \approx \frac{1}{3}\rhoz L \;.
 \label{PvsL}
\end{equation}

Finally, one can go a step further and apply the above formalism to
neutron star matter, which we assume to consist of neutrons, protons,
electrons, and muons in $\beta$ equilibrium. Note that no ``exotic''
degrees of freedom---such as hyperons, meson condensates, or
quarks---are included in the formalism. At the densities at which
neutron star matter is uniform, electrons and muons may be treated as
relativistic Fermi gases that contribute to the total energy density and
pressure of the system. In $\beta$ equilibrium only the baryon density
needs to be specified, as the neutron-proton asymmetry is adjusted to
minimize the total energy density of the system. Given that uniform
neutron-rich matter is unstable against cluster formation, we
supplement our RMF predictions for the EOS with the standard
parametrization for the outer crust by Baym, Pethick, and
Sutherland\,\cite{Baym:1971pw}. Finally, we resort to a polytropic EOS
to interpolate between the solid outer crust and the uniform liquid
core\,\cite{Link:1999ca,Carriere:2002bx}. Given that the EOS is the only 
input required to solve the Tolman-Oppenheimer-Volkoff equation, 
one can predict a variety of neutron star properties that can then be 
compared against observation. Particularly relevant in this work will be 
the predictions for the maximum stellar mass and the radius of a 
``canonical'' 1.4\,$M_{\odot}$ neutron star.

\subsection{An Insightful Transformation}

The main goal of the present work is the accurate calibration of 
a relativistic EDF by relying exclusively on measured properties 
of finite nuclei and neutron stars. The fitting protocol requires
both the specification of a theoretical model and the selection 
of physical observables to constrain the fit. The conventional 
approach to the calibration of the EDF consists of first 
minimizing the objective function and then validating the model 
against observables not included in the fit. Traditionally, the 
optimization of the model is carried out in parameter space. That
is, one searches for those model parameters (e.g., $g_{\rm s},
g_{\rm v},\ldots$) that minimize the objective function. Given that 
the connection between the model parameters and our physical
intuition is tenuous at best, the searching algorithm often ends 
up wandering aimlessly in search of the minimum. A remarkable,
but little known, fact in the framework of the RMF theory is that
many of the model parameters can be expressed in terms of a 
few bulk properties of infinite nuclear matter\,\cite{Glendenning:2000}.
Although relatively new, it appears that such a transformation 
between the model parameters and the bulk properties of infinite 
nuclear matter (or ``pseudo data'') is better known in the case
of the non-relativistic Skyrme interaction\,\cite{Agrawal:2005ix,
Chen:2010qx,Kortelainen:2010hv}. To avoid interrupting the flow 
of the narrative, we only summarize here the central points of the
transformation. A detailed account of the transformation has been 
reserved to the appendix. 

For the Lagrangian density given in Eq.\,(\ref{LDensity}), we identify
five isoscalar ($g_{\rm s},g_{\rm v},\kappa,\lambda$, and $\zeta$) and
two isovector ($g_{\rho}$ and $\Lambda_{\rm v}$) parameters. Note that
in a mean-field approximation, the properties of infinite nuclear
matter are only sensitive to the combinations 
$g_{\rm s}^{2}/m_{\rm s}^{2}$, $g_{\rm v}^{2}/m_{\rm v}^{2}$, and 
$g_{\rho}^{2}/m_{\rho }^{2}$. The transformation starts in the isoscalar 
sector and links the first four isoscalar parameters listed above with 
four bulk properties of symmetric nuclear matter; these are the 
density $\rhoz$, the binding energy per nucleon $\epsz$, the effective 
nucleon mass $M^{\ast}$, and the incompressibility coefficient $K$---all 
evaluated at saturation density. The fact that the pressure of SNM vanishes 
at saturation density implies, through the Hugenholtz-van Hove theorem, 
that the energy per nucleon must equal the nucleon Fermi energy. This fact, 
together with the classical equation of motion for the vector field, is 
sufficient to determine $g_{\rm v}^{2}/m_{\rm v}^{2}$, for a given value of 
$\zeta$. Note that $\zeta$ will remain as a model parameter throughout 
the optimization. To determine the remaining three scalar parameters 
($g_{\rm s},\kappa,\lambda$) one requires three pieces of information. 
These are (a) the binding energy per nucleon at saturation, (b) the
classical equation of motion for the scalar field, and (c) the 
incompressibility coefficient. Although the algebraic manipulations
are involved, they ultimately yield a system of three simultaneous 
linear equations\,\cite{Glendenning:2000}. That is, the solution is 
\emph{unique}. Once the transformation has been completed in the
isoscalar sector, one may proceed to determine the two remaining 
(isovector) parameters $g_{\rho}^{2}/m_{\rho }^{2}$ and $\Lambda_{\rm v}$
in terms of the value of symmetry energy $J$ and its slope $L$ at
saturation density. This derivation---that to our knowledge has never
been published in the literature---benefits greatly from the fact that 
the symmetry energy has a relatively simple analytic 
form\,\cite{Horowitz:2001ya}; for further details see the appendix.

In summary, we have carried out for the first time a transformation 
between the model parameters defining the Lagrangian density and 
various bulk parameters of infinite nuclear matter. Assuming that the 
nucleon mass as well as
the masses of the two vector mesons in free space are fixed at their
experimental value, i.e., $M\!=\!939$\, MeV, $m_{\rm v}\!=\!782.5\,$MeV 
and $m_{\rho}\!=\!763\,$MeV, a point in an 8-dimensional Lagrangian 
parameter space may be written as follows: ${\bf q}\!=\!(m_{\rm s},
g_{\rm s}^{2}/m_{\rm s}^{2}, g_{\rm v}^{2}/m_{\rm v}^{2},g_{\rho}^{2}/m_{\rho}^{2},
\kappa,\lambda, \Lambda_{\rm v},\zeta)$. As already mentioned, in a 
mean-field approximation the bulk properties of infinite nuclear matter 
are only sensitive to the combination $g_{\rm s}^{2}/m_{\rm s}^{2}$.  Hence, 
the range of the intermediate-range attraction, expressed as the Compton 
wavelength of the scalar meson $r_{s}\!=\!\hbar c/m_{\rm s}c^{2}$, can only
be determined from the properties of finite nuclei, primarily from
their charge radii.  Moreover, given that most bulk properties of
infinite nuclear matter at saturation density depend weakly on the
value of $\zeta$\,\cite{Mueller:1996pm}, the value of $\zeta$ must be 
determined from observables sensitive to the high-density component 
of the EOS, such as the maximum neutron star mass. In this way, the 
transformation enables one to write a point in the space of pseudo data 
as: ${\bf p}\!=\!(m_{\rm s},\rhoz,\epsz,M^{\ast},K,J,L,\zeta)$. Note that the
very existence of such transformation allows us to perform the model
optimization in the space of pseudo data rather than in the Lagrangian
parameter space.

There are several advantages to represent a point in parameter space
in terms of ${\bf p}$ rather than ${\bf q}$. First, that a unique 
algebraic solution exists for the Lagrangian parameters in terms of 
bulk properties of nuclear matter makes the theory well defined.
Second, the parameters have evolved from abstract coupling constants 
to quantities with a precise physical meaning and with values narrowed 
down by experiment to a fairly small range. Thus, running the optimization
in the space of pseudo data increases significantly the efficiency of the 
searching algorithm. Finally, given that the fitting protocol relies exclusively
on experimental and observational data, the culmination of the optimization 
procedure provides bona-fide \emph{theoretical predictions} for all bulk 
properties with meaningful error bars. These predictions may be compared 
against other theoretical approaches that could provide a bridge between 
{\it ab initio} calculations and phenomenological approaches.
\vfill 

\subsection{Optimization and Covariance Analysis} 
The aim of the optimization procedure is to determine the
set of model parameters that minimizes the objective function,
or \emph{goodness-of-fit parameter} $\chisq$, that is defined 
as follows:
\begin{equation}
 \chisq({\bf p}) \equiv \sum_{n=1}^{N}
 \frac{\Big(\mc{O}_{n}^{\rm (th)}({\bf p})-\mc{O}_{n}^{\rm (exp)}\Big)^{2}}
 {\Delta\mc{O}_{n}^{2}} \,,
 \label{ChiSquare}
\end{equation}
where ${\bf p}\!=\!(p_{1},\ldots,p_{F})$ is a point in the
$F$-dimensional parameter space, $\mc{O}_{n}^{\rm (exp)}$ is the
measured experimental value of the $n$-th observable (out of a total
of $N$), and $\mc{O}_{n}^{\rm (th)}({\bf p})$ the corresponding
theoretical prediction. Although in principle the adopted error
$\Delta\mc{O}_{n}$ is associated with the experimental uncertainty, in
practice it must be supplemented by a ``theoretical''
contribution. The main reason for adding a theoretical error is that
the objective function is weighted by the error associated with each
observable: the smaller the error the larger the weight. Given that
certain observables, such as nuclear binding energies, are known with
enormous precision, the minimization of the objective function could
be biased by such observables.  However, it is important to
recognize that no universal protocol exists for the selection of
theoretical errors, although Ref.\,\cite{Dobaczewski:2014jga} provides
a useful guiding principle. Most of the formalism required for the use 
of \emph{ information and statistics} in theoretical nuclear physics 
may be found in\,\cite{Reinhard:2010wz,Fattoyev:2011ns,
Fattoyev:2012rm,Dobaczewski:2014jga,Piekarewicz:2014kza} and 
in references contained therein. In turn, most of the central ideas 
presented in those references are contained in the two excellent 
texts by Brandt\,\cite{Brandt:1999} and Bevington\,\cite{Bevington2003}.
However, in the interest of clarity we present a succinct summary
of the main concepts. 

A concept of great pedagogical significance and closely connected
to the objective function is the \emph{likelihood function}:
\begin{equation}
{\mc L}({\bf p}) = \mbox{\large{$e$}}^{-\frac{1}{2}\chi^2({\bf p})} \;.
 \label{Likelihood}
\end{equation}
Clearly, minimizing the objective function $\chisq({\bf p})$ is fully
equivalent to maximizing the likelihood function ${\mc L}({\bf p})$.
However, the great merit of the likelihood function is that it may be
regarded as a probability distribution. That is, given two arbitrary
parameter sets (or ``models'') $\psub{1}$ and $\psub{2}$, the
likelihood function provides the \emph{relative} probability that the
given models reproduce the given experimental data. In particular, the
optimal (or most likely) parameter set is the one that
maximizes the likelihood function. Using the probabilistic nature of
the likelihood function one can efficiently sample the full parameter
space via, for example, a standard Metropolis Monte Carlo
algorithm. Averages, variances, and correlation coefficients can then
be computed in a standard fashion. For example, if
$\{\psub{1},\psub{2},\ldots,\psub{M}\}$ represent the $M$ models
generated by the sampling algorithm, then the average of a generic 
observable $A$ is simply given by
\begin{equation}
 \langle A \rangle = \lim_{M\rightarrow\infty}
  \frac{1}{M}\!\sum_{m=1}^{M} A(\psub{m}) \,.
 \label{AvgA}
\end{equation}

Although the method of maximum likelihood along with a sampling 
algorithm is simple and insightful,  generating a large set of model 
parameters, except in a few simple cases, is highly impractical.
 Indeed, certain observables adopted in the fit, such
as giant monopole energies, are computationally expensive to evaluate.
For such cases one must resort to other methods to minimize
the objective function, so we rely on the well-known gaussian approximation
where the parameter exploration is limited to the immediate vicinity of the 
$\chisq$ minimum. Denoting by $\psub0$ the \emph{optimal} parameter 
set, the gaussian approximation consists of studying the small (quadratic) 
oscillations around the $\chisq$ minimum. That is,
\begin{equation}
  \chisq({\bf p}) \approx \chisq(\psub0) + \frac{1}{2}\sum_{i,j=1}^{F}
  ({\bf p}-\psub0)_{i} ({\bf p}-\psub0)_{j}
  \left(\frac{\partial^{2}\chisq}{\partial p_{i} \partial p_{j}}\right)_{\!\!0} 
  \equiv \chisq_{0} + {\bf x}^{T}{\hat{\mc M}}_{0}\,{\bf x} \;,
 \label{Taylor1}
\end{equation}
where we have introduced the following dimensionless scaled variables:
\begin{equation}
  x_{i} \equiv \frac{({\bf p}-\psub0)_{i}}{(\psub0)_{i}} \;.
 \label{xDef}
\end{equation}
Note that the behavior of the $\chisq$ function around its minimum value
is determined by the \emph{curvature matrix} $\hat{{\mc M}}_{0}$, whose 
matrix elements are defined in terms of its second derivatives evaluated 
at the optimal point. That is,
\begin{equation}
 ({\mc M}_{0})_{ij} \equiv \frac{1}{2}\left(\frac{\partial^2\chisq}
 {\partial x_{i}\partial x_{j}}\right)_{\!\!0} .
 \label{MMatrix}
\end{equation}

In this work we employ the \emph{Levenberg-Marquardt} 
method\,\cite{NumericalRecipes} to minimize the objective function.
Initially the algorithm uses the inverse Hessian method and then switches 
continuously to the method of steepest decent on its way toward the minimum. 
Furthermore, we take advantage of the fact that the objective function to be 
minimized is neither arbitrary nor totally unknown. Rather, it is defined directly 
in terms of the physical observables appearing in the definition of the objective 
function given in Eq.\,(\ref{ChiSquare}). This fact enables us to write the curvature 
matrix---which is essential for both the optimization and the covariance 
analysis---as follows:
\begin{equation}
 {\mc M}_{ij} = \sum_{n=1}^{N} \frac{1}{\Delta {\mc O}_{n}^{2}}
 \left[
 \left(\frac{\partial{\mc O}_{n}^{\rm (th)}}{\partial x_{i}}\right)\! 
 \left(\frac{\partial{\mc O}_{n}^{\rm (th)}}{\partial x_{j}}\right) +
 \Big( {\mc O}_{n}^{\rm (th)}\!-\!{\mc O}_{n}^{\rm (exp)}\Big)
 \left(\frac{\partial^{2}{\mc O}_{n}^{\rm (th)}}{\partial x_{i} \partial x_{j}}\right)
\right].
 \label{CurvMatrx0}
\end{equation}
Notice that $\big({\mc O}_{n}^{\rm (th)}\!-\!{\mc O}_{n}^{\rm (exp)}\big)$ 
in the above expression represents the difference between the experimental 
value and the theoretical prediction of a given observable. Assuming that 
the model is rich enough to reasonably describe the set of observables 
included in the fit, then this term should be small. Moreover, we may 
expect that such a deviation is not only small but also random in sign. 
Therefore, the contributions from all observables to the second 
term in Eq.\,(\ref{CurvMatrx0}) will tend to cancel each other and the
curvature matrix may be computed without ever having to evaluate any
second derivative of ${\mc O}_{n}^{\rm (th)}$. That is, in the linear
approximation one obtains\,\cite{Brandt:1999,Bevington2003}
\begin{equation}
 {\mc M}_{ij} \approx \sum_{n=1}^{N} \frac{1}{\Delta {\mc O}_{n}^{2}}
 \left(\frac{\partial{\mc O}_{n}^{\rm (th)}}{\partial x_{i}}\right)\! 
 \left(\frac{\partial{\mc O}_{n}^{\rm (th)}}{\partial x_{j}}\right).
 \label{CurvMatrx1}
\end{equation}
The Levenberg-Marquardt method along with this simplified expression
for the curvature matrix has been shown to be very stable and
efficient, and has become one of the standard routines for nonlinear
optimization\,\cite{NumericalRecipes}.

As mentioned earlier in the Introduction, the importance of including
theoretical uncertainties in the prediction of physical quantities is
gaining significant momentum. Knowledge of the curvature matrix is all
that is needed to compute any statistical quantity, at least in the
gaussian approximation.  For example, the \emph{covariance} between
any two observables $A$ and $B$ is given by
\begin{equation}
 {\rm cov}(A,B) = {\rm cov}(B,A) =
 \sum_{i,j=1}^{F}
 \left(\frac{\partial A}{\partial x_{i}}\right)_{\!\!0}
  \Sigma_{ij}
 \left(\frac{\partial B}{\partial x_{j}}\right)_{\!\!0} \,,
\label{CovAB}
\end{equation}
where the \emph{covariance matrix} $\hat{\Sigma}\!=\!\hat{{\mc M}}_{0}^{-1}$ 
is equal to the inverse of the curvature matrix evaluated at the optimal point. 
In the case in which $A\!=\!B$, this equation gives the \emph{variance} of $A$ 
which equals the square of its uncertainty. That is, 
${\rm cov}(A, A)\!\equiv\!{\rm var}(A)\!=\!\sigmasub{A}^{2}$.  Note 
that the theoretical errors ($\sigmasub{A}$) that will be reported in the
next section have been computed in precisely this manner. Finally,
given the covariance between $A$ and $B$ and their corresponding
variances, the \emph{Pearson product moment correlation coefficient}
(or simply the correlation coefficient) is given
by\,\cite{Brandt:1999}
\begin{equation}
  \rho(A,B) = \frac{{\rm cov}(A, B)}{\sigmasub{A}\sigmasub{B}} \,.
\label{CorrAB}
\end{equation}
In identifying a connection between two observables, the correlation
coefficient provides a unique opportunity to infer the value of an
observable that may not be accessible in either experiments or
observations. Moreover, the correlation coefficient has an intuitive
geometric interpretation. Suppose that a large number of $M$ values
for both $A$ and $B$ are generated according to the likelihood
function ${\mc L}$. Then, by defining the following two unit vectors
in $M$-dimensions
\begin{equation}
 {a}_{m} \equiv \frac{1}{\sqrt{M}}
 \left(\frac{A_{m}-\langle A \rangle}{\sigmasub{A}}\right)
 \;\;{\rm and}\;\; {b}_{m} \equiv \frac{1}{\sqrt{M}}
 \left(\frac{B_{m}-\langle B \rangle}{\sigmasub{B}}\right) ,
 \label{CosAB}
\end{equation}
the correlation coefficient becomes equal to the cosine of the angle
between these two unit vectors. That is,
\begin{equation}
  {\rho(A,B)}= \hat{a}\cdot\hat{b}
  \equiv\cos(\hat{a},\hat{b}) \;.
 \label{Corrab}
\end{equation}
In particular, a value of $\rho(A,B)\!=\!\pm1$ implies that the two
observables are fully correlated/anti-correlated, whereas a value of
$\rho(A,B)\!=\!0$ means that the observables are totally uncorrelated.
In the next section we will implement a covariance analysis to
estimate theoretical uncertainties (i.e., ``errors'') in the model
parameters, the fitting observables, as well as a variety of
observables that were not included in the calibration
procedure. Moreover, we will examine correlations between: (i)
observables, (ii) model parameters, and (iii) observables and model
parameters. All three sets of correlations are insightful and provide
complementary information on the strengths and weaknesses of the
model. In the first case, a strong correlation between two
experimentally accessible observables prevents redundancy. However, if
one of the observables is not accessible either experimentally or
observationally, a strong correlation provides a clear path for its
determination. In the case of correlations among model parameters the
situation is vastly different. Indeed, rather than suggesting
redundancy, a strong correlation between model parameters underscores
the need for both. Finally, correlations between observables and
model parameters reveal the sensitivity of the parameters to a
particular kind of physics. Relying on such a covariance analysis makes
possible to connect a variety of physical phenomena to the underlying
microscopic theory and provides a unique and powerful tool for
improving the quality of the models.

\section{Results}
\label{Results}

Having developed in the previous section most of the required formalism,
we are now in a position to implement the calibration of a new relativistic
energy density functional. We provide details that involve the optimization
and the subsequent covariance analysis. Whenever appropriate, we supplement 
our results with properly estimated theoretical errors. Moreover, in a few
instances, we provide correlation coefficients involving both observables
and model parameters. The new relativistic EDF may be regarded as an 
improvement over the almost one-decade old FSUGold 
parametrization\,\cite{Todd-Rutel:2005fa}. Accordingly, we name this 
newer version as \emph{FSUGold\,2}.

\subsection{FSUGold\,2: An accurately calibrated interaction
for finite nuclei and neutron stars}
\label{FSUGold2}

Based on the relativistic Lagrangian density given in Eq.\,(\ref{LDensity}),
there are a total of 11 model parameters: seven coupling constants, one
nucleon mass, and three meson masses. The mass of the nucleon will be
fixed at its free space value of 
$M\!=\!(M_{p}\!+\!M_{n})/2\!\approx\!939\,{\rm MeV}$. Given the effective
character of the theory, the three meson masses should in principle be
treated as model parameters that should be determined by the fitting
procedure. However, we have found---as many others have found
before us---that with the exception of the scalar meson, the masses 
of the two vector mesons ($m_{\rm v}$ and $m_{\rho}$) may be fixed near
their experimental values: $m_{\rm v}\!\approx\!782.5\,{\rm MeV}$ and
$m_{\rho}\!\approx\!763\,{\rm MeV}$. Note that the mass of the scalar
meson controls the range of the scalar attraction and is therefore critical 
for an accurate reproduction of charge radii\,\cite{Serot:1984ey}. As
mentioned earlier, having fixed the masses of the vector mesons, the
transformation between model parameters {\bf q} and pseudo data 
{\bf p} may be represented as follows:
${\bf q}\!=\!(m_{\rm s},g_{\rm s}^{2}/m_{\rm s}^{2}, g_{\rm v}^{2}/m_{\rm v}^{2},
g_{\rho}^{2}/m_{\rho}^{2},\kappa,\lambda, \Lambda_{\rm v},\zeta)
\!\leftrightarrow\!{\bf p}\!=\!(m_{\rm s},\rhoz,\epsz,M^{\ast},K,J,L,\zeta)$.
In essence, the objective function $\chisq({\bf p})$ is a function of the 
pseudo data, but the theoretical predictions depend on the model
parameters {\bf q}. The transformation outlined in the appendix 
uniquely determines {\bf p} in terms of {\bf q}, and {\it vice versa}. 

Having defined the parameters that must be optimized, we must now
introduce the experimental and observational data that will be used 
to constrain the fit. The fitting observables that we use in the
optimization include (a) binding energies, (b) charge radii, and (c)
giant monopole resonance (GMR) of semi- and doubly-magic nuclei 
across the nuclear chart, and (d) the maximum neutron star mass 
observed up to date. Note that all these observables are genuine 
experimental or observational quantities; no properties of infinite
nuclear matter are incorporated in the definition of the objective
function. The ground-state properties and collective 
excitations of finite nuclei are effective in constraining the EOS of 
nuclear matter around saturation density with small to moderate
values of the neutron-proton (i.e., isospin) asymmetry. On the other 
hand, neutron star properties---such as the maximum neutron star
mass---may be used to constrain the high-density component of the 
EOS of neutron-rich matter. We believe that no laboratory experiment 
may constrain the EOS of cold, fully catalyzed, nuclear matter at high 
densities. 

One of the greatest challenges involved in the definition of the 
$\chisq$ function introduced in Eq.\,(\ref{ChiSquare}) is the 
choice of  errors $\Delta{\mc O}_{n}$ associated with each 
observable ${\mc O}_{n}$. Given the remarkable precision that 
has been achieved in measuring binding energies and charge 
radii, the $\chisq$ function would be dominated by the terms 
associated with these two sets of observables if we naively
adopt their associated experimental errors. Although the 
optimization could still be carried out in such a case, 
binding energies and charge radii would be well reproduced 
at the expense of all remaining observables. Therefore, in order 
to mitigate this deficiency, one should manipulate the errors in 
such a way that the relative weights of all observables be 
commensurate with each other. By necessity, this implies 
some ``trial and error'' as there is no clear choice for the optimal 
protocol\,\cite{Dobaczewski:2014jga}. The choice of error for
each observable adopted in the fit is discussed below.

Once the objective function has been properly defined by specifying 
a theoretical model and a set of observables with properly defined
errors, the Levenberg-Marquardt method was used to obtain the 
optimal set of parameters 
${\bf p}\!=\!(m_{\rm s},\rhoz,\epsz,M^{\ast},K,J,L,\zeta)$. In turn, the 
model parameters {\bf q} may be obtained from the transformation 
outlined in the appendix. The resulting set of model parameters for 
the newly built functional FSUGold\,2 (or ``\FSUt'' for short) are 
displayed in Table\,\ref{Table1}. Also shown for comparison are 
two canonical sets of parameters, NL3\,\cite{Lalazissis:1996rd} and 
FSUGold (or ``FSU'' for short)\,\cite{Todd-Rutel:2005fa}. Given that
the EOS for symmetric nuclear matter and the symmetry energy are
both stiff in the case of NL3 and both soft for FSU, such a comparison
is very informative. However, when comparing these models, one should 
keep in mind that different models are calibrated using different sets of 
observables and associated errors. This introduces some inherent biases 
into the models that ultimately become an important source of systematic 
errors.

\begin{widetext}
\begin{center}
\begin{table}[h]
\begin{tabular}{|l||c|c|c|c|c|c|c|c|c|c|}
\hline\rule{0pt}{2.5ex}   
\!\!Model   &  $m_{\rm s}$  &  $m_{\rm v}$  &  $m_{\rho}$  &  $g_{\rm s}^2$  &  $g_{\rm v}^2$  &  $g_{\rho}^2$  
                  &  $\kappa$       &  $\lambda$    &  $\zeta$       &   $\Lambda_{\rm v}$  \\
\hline
\hline
NL3        & 508.194 & 782.501 & 763.000 & 104.3871 & 165.5854 &  79.6000 & 3.8599  & $-$0.015905  & 0.0000  & 0.000000  \\
FSU         & 491.500 & 782.500 & 763.000 & 112.1996 & 204.5469 & 138.4701 & 1.4203  & $+$0.023762 & 0.0600  & 0.030000  \\
\FSUt      & 497.479 & 782.500 & 763.000 & 108.0943 & 183.7893 &  80.4656 & 3.0029  & $-$0.000533  & 0.0256  & 0.000823   \\ 
\hline
\end{tabular}
\caption{Model parameters for the newly optimized FSUGold\,2 relativistic EDF along with two accurately 
calibrated RMF models: NL3\,\cite{Lalazissis:1996rd} and FSUGold\,\cite{Todd-Rutel:2005fa}. The 
parameter $\kappa$ and the meson masses $m_{\rm s}$, $m_{\rm v}$, and $m_{\rho}$ are all given in MeV. 
The nucleon mass has been fixed at  $M\!=\!939$ MeV in all the models.}
\label{Table1}
\end{table}
\end{center}
\end{widetext}

\subsection{Ground-State Properties}
\label{GSP}

We start this section by displaying in Table\,\ref{Table2} 
ground-state binding energies and charge radii for all the 
nuclei involved in the optimization.
Experimental data for these observables were obtained from the 
latest atomic mass evaluation\,\cite{Wang:2012} and charge radii
compilation\,\cite{Angeli:2013}, respectively. In turn, the errors
assigned to the binding energies and charge radii are 0.1\% and
0.2\%, respectively. As mentioned earlier, these adopted errors 
are several orders of magnitude larger than the quoted experimental
uncertainties\,\cite{Wang:2012,Angeli:2013}. Only by doing so
one can prevent the optimization from being dominated by these two
ground-state observables. Also displayed in Table\,\ref{Table2}
are the theoretical predictions from all three models. Note that the
theoretical errors predicted by {\FSUt} (of about one part in a thousand) 
are too small to be displayed in the table. Also note that the quoted theoretical 
value for the charge radius was obtained by adding to the extracted nuclear 
point proton radius the intrinsic charge radius of the proton 
$r_{\!\rm p}\!=\!0.8783(86)\,{\rm fm}$\,\cite{Angeli:2013}. That is, 
$R_{\rm ch}\!=\!({R_{\rm p}^{2} + r_{\!\rm p}^{2}})^{1/2}$. We can see that both
the binding energies and charge radii are very well reproduced by all
the models. In the particular case of \FSUt, with the exception of the
charge radius of ${}^{16}$O, the discrepancy relative to experiment is
less than 0.5\%.  The slightly larger than 1\% deviation in the case
of ${}^{16}$O should not come as a surprise, as with only 16 nucleons
oxygen barely qualifies as a ``mean-field'' nucleus. It is important
to stress that neither binding energies nor charge radii have a
significant impact on the stiffness of the EOS. Indeed, NL3 and FSU
predict significantly different stiffness for the EOS (see below) yet 
they both reproduce fairly accurately the experimental results for 
these two observables.

\begin{widetext}
\begin{center}
\begin{table}[h]
\begin{tabular}{|l||c|c|c|c|c|}
\hline\rule{0pt}{2.5ex} 
\!\!Nucleus            &   Observable   &   Experiment   &   NL3   &   FSU   &   \FSUt   \\  
 \hline
 \hline\rule{0pt}{2.5ex} 
 \!\!${}^{16}$O  &   $B/A$   &   7.98    &   8.06   &   7.98   &   8.00   \\
                        &   $R_{\rm ch}$   &   2.70   &   2.75   &   2.71   &   2.73   \\ 
 ${}^{40}$Ca      &   $B/A$   &   8.55   &   8.56   &   8.54   &   8.54   \\
                        &   $R_{\rm ch}$   &   3.48   &   3.49   &   3.45   &   3.47   \\   
 ${}^{48}$Ca      &   $B/A$   &   8.67   &   8.66   &   8.58   &   8.63   \\
                        &   $R_{\rm ch}$   &   3.48   &   3.49   &   3.48   &   3.47   \\   
 ${}^{68}$Ni      &   $B/A$   &   8.68   &   8.71   &   8.66   &   8.69   \\
                        &   $R_{\rm ch}$   &   ---   &   3.88   &   3.88   &   3.86   \\   
 ${}^{90}$Zr      &   $B/A$   &   8.71   &   8.70   &   8.68   &   8.69   \\
                        &   $R_{\rm ch}$   &   4.27   &   4.28   &   4.27   &   4.26   \\   
${}^{100}$Sn    &   $B/A$   &   8.25   &   8.30   &   8.24   &   8.28   \\
                        &   $R_{\rm ch}$   &   ---   &   4.48   &   4.48   &   4.47   \\   
${}^{116}$Sn    &   $B/A$   &   8.52   &   8.50   &   8.50   &   8.49   \\
                        &   $R_{\rm ch}$   &   4.63   &   4.63   &   4.63   &   4.61   \\   
${}^{132}$Sn    &   $B/A$   &   8.36   &   8.38   &   8.34   &   8.36   \\
                        &   $R_{\rm ch}$   &   4.71   &   4.72   &   4.74   &   4.71   \\   
${}^{144}$Sm   &   $B/A$   &   8.30   &   8.32   &   8.32   &   8.31   \\
                        &   $R_{\rm ch}$   &   4.95   &   4.96   &   4.96   &   4.94   \\   
${}^{208}$Pb    &   $B/A$   &   7.87   &   7.90   &   7.89   &   7.88   \\
                        &   $R_{\rm ch}$   &   5.50   &   5.53   &   5.54   &   5.51   \\   
\hline
\end{tabular}
\caption{Experimental data for the binding energy per nucleon 
(in MeV)\,\cite{Wang:2012} and charge radii (in fm)\,\cite{Angeli:2013}
for all the nuclei involved in the optimization. Also displayed are the 
theoretical results obtained with NL3\,\cite{Lalazissis:1996rd},
FSUGold\,\cite{Todd-Rutel:2005fa}, and FSUGold2.}
\label{Table2}
\end{table}
\end{center}
\end{widetext}

\subsection{Giant Monopole Resonances}
\label{GMR}

\begin{figure}[h]
\vspace{-0.05in}
\includegraphics[width=0.5\columnwidth,angle=0]{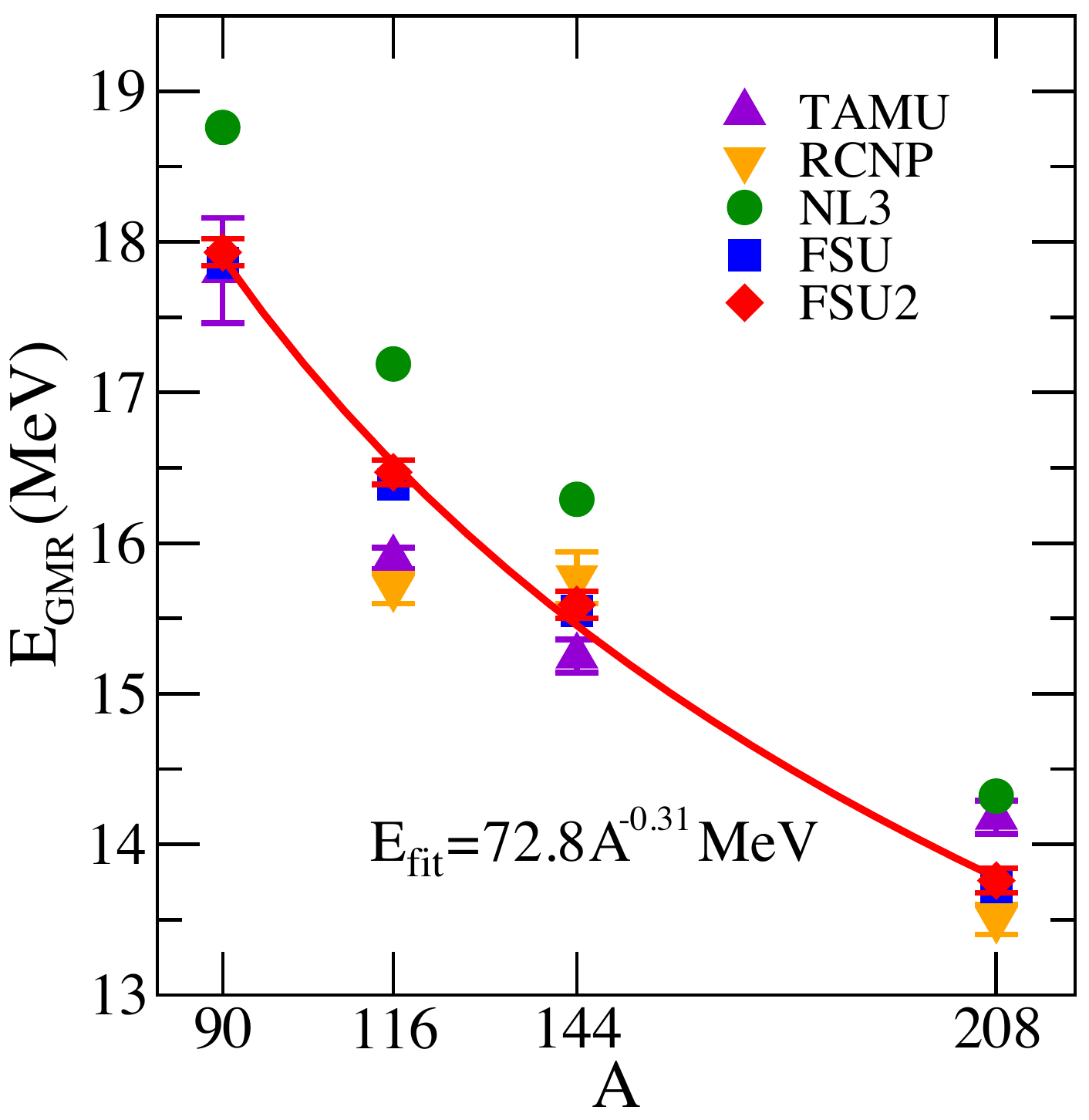}
\caption{(Color online) Constrained giant monopole energies for ${}^{90}$Zr,
${}^{116}$Sn, ${}^{144}$Sm, and ${}^{208}$Pb. Experimental data 
were obtained from experiments carried out at 
TAMU\,\cite{Youngblood:1999} and 
RCNP\,\cite{Uchida:2004bs,Li:2007bp,Li:2010kfa,Patel:2013uyt,
PatelPC:2013}. Theoretical predictions are presented for 
NL3\,\cite{Lalazissis:1996rd}, FSUGold\,\cite{Todd-Rutel:2005fa}, 
and FSUGold\,2 supplemented with theoretical errors. The red 
solid line represents a best fit to the FSUGold\,2 predictions of 
the form $E_{\rm fit}\!=\!72.8A^{-0.31}$\,MeV.}
\label{Fig1}
\end{figure}

In optimizing the FSUGold\,2 functional, we have also incorporated 
GMR energies for ${}^{90}$Zr, ${}^{116}$Sn, ${}^{144}$Sm, and ${}^{208}$Pb.
In Table\,\ref{Table3} we display \emph{constrained} GMR energies 
$E_{\rm GMR}\!=\!\sqrt{m_{1}/m_{-1}}$ extracted from measurements 
at the Texas A\&M University (TAMU) cyclotron facility\,\cite{Youngblood:1999} 
and at the Research Center for Nuclear Physics (RCNP) in Osaka,
Japan\,\cite{Uchida:2004bs,Li:2007bp, Li:2010kfa,Patel:2013uyt,
PatelPC:2013}. Here $m_{1}$ and $m_{-1}$ are suitable moments of the 
strength distribution that represent the energy weighted and inverse 
energy weighted sums, respectively. The theoretical results listed on
the table were obtained by following the constrained RMF formalism 
developed in Ref.\,\cite{Chen:2013tca}. The same information has 
been displayed in graphical form in Fig.\,\ref{Fig1}. Note that the red 
solid line in the figure represents a fit to the {\FSUt} predictions of the 
form $E_{\rm fit}\!=\!72.8A^{-0.31}$\,MeV; this compares favorably
against the macroscopic expectation of
$E_{\rm GMR}\!\approx\!80A^{-1/3}$\,MeV\,\cite{Ring:2004,Harakeh:2001}.
We find both intriguing and unsettling that the TAMU and RCNP 
data---particularly for ${}^{208}$Pb---are inconsistent with each other.
Given the critical nature of this information, we trust that the discrepancy 
may be resolved in the near future. In the meantime, and to account for the 
experimental discrepancy, we have adopted slightly larger errors in the 
optimization of the functional, namely, 2\% for ${}^{90}$Zr and 1\% for 
the rest. 

Our results indicate that the predictions from FSU and {\FSUt} are
compatible with each other. This is consistent with the notion that
GMR energies probe the incompressibility coefficient of SNM, that is,
$K$ (see Table\,\ref{Table4}). Moreover, with the
exception of ${}^{116}$Sn, both FSU and {\FSUt} reproduce the
experimental data, although they both favor the smaller RCNP 
measurement in the case of ${}^{208}$Pb. Note that the answer 
to the question of \emph{Why is Tin so soft?''}\,
\cite{Li:2007bp,Li:2010kfa,Piekarewicz:2008nh} continues to 
elude us to this day\,\cite{Piekarewicz:2007us,Sagawa:2007sp,
Avdeenkov:2008bi,Piekarewicz:2009gb,Cao:2012dt,Vesely:2012dw,
Piekarewicz:2013bea,Chen:2013jsa}. By the same token NL3, with 
a significantly larger value of $K$ than both FSU and \FSUt, 
overestimates the experimental data---except in the case of the 
TAMU data for ${}^{208}$Pb\,\cite{Piekarewicz:2003br}. Although in 
principle GMR energies of neutron-rich nuclei probe the incompressibility 
coefficient of \emph{neutron-rich matter}\,\cite{Piekarewicz:2008nh}, in 
practice the neutron-proton asymmetry for these nuclei is simply too 
small to provide any meaningful constraint on the density dependence 
of the symmetry energy. This is the main reason behind the agreement 
between FSU and {\FSUt}, even though they predict radically different 
values for the slope of the symmetry energy $L$ (see Table\,\ref{Table4}).

\begin{table}[h]
  \begin{tabular}{|c|c|c|c|c|c|c|}
    \hline\rule{0pt}{2.5ex} 
    \!\!Nucleus &  TAMU & RCNP & NL3 & FSU  & \FSUt \\
    \hline
    \hline\rule{0pt}{2.5ex}     
    \!\!${}^{90}$Zr   & $17.81\pm0.35$ & --- & $18.76$ & $17.86$ & $17.93\pm0.09$ \\    
       ${}^{116}$Sn   & $15.90\pm0.07$  & $15.70\pm0.10$ & $17.19$ & $16.39$ & $16.47\pm0.08$\\    
       ${}^{144}$Sm  & $15.25\pm0.11$ & $15.77\pm0.17$ & $16.29$ & $15.55$ & $15.59\pm0.09$\\  
        ${}^{208}$Pb  & $14.18\pm0.11$ & $13.50\pm0.10$ & $14.32$ & $13.72$ & $13.76\pm0.08$ \\
    \hline
  \end{tabular}
 \caption{Constrained energies $E_{\rm GMR}\!=\!\sqrt{m_{1}/m_{-1}}$ (in MeV) for the giant monopole resonance
  in ${}^{90}$Zr, ${}^{116}$Sn, ${}^{144}$Sm, and ${}^{208}$Pb obtained from experiments at
  TAMU\,\cite{Youngblood:1999}  and RCNP\,\cite{Uchida:2004bs,Li:2007bp,Li:2010kfa,
  Patel:2013uyt,PatelPC:2013}. Theoretical results were obtained by following the 
  constrained RMF formalism developed in Ref.\,\cite{Chen:2013tca}.}
  \label{Table3}
 \end{table}

\subsection{Neutron Star Structure}
\label{NS}

\begin{figure}[ht]
\vspace{-0.05in}
\includegraphics[width=0.55\columnwidth,angle=0]{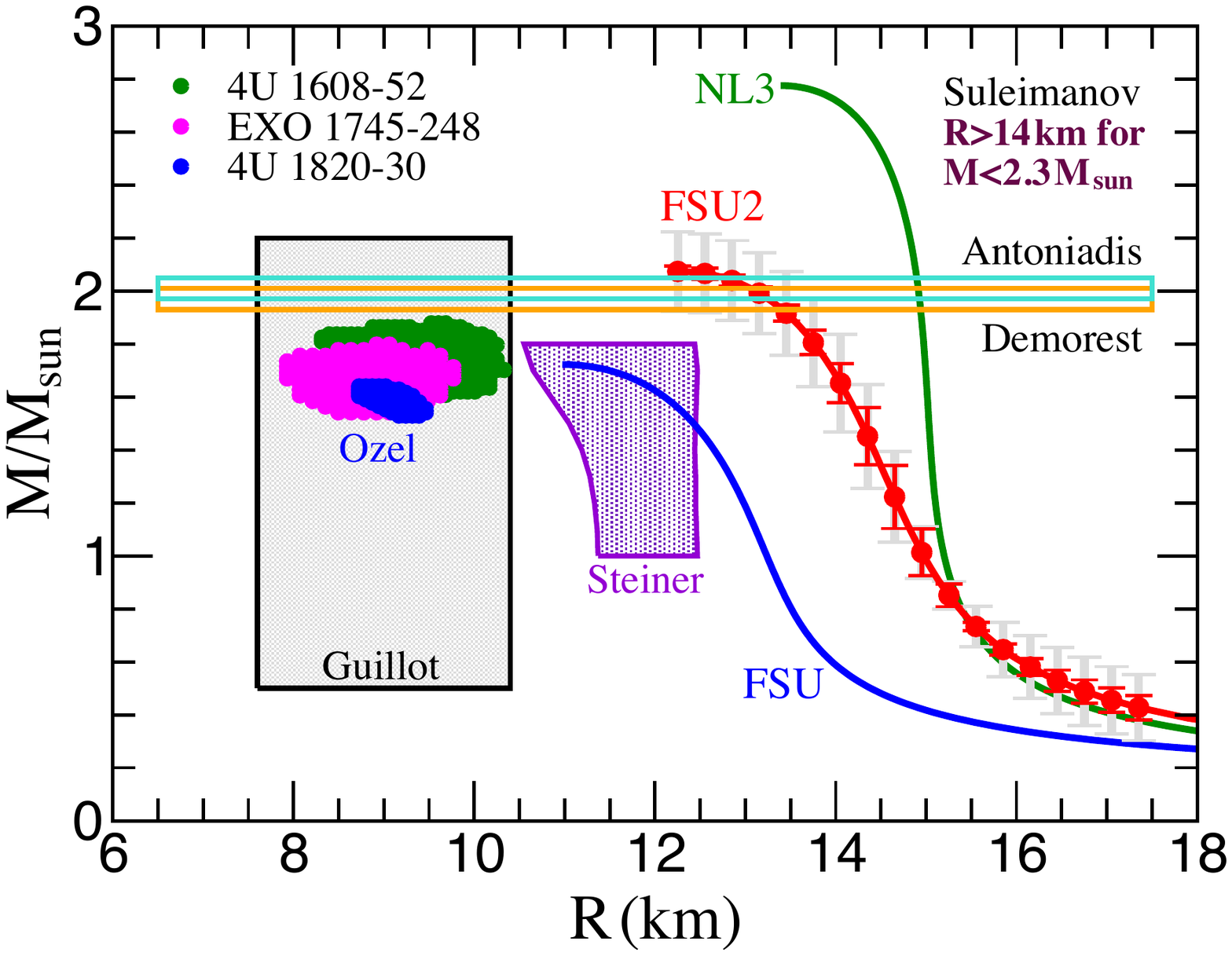}
\caption{(Color online) Mass-vs-radius relation predicted by 
the three models considered in the text: NL3\,\cite{Lalazissis:1996rd}, 
FSUGold~\cite{Todd-Rutel:2005fa}, and FSUGold\,2. Also shown 
are recent observational constraints on neutron star 
masses\,\cite{Demorest:2010bx,Antoniadis:2013pzd} and 
radii\,\cite{Ozel:2010fw,Steiner:2010fz,Suleimanov:2010th,
Guillot:2013wu}. The FSUGold\,2 results are supplemented 
with two sets of theoretical errors: one (red) in which the 
maximum neutron star mass was included in the calibration 
of the functional and the other (grey) estimated also using 
FSUGold\,2, but with the impact of the maximum neutron 
star mass removed from the curvature matrix, as explained 
in the text.} 
\label{Fig2}
\end{figure}

The last observable that was included in the calibration of the 
new {\FSUt} functional was the maximum neutron star 
mass. Displayed in Fig.\,\ref{Fig2} with horizontal bars are the 
two most massive, and accurately measured, neutron 
stars observed to date\,\cite{Demorest:2010bx,Antoniadis:2013pzd}. 
Clearly, those observations place stringent constraints on the 
high-density component of the EOS, as models that predict
limiting masses below $2\,M_{\odot}$---such as 
FSUGold---must be stiffened accordingly. Therefore, for the
optimization of the {\FSUt} functional, we have adopted 
a value of $M_{\rm max}\!=\!2.10\,M_{\odot}$ with a 
relatively small 1$\%$ error. If an even larger mass is
discovered in the future, such new input can be easily
incorporated into the calibration procedure.

Also displayed in Fig.\,\ref{Fig2} are theoretical predictions for 
the mass-vs-radius (M-R) relations for the three models considered 
in the text. As alluded earlier, with a stiff EOS,  NL3 predicts large 
stellar radii and a maximum neutron star mass of almost 
$3\,M_{\odot}$. In contrast, FSUGold with a relatively soft EOS 
predicts smaller values for both. The new FSUGold\,2 functional
displays a M-R relation that appears intermediate between NL3
and FSUGold. In particular, after the optimization we obtain a 
maximum stellar mass of $M_{\rm max}\!=\!(2.07\!\pm0.02)\,M_{\odot}$, 
safely within the bounds set by observation. Given the large impact 
that the quartic vector coupling constant $\zeta$ has on the EOS 
at high densities, these results are totally consistent with our
expectations (see Table\,\ref{Table1}). On the other hand, stellar 
radii seem to be controlled by the density dependence of the 
symmetry energy in the immediate vicinity of saturation 
density\,\cite{Lattimer:2006xb}. Thus models with large 
values of $L$ tend to predict neutron stars with large 
radii\,\cite{Horowitz:2001ya}. This is the main reason behind 
the relatively uniform ``shift'' between FSU and {\FSUt} (see
Table\,\ref{Table4}.) It is important to realize that no observable
highly sensitive to the density dependence of the symmetry 
energy, such as the neutron-skin thickness of ${}^{208}$Pb
or stellar radii, was used in the calibration of \FSUt. Such a choice
was deliberate, as at present there are no stringent experimental
or observational constraints on the isovector sector of the nuclear
density functional. Although the Lead Radius Experiment (``PREX'') 
at the Jefferson Laboratory has provided the first model-independent 
evidence on the existence of a neutron-rich skin in 
${}^{208}$Pb\,\cite{Abrahamyan:2012gp,Horowitz:2012tj}, the 
determination came with an error that is too large to impose any
significant constraint. That is,
\begin{equation}
 R_{\rm skin}^{208}\!=\!{0.33}^{+0.16}_{-0.18}\,{\rm fm} \,. 
 \label{Rskin208}
\end{equation}
In the case of stellar radii, the present situation is highly unsatisfactory 
as further illustrated in Fig.\,\ref{Fig2}. First, an initial attempt by \"Ozel 
and collaborators to determine simultaneously the mass and radius of 
three x-ray bursters resulted in predictions for stellar radii between 
$8$ and $\!10$\,km\,\cite{Ozel:2010fw}. Shortly after, Steiner et al.
supplemented \"Ozel's study with three additional neutron stars and 
concluded that systematic uncertainties make the most probable 
radii lie in the 11-12 km region\,\cite{Steiner:2010fz}. However, even 
this more conservative estimate has been put into question by Suleimanov 
and collaborators, who suggested a \emph{lower limit} on the stellar radius 
of 14\,km on neutron stars with masses below 
$2.3\,M_{\odot}$\,\cite{Suleimanov:2010th}. That is, three different analyses 
of (mostly) the same sources seem to differ in their conclusions by more
than 5\,km in the radius of a typical neutron star. Recognizing this unfortunate
situation and the many challenges posed by the study of x-ray bursters, 
Guillot and collaborators concentrated on the determination of stellar radii by 
studying five quiescent low mass x-ray binaries (qLMXB) in globular 
clusters. By clearly and explicitly stating all their assumptions, some
of them apparently not without controversy\,\cite{Lattimer:2013hma},
Guillot et al. were able to determine a rather small neutron star 
radius of\,\cite{Guillot:2013wu}: 
\begin{equation}
 R_{0}=9.1^{+1.3}_{-1.5}\,{\rm km}\,.
\label{RGuilliot}
\end{equation}
Note that this value represents the ``common'' radius of \emph{all} neutron stars, 
a critical assumption in the analysis of Ref.\,\cite{Guillot:2013wu}. Based on such 
a confusing state of affairs concerning stellar radii, we have then decided against 
including such information into the calibration of FSUGold\,2. 

Of course, this does not prevent us from offering {\FSUt} predictions for stellar radii, 
as displayed in Fig.\,\ref{Fig2}. In particular, we find the radius of a ``canonical" 
$1.4\,M_{\odot}$ neutron star to be $R_{1.4}\!=\!(14.42\pm0.26)\,{\rm km}$. Note 
that the large stellar radii predicted by {\FSUt} satisfy the constraint set by 
Suleimanov et al., but only for neutron stars with masses below 
$\simeq\!1.8\,M_{\odot}$. Moreover, we should mention that although no assumptions 
on either the neutron-skin thickness of ${}^{208}$Pb or stellar radii were incorporated 
into the calibration of FSUGold\,2, a manuscript that contemplates various 
possible scenarios is in preparation.

We close this section by exploring the impact of the 
maximum neutron star mass $M_{\rm max}$ on the estimation of errors.
Recall that $M_{\rm max}$ is the only observable included in the calibration
that is sensitive to the high-density component of the EOS. Although we 
preserve the same optimal set of parameters as FSUGold\,2, we assess 
the impact of $M_{\rm max}$ by removing its contribution to the curvature 
matrix. This invariably results in some flattening of certain directions in 
parameter space. In particular, the additional set of theoretical errors 
displayed (in grey) in Fig.\,\ref{Fig2} were estimated in precisely this manner. 
As expected, the (grey) theoretical ``error band'' becomes significantly thicker 
when the maximum neutron star mass is removed from consideration. 
Particularly, the uncertainty in $M_{\rm max}$ is increased significantly 
from 0.02 to 0.15 $M_{\odot}$ and the error in the radius of a 1.4 $M_{\odot}$ 
neutron star becomes almost three times as large. It is clear that the inclusion 
of $M_{\rm max}$ in the calibration of the functional is essential to constrain 
the high-density component of the EOS. Indeed, we believe that no terrestrial 
experiment can reliably constrain the EOS of neutron star matter.

\subsection{Predictions and Correlations}
\label{Predictions}

With the exception of stellar radii, up until now we have concentrated 
on physical observables that were included in the calibration of the 
density functional. In the present
section we shift our attention to genuine theoretical predictions of 
a variety of observables that were not incorporated into the fit. We 
start by displaying in Table\,\ref{Table4} a few bulk properties of 
nuclear matter at saturation density. These properties are of critical 
importance in constraining the EOS of neutron-rich matter and the 
covariance analysis developed here serves to determine whether the 
physical observables incorporated into the fit impose meaningful 
constraints on these properties. We note that the four isoscalar
properties that characterize the EOS of SNM
(i.e., $\rhoz$, $\epsz$, $M^{\ast}/M$, and $K$) are all accurately
determined (to about 1\%). In particular, we attribute the small 
theoretical error associated with the incompressibility coefficient 
($K\!=\!238.0\pm2.8$\,MeV) to the inclusion of GMR energies
into the calibration of FSUGold\,2. Moreover, we find good 
agreement with the isoscalar predictions from both NL3 and FSU 
except in the case of $K$ for NL3. 

\begin{widetext}
\begin{center}
\begin{table}[h]
\begin{tabular}{| l | | c | c | c | c | c | c |}
 \hline\rule{0pt}{2.5ex} 
\!\!Model  &  $\rhoz ({\rm fm}^{-3})$ &  $\epsz\,{\rm (MeV)}$ 
                 &  $M^{\ast}\!/M$ & $K\,{\rm (MeV)}$ 
                 &  $J\,{\rm (MeV)}$ & $L\,{\rm (MeV)}$ \\  
\hline
 \hline\rule{0pt}{2.5ex} 
\!\!NL3    &  0.1481 &   $-$16.24   &   0.595   &   271.5    &   37.28   &   118.2   \\
     FSU     &   0.1484 &   $-$16.30  &   0.610   &   230.0    &   32.59   & \phantom{0}60.5   \\
   \FSUt    &   $0.1505\pm0.0007$  &   $-16.28\pm0.02$ &  $0.593\pm0.004$ 
                &  $238.0\pm2.8$   &   $37.62\pm1.11$   &   $112.8\pm16.1$   \\ 
\hline
\end{tabular}
\caption{Bulk properties of nuclear matter as predicted by the three models
considered in the text: NL3\,\cite{Lalazissis:1996rd}, FSUGold\,\cite{Todd-Rutel:2005fa}, 
and FSUGold\,2 supplemented with theoretical errors.}
\label{Table4}
\end{table}
\end{center}
\end{widetext}

However, the situation is radically different in the isovector sector.
Although the ground-state properties of neutron-rich nuclei, such 
as ${}^{48}$Ca, ${}^{132}$Sn, ${}^{208}$Pb, are able to constrain the
value of the symmetry energy $J$ to about 3\%, its slope $L$ remains
poorly constrained (to about 15\%). We attribute this situation to the
lack of well measured isovector observables, such as the neutron skin
of heavy nuclei. We reiterate that when relativistic models of the kind 
given in Eq.\,(\ref{LDensity}) do not incorporate strong isovector constraints, 
they tend to generate a fairly stiff symmetry energy. However, note that 
although the density dependence of the symmetry energy remains rather 
uncertain, all three models considered in the table seem to agree on its
value at a sub-saturation density of 
$\widetilde{\rhoz}\!\approx\!0.10\,{\rm fm}^{-3}\!\approx\!2\rhoz/3$.
Indeed, according to Eq.\,(\ref{EandSb}) one obtains
\begin{equation}
\tilde{J} \equiv {\cal S}(\widetilde{\rhoz}) \approx 
 J + L \frac{(\widetilde{\rhoz}\!-\!\rhoz)}{3\rhoz} 
 \approx \left(J\!-\!\frac{L}{9}\right) \approx (25\!-\!26)\,{\rm MeV} \,.
\end{equation}
This point has been emphasized repeatedly in various
references\,\cite{Farine:1978,Brown:2000,Horowitz:2000xj,
Furnstahl:2001un,Zhang:2013wna,Brown:2013mga,Horowitz:2014bja}.  
That is, the above correlation between $J$ and $L$ that emerges from 
the masses of neutron-rich nuclei determines rather accurately the 
value of the symmetry energy at an average between the central nuclear 
density $\rhoz$ and some characteristic density at the surface. Clearly, 
more information is required to determine uniquely both $J$ and $L$.

\begin{figure}[ht]
\vspace{-0.05in}
\includegraphics[height=8cm,angle=0]{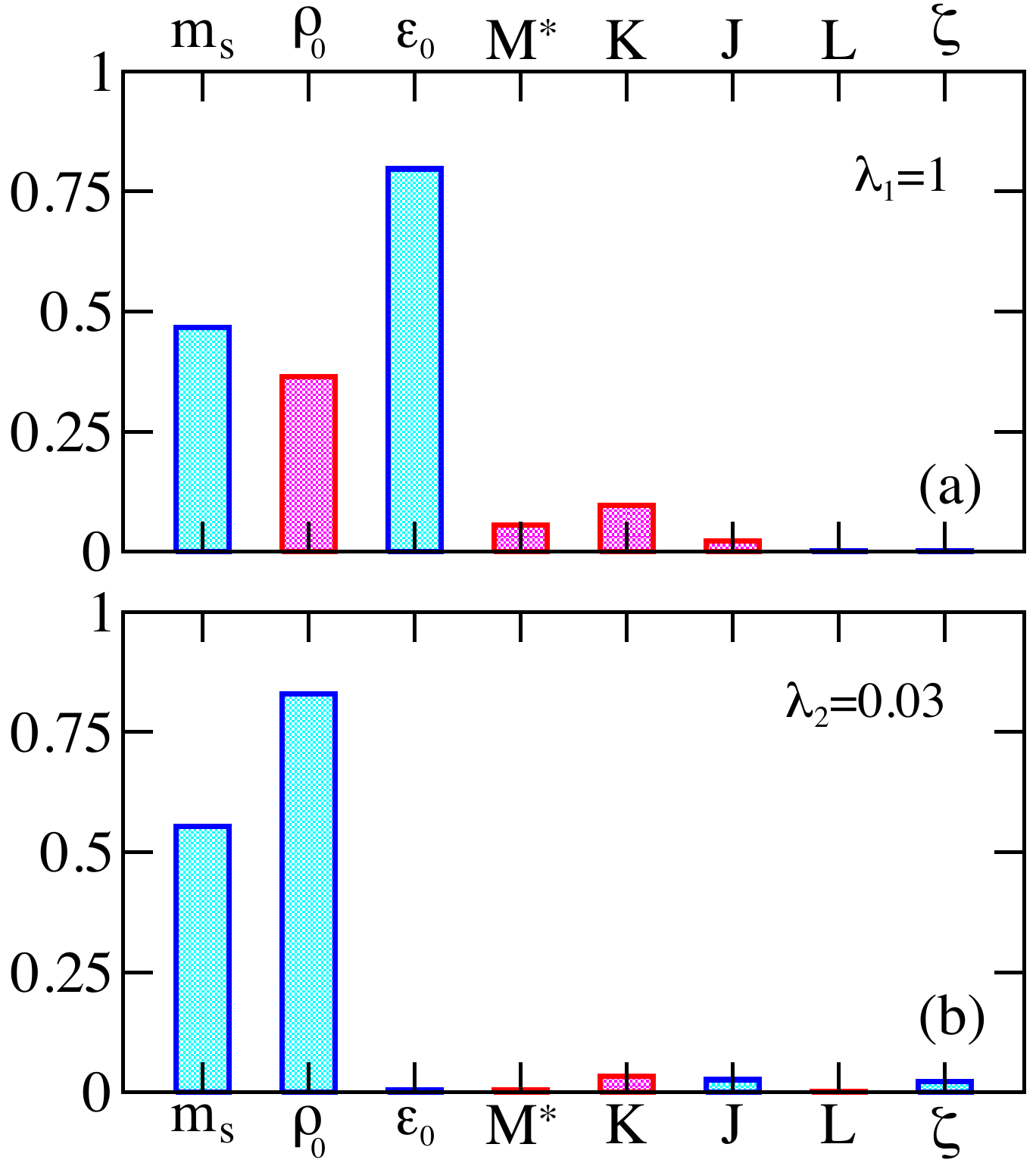}
 \hspace{2pt}
\includegraphics[height=8cm,angle=0]{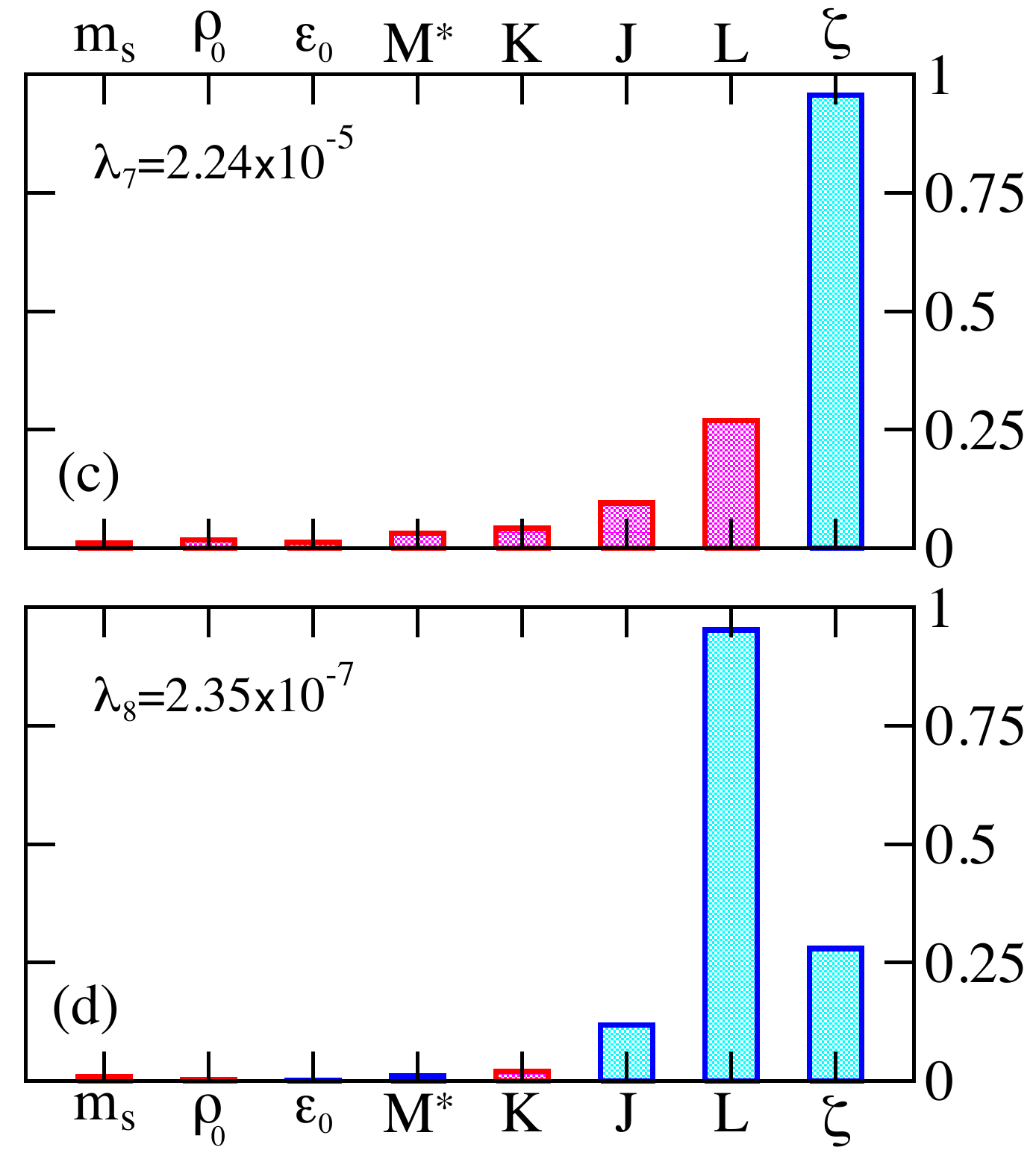}
\caption{(Color online) Amplitude decomposition of the eigenvectors 
of the curvature matrix corresponding to the two largest [(a) and (b)]
and the two smallest [(c) and (d)] eigenvalues, with the largest 
eigenvalue normalized to one. The two different colors (blue and red) 
indicate that the amplitudes contribute with opposite signs.}
\label{Fig3}
\end{figure}

The large theoretical error attached to the prediction of $L$ suggests
that relatively large changes in $L$ from its average value 
produce a mild deterioration in the quality of the fit. This indicates 
that there are directions in the model space that are relatively ``soft'' 
or ``flat''. A highly intuitive way to illustrate this effect is to diagonalize 
the $8\!\times\!8$ curvature matrix $\hat{\cal M}$ defined in 
Eq.\,(\ref{CurvMatrx1}), which then 
becomes effectively a \emph{small-oscillations} problem. In particular, 
each eigenvalue $\lambda_{i}$ of $\hat{\cal M}$ controls the 
deterioration in the quality of the fit as one moves along a direction 
defined by its corresponding eigenvector\,\cite{Fattoyev:2011ns}. 
A ``flat'' direction, characterized by a small eigenvalue $\lambda_{i}$, 
involves a particular linear combination of parameters that is poorly 
constrained by the choice of observables included in the calibration 
of the functional. We depict such a behavior in Fig.\,\ref{Fig3} by
displaying the components of four of the eigenvectors along  the 
original directions in the pseudo-parameter space. Note 
that we have considered only those eigenvectors having the two 
largest and two smallest eigenvalues, with the largest eigenvalue 
being normalized arbitrarily to one. The blue and red rectangles serve 
to indicate component having opposite signs. The eigenvectors 
associated with the two largest eigenvalues determine the two
stiffest directions in parameter space. Small departures from the
minimum along those two eigenvectors result in a rapid 
deterioration of the quality of the fit. Perhaps not surprisingly 
given the importance of ground-state energies and charge radii (see
Table\,\ref{Table2}), the scalar-meson mass, the saturation density,
and the binding energy per nucleon are the most accurately determined
parameters. Note that the scalar mass was determined with a small
0.3\% theoretical error: $m_{\rm s}\!=\!(497.479\pm1.492)\,{\rm MeV}$. 
In stark contrast, the eigenvalues associated with the two softest
directions are down by five to seven orders of magnitude. These two
directions are represented by almost ``pure'' eigenvectors with 
amplitudes in excess of 0.95 along the original $\zeta$ and $L$
directions, respectively. The reason for $L$ to remain poorly
constrained has already been discussed earlier. However, the
reason for $\zeta$ to remain largely undetermined is slightly
more subtle. From the work of M\"uller and Serot it is already
known that the value of $\zeta$ is insensitive to ground-state
properties of finite nuclei that probe densities near 
nuclear matter saturation\,\cite{Mueller:1996pm}. On 
the other hand, M\"uller and Serot showed that the value of 
$\zeta$ may be efficiently tuned to control the high-density
component of the EOS, and ultimately the maximum 
neutron star mass $M_{\rm max}$. Naively then, one would have 
expected a better constraint on $\zeta$ from the inclusion of 
$M_{\rm max}$ in the calibration of the functional. We believe that 
the poor determination of $\zeta$ may be attributed to the large 
value of $L$ suggested by FSUGold\,2 (see Table\,\ref{Table4}).
Indeed, when $L$ is small as in the case of FSUGold, the
high-density component of the EOS needs to be stiffened to account
for the existence of massive stars. And this can be efficiently 
done by only tuning $\zeta$, as was done in 
Ref.\,\cite{Fattoyev:2010mx}. However, if the symmetry energy 
is already stiff and no isovector constraints are available, then it 
appears that only a linear combination of $L$ and $\zeta$ can be 
constrained. This analysis reinforces the urgent need for well 
measured isovector observables.

\begin{figure}[ht]
\vspace{-0.05in}
\includegraphics[height=8.5cm,angle=0]{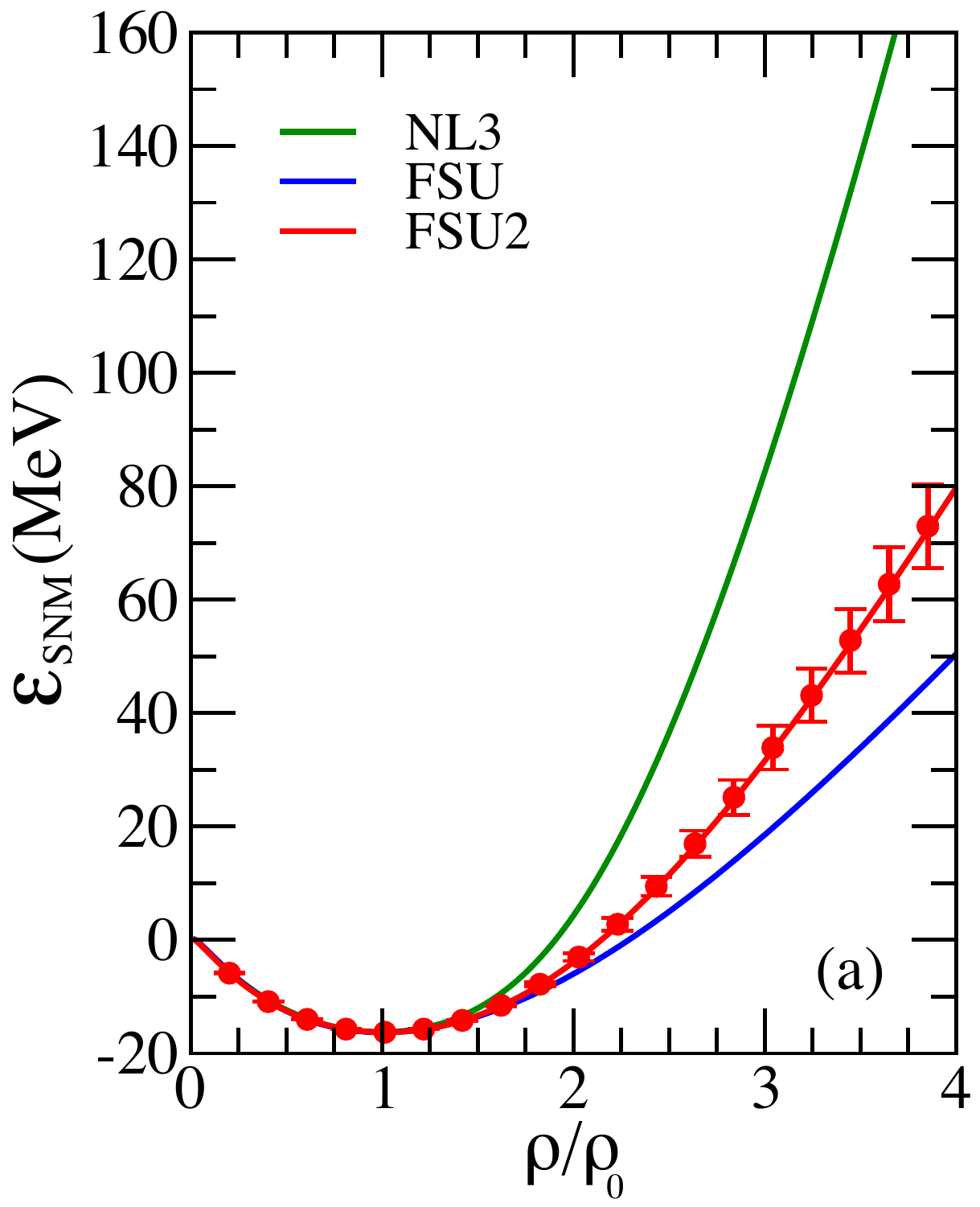}
 \hspace{2pt}
\includegraphics[height=8.5cm,angle=0]{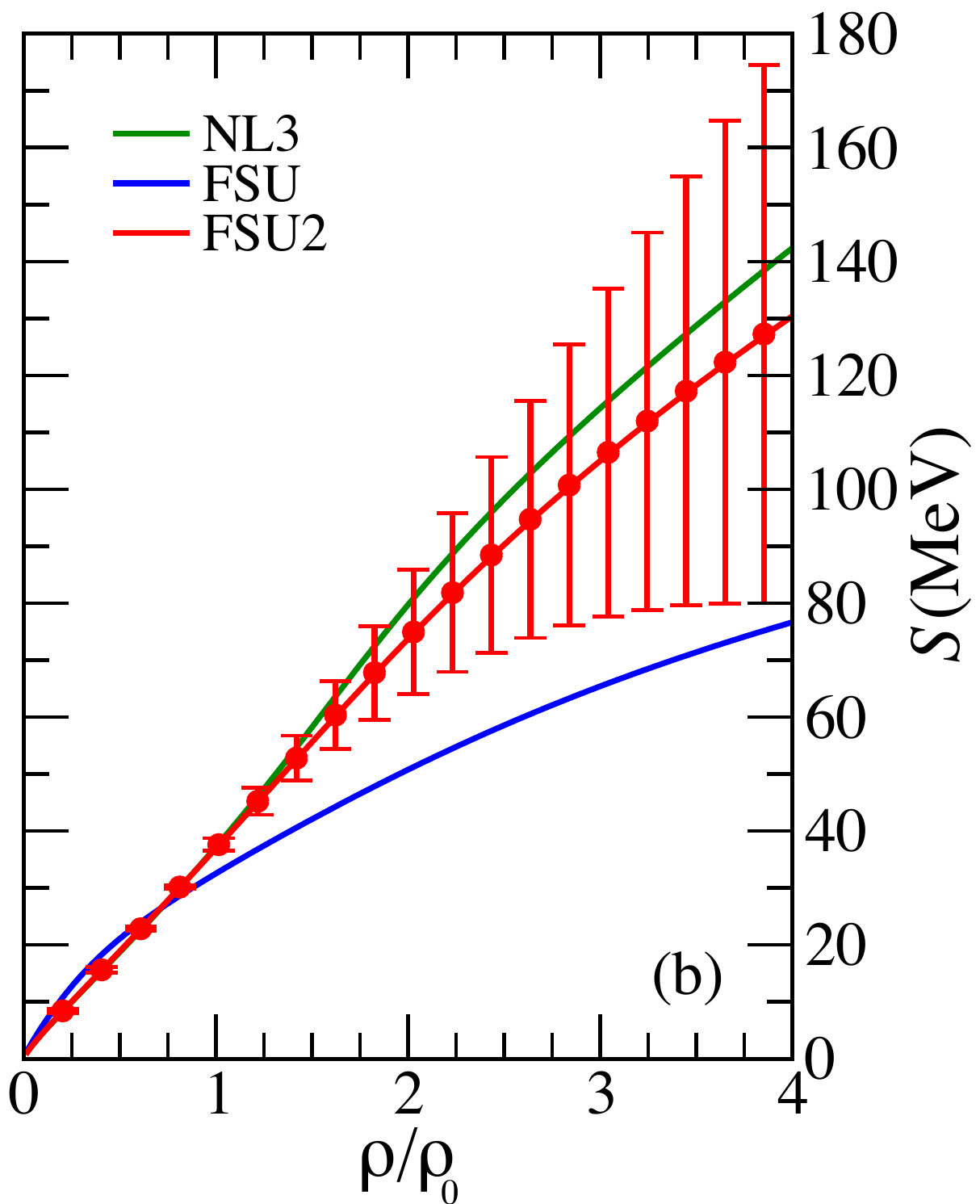}
\caption{(Color online) (a) Binding energy per nucleon of symmetric
nuclear matter and (b) symmetry energy as a function of density in 
units of nuclear matter saturation density 
$\rho_{{}_{0}}\!= 0.148\,{\rm fm}^{-3}$. Predictions are included from
the three models discussed in the text: NL3\,\cite{Lalazissis:1996rd}, 
FSUGold\,\cite{Todd-Rutel:2005fa}, and FSUGold\,2 supplemented 
with theoretical errors.} 
\label{Fig4}
\end{figure}

A more detailed view of the behavior of infinite nuclear matter
is given in Fig.\,\ref{Fig4} where predictions for the EOS of SNM
(left panel) and the symmetry energy (right panel) are displayed for
the three RMF models considered in this work.  Due to the inclusion of
GMR energies into the calibration of FSUGold\,2, the incompressibility
coefficient was fairly accurately determined (see Table\,\ref{Table4})
and this, in turn, generates small theoretical errors on the EOS up to
2-3 times saturation density. The larger theoretical uncertainty with
increasing density is a reflection of the inability of ground-state
properties and GMR energies to constrain the high-density behavior of
the EOS. In principle, the inclusion of a maximum neutron star mass
$M_{\rm max}$ into the fit should have served to constrain the EOS at
high density.  However, given that the symmetry energy is stiff (see
right-hand panel) one can satisfy the $M_{\rm max}$ constraint without
imposing stringent limits on the EOS of SNM at high densities.
The situation appears to be radically different in the case of the
symmetry energy, as the model has lost its predicability at densities
only slightly above saturation density. Although we expect to mitigate
this situation once strong isovector observables, such as neutron
skins and stellar radii, are incorporated into the calibration of the
density functional, our results underscore the importance of including
theoretical uncertainties. Whereas the symmetry energy predicted by
FSUGold\,2 is stiff at saturation density, it is consistent at the
1$\sigma$ level with a symmetry energy almost as soft as FSUGold 
and as stiff as (or even stiffer than) NL3 at high densities. The impact
of a stiff symmetry energy on the neutron-skin thickness of all the
nuclei used in the calibration procedure is displayed in Table\,\ref{Table5}.
These results help to reinforce the recent claim that at present there is 
no compelling reason to rule out models with large neutron 
skins\,\cite{Fattoyev:2013yaa}.
We close this part of the discussion with a brief comment on the EOS of pure 
neutron matter. Given that the EOS of PNM may be approximated as that of 
SNM plus the symmetry energy, the EOS of PNM at low densities for
FSUGold\,2 strongly resembles the one for NL3. Although PNM is not
experimentally accessible, there are important theoretical constraints
that have emerged from the universal behavior of dilute Fermi gases in
the unitary limit\,\cite{Schwenk:2005ka}. As already mentioned, without 
additional isovector constraints the symmetry energy predicted by RMF 
models tends to be fairly stiff. Therefore, whereas FSUGold is consistent 
with most theoretical
constraints\,\cite{Schwenk:2005ka,Gezerlis:2007fs,
Gezerlis:2009iw,Vidana:2009is}, both FSUGold\,2 and NL3 are not.

\begin{widetext}
\begin{center}
\begin{table}[h]
\begin{tabular}{|l||r|r|r|}
\hline\rule{0pt}{2.5ex} 
\!\!Nucleus    &   \hfill NL3 \hfill  &  \hfill  FSU \hfill  &  \hfill\hfill  FSU\,2 \hfill\hfill    \\  
 \hline
 \hline\rule{0pt}{2.5ex} 
 \!\!${}^{16}$O  &   $-0.028$   &   $-0.029$   &   $-0.028\pm0.005$   \\
 ${}^{40}$Ca      &   $-0.049$   &   $-0.051$   &   $-0.050\pm0.004$   \\
 ${}^{48}$Ca      &         0.226   &        0.197   &      $0.232\pm0.008$   \\
 ${}^{68}$Ni      &          0.261   &        0.211   &      $0.268\pm0.010$   \\ 
 ${}^{90}$Zr      &          0.114   &        0.088   &      $0.117\pm0.008$   \\
${}^{100}$Sn     &   $-0.076$   &   $-0.080$   &    $-0.077\pm0.008$   \\
${}^{116}$Sn     &         0.167   &         0.122   &      $0.172\pm0.011$   \\
${}^{132}$Sn     &         0.346   &         0.271   &      $0.354\pm0.019$   \\
${}^{144}$Sm    &         0.145   &         0.103   &      $0.149\pm0.011$   \\
${}^{208}$Pb     &         0.278   &         0.207   &      $0.287\pm0.020$   \\
\hline
\end{tabular}
\caption{Predictions for the neutron skins, $R_{\rm skin}\!\equiv\!R_{\rm n}\!-\!R_{\rm p}$,
(in fm) of all the nuclei included in the calibration procedure for NL3\,\cite{Lalazissis:1996rd},
FSUGold\,\cite{Todd-Rutel:2005fa}, and FSUGold\,2 supplemented with theoretical error bars.}
\label{Table5}
\end{table}
\end{center}
\end{widetext}

\begin{figure}[ht]
\vspace{-0.05in}
\includegraphics[width=0.55\columnwidth,angle=0]{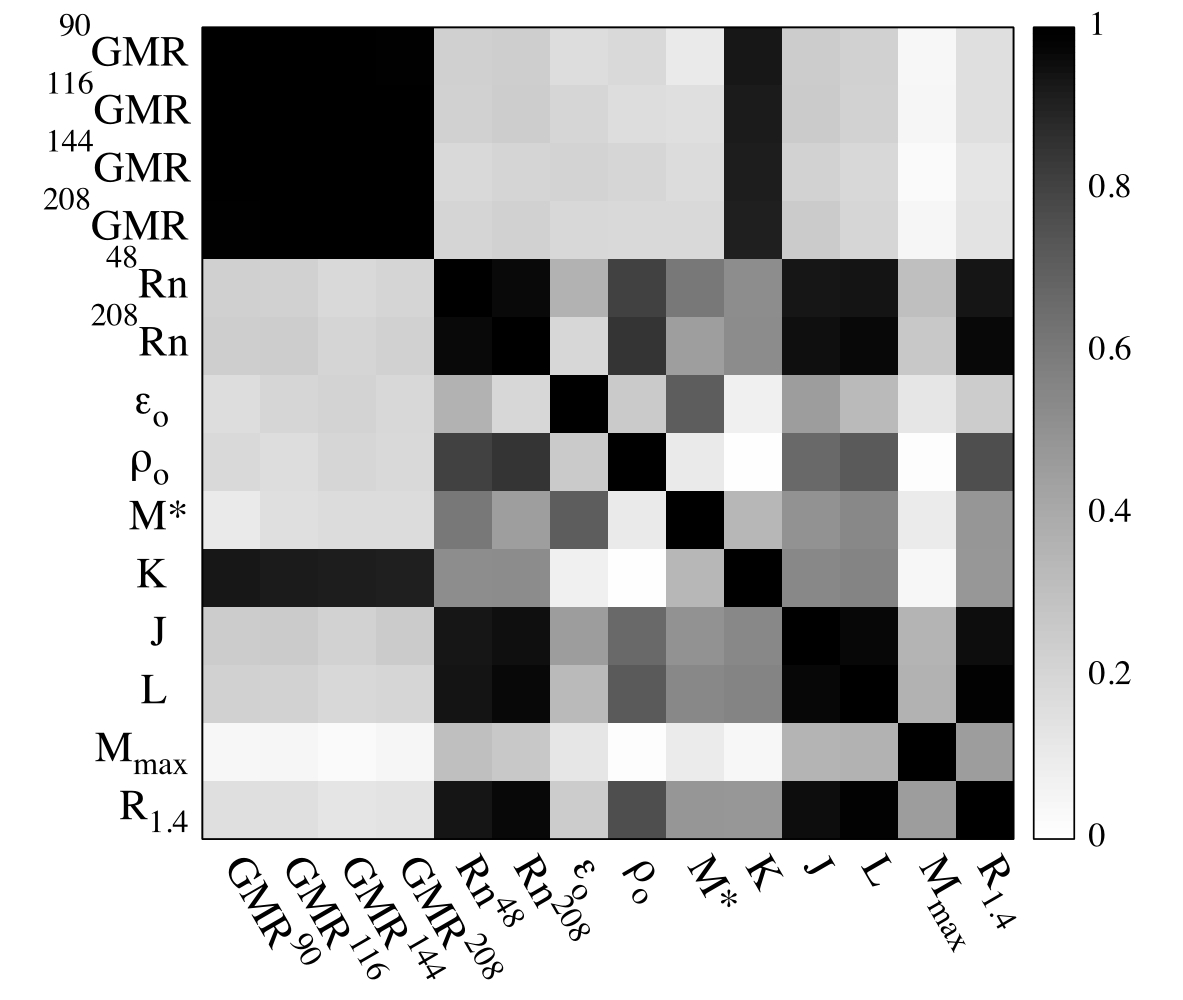}
\caption{Correlation coefficients (in absolute value) depicted in 
graphical form for a representative set of observables. The set 
includes four GMR energies (for ${}^{90}$Zr, ${}^{116}$Sn, 
${}^{144}$Sm, and ${}^{208}$Pb), two neutron radii (for 
${}^{48}$Ca and ${}^{208}$Pb), several bulk properties of 
nuclear matter ($\varepsilon_{{}_{0}}$, $\rho_{{}_{0}}$, 
$M^{\ast}$, $K$, $J$, and $L$), and two neutron star 
observables (the maximum mass $M_{\rm max}$ and 
the radius of a 1.4\,$M_{\odot}$ neutron star $R_{1.4}$).}
\label{Fig5}
\end{figure}

So far we have discussed the results from the optimization and the
theoretical errors associated to a large number of physical
quantities. We now turn the discussion to the important topic of
correlations [see Eqs.\,(\ref{CovAB}) and\,(\ref{CorrAB})].  We
start in Fig.\,\ref{Fig5} by displaying correlation coefficients in
graphical form for various physical quantities. From these, only GMR
energies and the maximum neutron star mass were included in the
calibration procedure. As anticipated, we find a strong correlation of
the GMR energies to the nuclear incompressibility coefficient $K$,
verifying the age-old idea of extracting a fundamental parameter of
the EOS from laboratory measurements of the breathing mode. To our
knowledge, this is the first time that GMR energies are directly
incorporated into the calibration of a relativistic EDF. In the case
of the two fundamental parameters of the symmetry energy $J$ and 
$L$, we observe a strong correlation with ``size'' parameters, specifically
with the neutron radius of ${}^{48}$Ca and ${}^{208}$Pb, as well as
with the radius of ``canonical'' 1.4\,$M_{\odot}$ neutron star. The
sensitivity of the size parameters to $L$ has a clear physical
underpinning. In the particular case of a nucleus, surface tension
favors the formation of a spherical drop of uniform equilibrium
density. However, if the nucleus has a significant neutron excess, it
may be energetically advantageous to move some of these neutrons from
the center of the nucleus to the dilute surface where the symmetry
energy is reduced. In particular, if the slope $L$ is large, then this
reduction is significant so it becomes favorable to move most of the
excess neutrons to the surface, thereby creating a thick neutron
skin\,\cite{Horowitz:2014bja}. And given that the same pressure that
pushes against surface tension in a nucleus pushes against gravity in
a neutron star, the larger the value of $L$ the larger the stellar
radius\,\cite{Horowitz:2000xj, Horowitz:2001ya}. However, whereas
the neutron skin is sensitive to the pressure around the saturation 
density, the neutron star radius also depends on the pressure at higher 
densities. This weakens slightly the correlation between the stellar radius 
and the neutron radius of the nucleus. Nevertheless, that a correlation 
between systems that differ in size by 18 orders of magnitude exists, is 
remarkable indeed.
Moreover, the correlation between the neutron-skin thickness of
${}^{208}$Pb and the radius of low mass neutron stars is even 
stronger\,\cite{Carriere:2002bx,Fattoyev:2012rm}. This suggests 
how a laboratory measurement may place a significant constraint
on an astronomical object, and {\it vice versa}. This example clearly 
illustrates the power of the covariance analysis.

\begin{figure}[ht]
\vspace{-0.05in}
\includegraphics[width=0.55\columnwidth,angle=0]{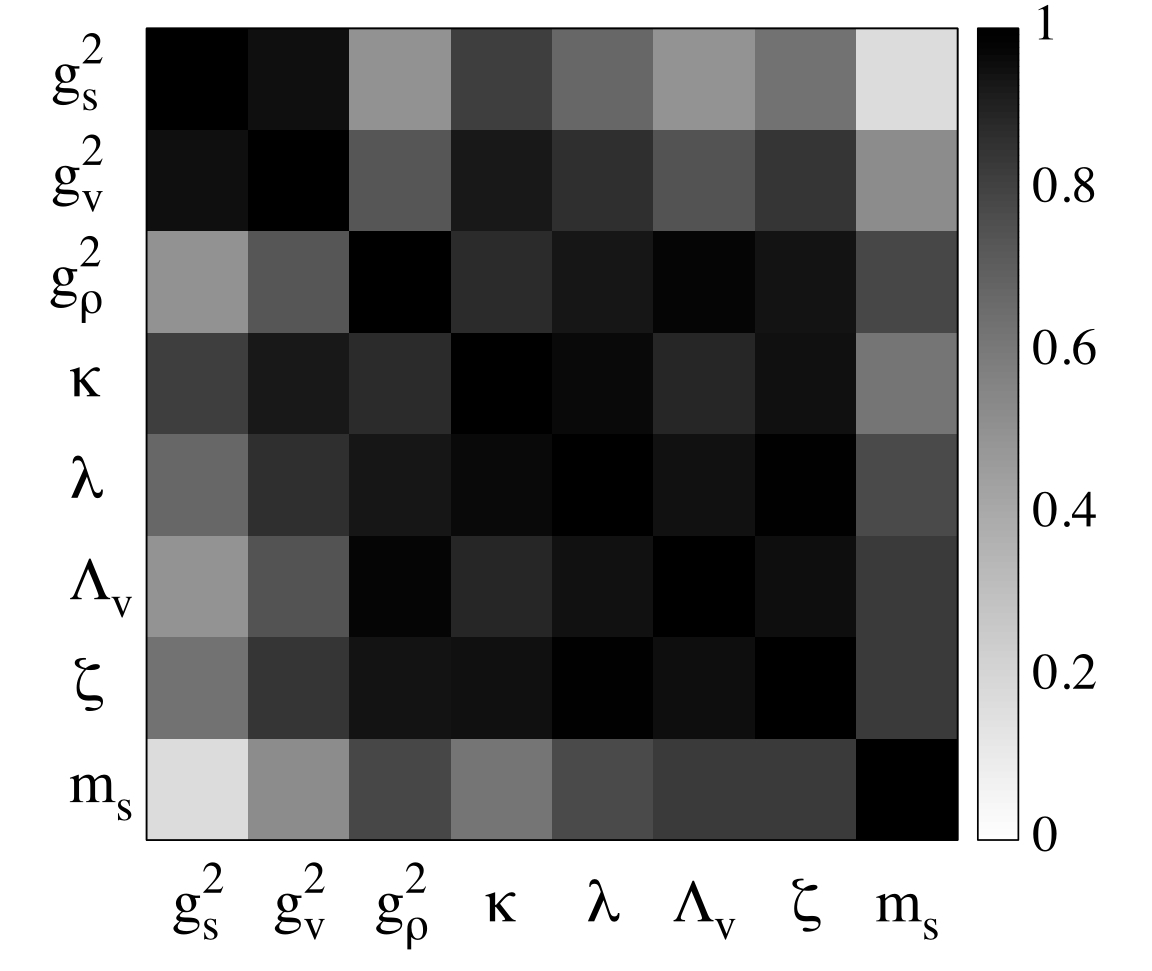}
\caption{Correlation coefficients (absolute values) between Lagrangian 
model parameters depicted in graphical form.}
\label{Fig6}
\end{figure}

We now proceed to display in Fig.\,\ref{Fig6} correlation coefficients 
between the Lagrangian model parameters. The prevalence of 
``dark patches'' suggests a strong correlation among several model 
parameters. A large correlation coefficient of $\big|\rho(A,B)\big|\simeq\!1$ 
between two observables may indicate ``redundancy'', in the sense 
that there may be little to gain by including both observables in the 
calibration procedure. This could alleviate the need for performing a 
complex experiment. Alternatively, a strong correlation may suggest 
an experiment that could constrain the value of an inaccessible quantity. 
However, in the case of the model parameters, a strong correlation does 
not imply redundancy, but quite the opposite. For example, a strong 
correlation between two well determined model parameters, such as 
$g_{\rm s}^{2}\!=\!108.0943\pm1.8376$ and 
$g_{\rm v}^{2}\!=\!183.7893\pm4.9623$ implies a strong interdependence. 
That is, if $g_{\rm s}^{2}$ is fixed at a certain value, then $g_{\rm v}^{2}$ 
must attain the precise value suggested by their correlation; otherwise the 
quality of the fit will deteriorate significantly. 

\begin{figure}[ht]
\vspace{-0.05in}
\includegraphics[width=8cm,height=9cm,angle=0]{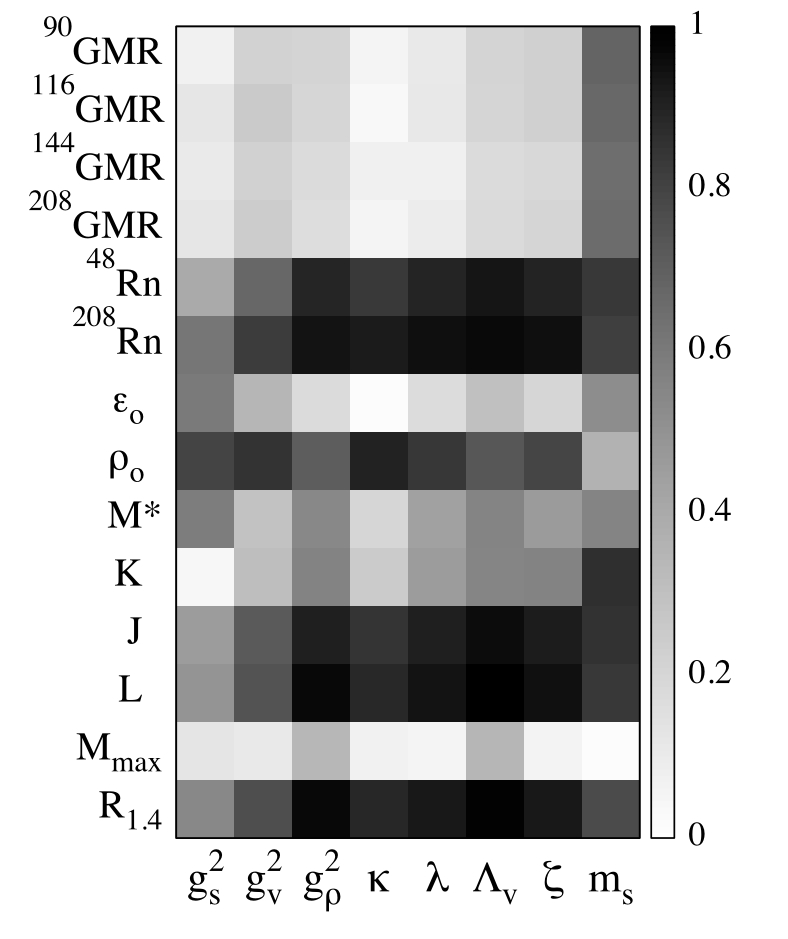}
\caption{Correlation coefficients (absolute values) between Lagrangian 
model parameters and a representative set of physical observables.
The set of observables are the same as those considered in
Fig.\,\ref{Fig5}.}
\label{Fig7}
\end{figure}

We conclude by displaying in Fig.\,\ref{Fig7} correlation coefficients 
between the Lagrangian model parameters and a representative set
of physical observables. Contrary to expectations, the strong 
correlation between $\zeta$ and the maximum neutron star mass is 
missing. As already explained, a large maximum neutron star mass 
may be generated by having either a stiff EOS for SNM or a stiff 
symmetry energy. If the symmetry energy is soft, as in the case of 
FSUGold, then one must stiffen the EOS of SNM, which may be
efficiently done by tuning $\zeta$. However, given that the symmetry 
energy predicted by FSUGold\,2 is stiff (see Fig.\,\ref{Fig4}) the
correlation between $\zeta$ and $M_{\rm max}$ weakens. Indeed, 
$M_{\rm max}$ displays the strongest correlation with the two 
isovector parameters $g_{\rho}^{2}$ and $\Lambda_{\rm v}$---although 
the correlation is fairly weak. This suggests that the maximum mass
constraint results from a competition between $\zeta$ and the slope 
of symmetry energy $L$. For instance, if $\zeta$ increases, thereby
softening the EOS of SNM, then $M_{\rm max}$ is reduced. Thus,
in order to maintain $M_{\rm max}$ at its specified value, the 
symmetry energy must stiffen accordingly. This implies a strong
and positive correlation between $\zeta$ and $L$, as precisely
indicated in Fig.\,\ref{Fig7}. An important lesson learned from the 
present discussion is that one must exercise caution in examining 
correlations among parameters and observables. For example, it
appears that certain bulk parameters of SNM, such as the binding 
energy per nucleon $\epsz$, the effective nucleon mass $M^{\ast}$, 
and the incompressibility coefficient $K$ are uncorrelated to the four 
isoscalar parameters $g_{\rm s}^{2}$, $g_{\rm v}^{2}$, $\kappa$, and 
$\lambda$. Such lack of correlation may come as a surprise in view 
that $\epsz$, $M^{\ast}$, $K$, and the saturation density $\rhoz$ 
\emph{uniquely} determine the value of the four isoscalar parameters 
(see appendix). The solution to this apparent contradiction lies in the 
fact that in generating the distribution of Lagrangian model parameters
all four isoscalar parameters become inextricably linked. In order to
isolate the proper correlation between a given observables (say 
$\epsz$) and a given model parameter (say $g_{\rm s}^{2}$) one 
should monitor the response of the observable to changes to only
that one parameter. That is, if one could provide suitable selection 
cuts to maintain the other parameters (say $g_{\rm v}^{2}$, $\kappa$, 
and $\lambda$) fixed, then the strong correlation between $\epsz$ and 
$g_{\rm s}^{2}$ will become manifest\,\cite{Piekarewicz:2014kza}.

\section{Summary and Outlook}
\label{Conclusions}

Finite nuclei, infinite nuclear matter, and neutron stars are strongly
interacting, nuclear many-body systems that span an enormous 
range of densities and isospin asymmetries. Lacking the tools to 
solve QCD in these regimes, DFT-based approaches, such as 
Skyrme and RMF models, provide the most powerful 
alternative for investigating such complex systems within a single 
unified framework. For the systematic study of such diverse nuclear
systems, we have developed a new RMF model, \emph{FSUGold\,2}, 
to describe the physics of both finite nuclei and neutron stars; 
objects that differ in size by 18 orders of magnitude.

The philosophy behind our calibration procedure adheres to two
guiding principles. First, the calibration relies exclusively on
genuine physical observables that can be measured either in the
laboratory or extracted from observation. Second, the optimization 
of the functional was implemented in the space of ``pseudo data'',
consisting mostly of bulk properties of infinite nuclear matter. This
has the enormous advantage that, unlike the Lagrangian model
parameters, the pseudo data have both a clear physical interpretation
and acceptable values that range over a fairly narrow interval. To our
knowledge, this is the first time that such a transformation between
model parameters and pseudo data is implemented in the relativistic
domain. We should note that in an effort to limit the input to only 
accurately measured physical observables, neither neutron skins 
of neutron-rich nuclei nor radii of neutron stars were included in 
the optimization. Hence, values for these observables become 
bona-fide model predictions. 
 
In addition to neutron skins and stellar radii, we provide predictions
for a variety of bulk properties of both symmetric nuclear matter and
the symmetry energy. Isoscalar properties, such as the density,
binding energy per nucleon, and incompressibility coefficient of SNM
at saturation are all determined with small theoretical errors and in
close agreement with their conventionally accepted values. In
particular, the incompressibility coefficient was determined with a
theoretical uncertainty of only 1\%. Such a small theoretical error was
obtained by the inclusion of GMR energies into the calibration
of FSUGold\,2. This too, we believe, has been done here for the first time.
The theoretical errors attached to the predictions of $\rhoz$ and
$\epsz$ are even smaller, indicating that the isoscalar sector is
well constrained by the binding energies, charge radii, and GMR
energies of finite nuclei.

The lack of well measured isovector observables in the calibration of
the functional has radically different consequences on the
determination of the bulk parameters of the symmetry energy,
especially in the case of its slope $L$. First, without stringent
isovector constraints, RMF models of the type used here tend to favor a
stiff symmetry energy. Indeed, we obtained a value for the slope of the symmetry
energy of $L\!=\!(112.8\pm16.1)\,{\rm MeV}$. In turn, this large slope yields
values of $R_{\rm skin}^{208}\!=\!(0.287\pm0.020)\,{\rm fm}$ and
$R_{1.4}\!=\!(14.42\pm0.26)\,{\rm km}$ for the neutron-skin
thickness of ${}^{208}$Pb and the radius of a 1.4\,$M_{\odot}$ neutron
star, respectively. Although both large, we underscore that at present
there is no conclusive experimental measurement nor astrophysical observation
that can rule out large neutron skins\,\cite{Fattoyev:2013yaa} or
large stellar radii. Thus, there is urgent need for the accurate
determination of both.

Following the optimization of the density functional, we proceeded to
explore the richness of the covariance analysis. This we did in two
stages. First, we provided predictions for a variety of observables
with properly estimated theoretical errors. This is particularly
critical when models are extrapolated to unknown regions. Second, we
explored correlations between both observables and model parameters.
A correlation analysis can reveal interdependences that may be of
great value. For example, a strong correlation between two observables
may eliminate the need to measure both. Further, if from these two
observables, e.g., $L$ and $R_{\rm skin}^{208}$, one of these is of
critical importance but inaccessible in the laboratory (e.g., $L$) one
could measure the latter to determine the former. Although there are
ambitious plans to experimentally constrain the isovector sector by
improving and expanding on previous measurements of both neutron 
skins and electric dipole polarizabilities, we will use some of the 
insights developed here to anticipate several different outcomes. We 
are planning to exploit the power and flexibility of the covariance analysis
to constrain the poorly determined isovector parameters $g_{\rho}^{2}$
and $\Lambda_{\rm v}$ by assuming a variety of scenarios involving
neutron skins of neutron-rich nuclei. For example, how precisely does
one have to measure the neutron radius of ${}^{208}$Pb in order to 
constrain $L$ to a given acceptable range? Is this precision attainable
with PREX-II? If not, what other neutron-rich nuclei should be used?
Or, is it better to measure the weak form factor of ${}^{208}$Pb at another
momentum transfer?  In this manner the development of an efficient 
modeling scheme is invaluable for the simulation of various scenarios.
Research along these lines is in progress and its results will be presented 
in a forthcoming publication.

\begin{acknowledgments}
We are grateful to Dr. F. J. Fattoyev for calling to our attention the transformation employed in the model 
building which greatly facilitates the optimization. This material is based upon work supported by the U.S. 
Department of Energy Office of Science, Office of Nuclear Physics under Award Number 
DE-FD05-92ER40750.
\end{acknowledgments}

\appendix*
\section{}
In this appendix we describe the connection between the coupling constants appearing 
in the Lagrangian density depicted in Eq.\,(\ref{LDensity}) and various bulk parameters 
of infinite nuclear matter. This connection has proved to be extremely useful. Indeed, 
expressing the objective function in terms of physically intuitive parameters provides
important insights on the quest for the optimal parametrization. For example, based
on the large experimental database of accurately measured nuclear masses, both the
saturation density and the energy per nucleon at saturation are fairly well known. In
turn, limiting the searches to a narrow region of parameter space increases 
significantly the efficiency of the Levenberg-Marquardt algorithm. We start by connecting
the isoscalar sector of the Lagrangian density with a few bulk parameters of symmetric
nuclear matter\,\cite{Glendenning:2000}. We then proceed to determine the two isovector
parameters of the Lagrangian density ($g_{\rho}^{2}$ and $\Lambda_{\rm v}$) from 
the value of the symmetry energy $J$ and its slope $L$ at saturation density. To our
knowledge, we are the first ones to establish such a connection in the isovector sector.

\subsection{Isoscalar sector}
Given the Lagrangian density of Eq.\,(\ref{LDensity}), the energy density (${\mathscr E}\!=\!E/V$) 
of infinite nuclear matter may be computed directly from the corresponding energy-momentum tensor
in the mean-field approximation. Note that only the zero-temperature limit will be addressed. 
Restricting ourselves to the isoscalar sector, the energy density of symmetric nuclear matter is
given by the following expression\,\cite{Serot:1984ey}:
\begin{align}
 {\mathscr E}(\rho) & = \gamma\int_{0}^{k_{\rm F}}\frac{d^{3}k}{(2\pi)^{3}}E_{k}^{{}^{(+)}} +
 \left(\frac{1}{2}\frac{m_{\rm s}^{2}}{g_{\rm s}^{2}}\Phi_{0}^{2}+\frac{\kappa}{6}\Phi_{0}^{3}
   +\frac{\lambda}{24}\Phi_{0}^{4}\right) - \left(\frac{1}{2}\frac{m_{\rm v}^{2}}{g_{\rm v}^{2}}W_{0}^{2}
   +\frac{\zeta}{24}W_{0}^{4}\right) \nonumber \\
   & = \frac{M^{\ast 4}}{4\pi^{2}}
 \left[\frac{k_{\rm F}E_{\rm F}^{\star}\left(k_{\rm F}^{2}+E_{\rm F}^{\star 2}\right)}
 {M^{\ast 4}}\!-\!\ln\left(\frac{k_{\rm F}+E_{\rm F}^{\star}}{M^{\ast}}\right)\right] +
 \left(\frac{1}{2}\frac{m_{\rm s}^{2}}{g_{\rm s}^{2}}\Phi_{0}^{2}+\frac{\kappa}{6}\Phi_{0}^{3}
   +\frac{\lambda}{24}\Phi_{0}^{4}\right) 
   + \left(\rho_{\rm v}W_{0}-\frac{1}{2}\frac{m_{\rm v}^{2}}{g_{\rm v}^{2}}W_{0}^{2}
   -\frac{\zeta}{24}W_{0}^{4}\right) \,.
 \label{Edensity}
\end {align}
where $\gamma\!=\!4$ is the spin-isospin degeneracy, 
$\rho_{\rm v}\!\equiv\!\rho=(2k_{\rm F}^{3})/(3\pi^{2})$ is the conserved baryon density,
$\Phi_{0}\!=g_{\rm s}\phi_{0}$, $W_{0}\!=g_{\rm v}V_{0}$, $M^{\ast}\!=\!M\!-\!\Phi_{0}$
is the effective nucleon mass, and 
$E_{k}^{{}^{(+)}}\!=\!E_{k}^{\ast}+W_{0}\!=\!\sqrt{k^{2}\!+\!M^{\ast 2}}+W_{0}$ is
the single-nucleon energy. Note that the classical equations of motion for the meson 
fields may be obtained directly from the Lagrangian density or equivalently, by demanding 
that the derivatives of $ {\mathscr E}(\rho)$ with respect to $\Phi_{0}$ and $W_{0}$ both 
vanish. That is,
\begin{subequations}
\begin{align}
 & \frac{\partial{\mathscr E}}{\partial\Phi_{0}} =
 \frac{m_{\rm s}^{2}}{g_{\rm s}^{2}}\Phi_{0}+\frac{\kappa}{2}\Phi_{0}^{2} 
 +\frac{\lambda}{6}\Phi_{0}^{3}-\rho_{\rm s}=0\, \label{EOMa} \,,\\
 & \frac{\partial{\mathscr E}}{\partial W_{0}} =
      \frac{m_{\rm v}^{2}}{g_{\rm v}^{2}}W_{0}
    +\frac{\zeta}{6}W_{0}^{3}-\rho_{\rm v}=0 \,. \label{EOMb} 
\end{align} 
\label{EOM}
\end{subequations}
Here $\rho_{\rm s}$ is the scalar density that is defined as follows:
\begin{equation}
 \rho_{\rm s}(M^{\ast})  = \gamma\int_{0}^{k_{\rm F}}\frac{d^{3}k}{(2\pi)^{3}}
 \frac{M^{\ast}}{E_{k}^{\star}} = \frac{M^{\ast 3}}{\pi^{2}}
 \left[\frac{k_{\rm F}E_{\rm F}^{\star}}{M^{\ast 2}} -
 \ln\left(\frac{k_{\rm F}+E_{\rm F}^{\star}}{M^{\ast}}\right)\right] \,.
 \label{Rhos}
\end {equation}
Note that the scalar density is not conserved and must be self-consistently
determined from the equations of motion.

At zero temperature the pressure of the system may be calculated from
its thermodynamic definition, i.e.,
\begin{equation}
 P = -\left(\frac{\partial E}{\partial V}\right)_{\!N} = 
 \rho\frac{\partial{\mathscr E}}{\partial\rho} - {\mathscr E} =
 \rho\left(E_{\rm F}^{{}^{(+)}} - \frac{E}{A}\right) \,, 
 \label{Pressure}
\end {equation}
where the last line follows from using $\partial{\mathscr E}/{\partial\rho}\!=\!E_{\rm F}^{{}^{(+)}}$,
an identity that should hold in any thermodynamically consistent many-body theory. Moreover,
note that at saturation density, the pressure vanishes and one obtains---in accordance with the 
Hugenholtz-van Hove theorem---that the energy per nucleon becomes equal to the Fermi energy. 
That is,
\begin{equation}
 E_{\rm F}^{{}^{(+)}} = \sqrt{k_{\rm F}^{2}+M^{\ast 2}}+W_{0} = \frac{E}{A} \,.
 \label{HvH}
\end {equation}
  
To make further progress, we now obtain an analytic expression for the incompressibility 
coefficient of symmetric nuclear matter $K$. As defined in Eq.\,(\ref{EandSa}), it is given 
by
\begin{equation}
 K= 9\rhoz^{2}\left[\frac{d^{2}(E/A)}{d\rho^{2}}\right]_{\!0} =
       9\rhoz^{2}\left[\frac{d}{d\rho}\!\left(\frac{P}{\rho^{2}}\right)\right]_{\!0} =
       9\rhoz\!\left(\frac{dE_{\rm F}^{{}^{(+)}}}{d\rho}\right)_{\!\!0} \,.
 \label{K0}
\end {equation}
Given that the Fermi energy depends in a complicated way on the density, i.e.,
both explicitly and implicitly through $M^{\ast}$ and $W_{0}$, there are three 
terms that need to be evaluated. That is,
\begin{align}
 \frac{K}{9\rhoz} = & \left(\frac{\partial E_{\rm F}^{{}^{(+)}}}{\partial\rho}\right)_{\!\!0}
                         + \left(\frac{\partial E_{\rm F}^{{}^{(+)}}}{\partial W_{0}}\right)_{\!\!0} 
 	                 \! \left(\frac{{\partial W_{0}}}{\partial\rho}\right)_{\!\!0} 
                        + \left(\frac{\partial E_{\rm F}^{{}^{(+)}}}{\partial M^{\ast}}\right)_{\!\!0} 
 	                 \! \left(\frac{{\partial M^{\ast}}}{\partial\rho}\right)_{\!\!0} \nonumber \\
                        = & \left(\frac{\partial E_{\rm F}^{{}^{(+)}}}{\partial\rho}\right)_{\!\!0}
                         + \left(\frac{{\partial W_{0}}}{\partial\rho}\right)_{\!\!0} 
                        + \left(\frac{M^{\ast}}{E_{\rm F}^{\ast}}\right)_{\!\!0} 
 	                 \! \left(\frac{{\partial M^{\ast}}}{\partial\rho}\right)_{\!\!0} \,.     		                  
 \label{K1}
\end {align}
We now proceed to evaluate each of the three terms. The first one is the simplest
and yields:
\begin{equation}
 \left(\frac{\partial E_{\rm F}^{{}^{(+)}}}{\partial\rho}\right)_{\!\!0} =
 \left(\frac{\pi^{2}}{2k_{\rm F}E_{\rm F}^{\ast}}\right)_{\!\!0}  \,. 	                          
 \label{Term1}
\end {equation}
We continue with the second term and make use of the equation of motion for $W_{0}$
[Eq.\,(\ref{EOMb})] to write:
\begin{equation}
  \left(\frac{{\partial W_{0}}}{\partial\rho}\right)_{\!\!0} = 
  \left(\frac{g_{\rm v}^{2}}{m_{\rm v}^{\ast 2}}\right)_{\!\!0} \,,
  \hspace{0.2cm} {\rm with} \hspace{0.15cm}
  m_{\rm v}^{\ast 2}\equiv m_{\rm v}^{2}+\frac{\zeta}{2}g_{\rm v}^{2}W_{0}^{2}\,.  
 \label{Term2}
\end {equation}
Using the previous two results we can rewrite Eq.\,(\ref{K1}) as follows:
\begin{equation}
 \left(\frac{{\partial M^{\ast}}}{\partial\rho}\right)_{\!\!0} = 
 \Bigg[\frac{E_{\rm F}^{\ast}}{M^{\ast}} \left(\frac{K}{9\rho} - 
 \frac{\pi^{2}}{2k_{\rm F}E_{\rm F}^{\ast}} - 
 \frac{g_{\rm v}^{2}}{m_{\rm v}^{\ast 2}}\right)\Bigg]_{\!0}\,.
 \label{Term3a}
\end {equation}
The left-hand side of the equation may be computed by invoking the scalar equation of motion 
[Eq.\,(\ref{EOMa})] and depends on the three isoscalar coupling constants. We obtain,
\begin{equation}
 \left(\frac{{\partial M^{\ast}}}{\partial\rho}\right)_{\!\!0} = -
 \Bigg[
 \frac{M^{\ast}}{E_{\rm F}^{\ast}} 
 \left(\frac{m_{\rm s}^{\ast 2}}{g_{\rm s}^{2}}+\rho_{\rm s}^{\,\prime}(M^{\ast})\right)^{-1}\Bigg]_{\!0}\,,
 \hspace{0.2cm} {\rm with} \hspace{0.15cm}
 \frac{m_{\rm s}^{\ast 2}}{g_{\rm s}^{2}}\equiv \frac{m_{\rm s}^{2}}{g_{\rm s}^{2}}+\kappa\Phi_{0}+
 \frac{\lambda}{2}\Phi_{0}^{2}\,. 
 \label{Term3b}
\end {equation}
Note that we have defined the derivative of the scalar density [Eq.\,(\ref{Rhos})]
with respect to $M^{\ast}$ as follows:
\begin{equation}
 \rho_{\rm s}^{\,\prime}(M^{\ast})  = 
 \left(\frac{\partial\rho_{\rm s}}{\partial M^{\ast}}\right) = 
 \frac{1}{\pi^{2}}
 \left[\frac{k_{\rm F}}{E_{\rm F}^{\star}}(E_{\rm F}^{\star 2}+2M^{\ast 2}) -
 3M^{\ast 2}\ln\left(\frac{k_{\rm F}+E_{\rm F}^{\star}}{M^{\ast}}\right)\right] .
 \label{Rhos1}
\end {equation}
This is all the formalism that is needed to establish the connection between the isoscalar
parameters appearing in the Lagrangian and a few bulk parameters of infinite nuclear
matter. In the isoscalar sector the four bulk parameters of infinite nuclear matter that
we consider here are as follows: (i) the density $\rho$, (ii) the binding energy per nucleon $E/A$,
(iii) the incompressibility coefficient $K$, and (iv) the effective nucleon mass 
$M^{\ast}$---all of them evaluated at saturation density. Specification of these four bulk 
parameters enables one to determine four out of the five isoscalar coupling constants,
namely, $g_{\rm v}^{2}/m_{\rm v}^{2}$, $g_{\rm s}^{2}/m_{\rm s}^{2}$, $\kappa$, and 
$\lambda$. The sole remaining coupling constant $\zeta$ is left intact as it is fairly 
insensitive to the properties of symmetric nuclear matter. Indeed, $\zeta$ is sensitive
to the high-density component of the EOS and can be easily tuned by specifying the 
maximum neutron star mass. Note that in the mean-field approximation the Yukawa 
meson couplings always appear in combination with the corresponding meson mass.

The vector coupling may be readily determined from the vanishing of the pressure at 
saturation density. Indeed, from Eq.\,(\ref{HvH}) one obtains the value of the vector 
field $W_{0}$ at saturation density. In turn, substituting this value in Eq.\,(\ref{EOMb}) 
determines (for a given $\zeta$) $g_{\rm v}^{2}/m_{\rm v}^{2}$. Given that the vector
mass has been fixed at its experimental value of $m_{\rm v}\!=\!782.5$\,MeV, this 
provides a determination of $g_{\rm v}^{2}$.

The specification of the three isoscalar parameters is significantly more involved 
and depends critically on knowledge of the effective nucleon mass $M^{\ast}$ at
saturation density. Further, it requires three independent pieces of information for
their determination. Perhaps surprisingly, such information is provided in the form 
of three \emph{simultaneous linear equations}. That is, the solution is unique. The 
first equation to be used involves the energy density of symmetric nuclear matter 
depicted in Eq.\,(\ref{Edensity}). Given that at saturation density
${\mathscr E}(\rhoz)\!=\!\rhoz(E/A)_{0}$, every term in such expression is
known---with the exception of  $m_{\rm s}^{2}/g_{\rm s}^{2}$, $\kappa$, and 
$\lambda$. The classical equation of motion for the scalar field Eq.\,(\ref{EOMa})
provides the second linear equation in these three parameters, since the scalar
density is fully specified in terms of the density and effective nucleon mass at
saturation. Finally, knowledge of the incompressibility coefficient $K$ at saturation
density supplies the third and last linear equation. Indeed, a comparison between
Eq.\,(\ref{Term3a}) and Eq.\,(\ref{Term3b}) indicates that the only unknown is the
quantity $m_{\rm s}^{\ast 2}/g_{\rm s}^{2}$, which again contains the three scalar
parameters of interest. Given that these equations provide a system of three 
simultaneous linear equations, the solution may be obtained by elementary means.

\subsection{Isovector sector}

In the previous section we concentrated on connecting the isoscalar parameters 
of the Lagrangian density to a few bulk parameters of symmetric nuclear matter. 
We now shift our focus to the isovector sector and show that the two isovector 
parameters $g_{\rho}^{2}/m_{\rho}^{2}$ and $\Lambda_{\rm v}$ may be determined
from knowledge of two quantities of central importance, namely, the symmetry 
energy $J$ and its slope at saturation density $L$. To our knowledge, this 
connection is established here for the first time.

For the Lagrangian density given in Eq.\,(\ref{LDensity}), an analytic expression 
for the density dependence of the symmetry energy was derived in 
Ref.\,\cite{Horowitz:2001ya}. One obtains,
\begin{equation}
 S(\rho) = \frac{k_{\rm F}^{2}}{6E_{\rm F}^{\star}} +
 \frac{g_{\rho}^{2}\rho}{8m_{\rho}^{\ast 2}} \,,
  \hspace{0.2cm} {\rm with} \hspace{0.15cm}
  \frac{m_{\rho}^{\ast 2}}{g_{\rho}^{2}} \equiv 
  \frac{m_{\rho}^{2}}{g_{\rho}^{2}}+2\Lambda_{\rm v}W_{0}^{2}\,. 
 \label{SymEnergy}
\end {equation}
We note that the density dependence of the symmetry energy given above consists
of a purely ``isoscalar'' term and a largely ``isovector'' term. That is, we define
\begin{equation}
  S_{0}(\rho) \!=\! \frac{k_{\rm F}^{2}}{6E_{\rm F}^{\star}} \;\; {\rm and} \;\;
  S_{1}(\rho) \!=\! \frac{g_{\rho}^{2}\rho}{8m_{\rho}^{\ast 2}} \,. 
\end{equation}
In particular, given that the isoscalar sector has already been fixed, $S_{0}(\rho)$
along with all its derivatives are known. In contrast, $S_{1}(\rho)$ depends on both
$g_{\rho}^{2}/m_{\rho}^{2}$ and $\Lambda_{\rm v}$ which are unknown. As already 
mentioned, critical to the determination of these two isovector parameters are the 
symmetry energy and its slope at saturation density, which according to 
Eq.\,(\ref{EandSb}) are given as follows:
\begin{equation}
 J = S(\rhoz) \;\; {\rm and} \;\;
 L = 3 \rhoz \left(\frac{d S}{d\rho}\right)_{\!\!0} \,. 
\end{equation}

The determination of the quantity $m_{\rho}^{\ast 2}/g_{\rho}^{2}$, which still depends 
on both isovector parameters, is fairly simple:
\begin{equation}
 J_{1} \equiv  \left(\frac{g_{\rho}^{2}\rho}{8m_{\rho}^{\ast 2}}\right)_{\!\!0} =
 \Big(J\!-\!J_{0}\Big) = J - \left(\frac{k_{\rm F}^{2}}{6E_{\rm F}^{\star}}\right)_{\!\!0} \,.
 \label{J1}
\end{equation}
In contrast, the determination of each individual isovector parameters is considerably 
more difficult and involves several of the same manipulations carried out in the 
isoscalar sector. In analogy with the above equation we write:
\begin{equation}
 L_{1} = 3 \rhoz\!\left(\frac{d S_{1}}{d\rho}\right)_{\!\!0}
         = \Big(L\!-\!L_{0}\Big) = L - 
            3 \rhoz\!\left(\frac{d S_{0}}{d\rho}\right)_{\!\!0}\;.
 \label{L1a}
\end{equation}
We start by computing the contribution to the slope from the isoscalar term.
That is,
\begin{align}
 L_{0} & = 3 \rhoz\!\left(\frac{d S_{0}}{d\rho}\right)_{\!\!0}
             = 3 \rhoz\!\left[
                   \left(\frac{\partial S_{0}}{\partial\rho}\right) +
                   \left(\frac{\partial S_{0}}{\partial M^{\ast}}\right)          
                   \left(\frac{\partial M^{\ast}}{\partial\rho}\right) \right]_{0}   \nonumber \\  	
         & = J_{0} \Bigg(1+\frac{M^{\ast 2}}{E_{\rm F}^{\ast 2}} 
                   \left[1-\!\frac{3\rho}{\,M^{\ast}}\!\left(\frac{\partial M^{\ast}}
                   {\partial\rho}\right)\right]\Bigg)_{\!0} \,.	                  
 \label{L0}
\end {align}
Note that this expression is given exclusively in terms of isoscalar parameters, 
so it is completely known. Also note that the answer has been left in terms of 
$(\partial M^{\ast}\!/\partial\rho)_{{}_{0}}$ which has already been calculated in 
the previous section. We now proceed to compute the isovector contribution to 
the slope of the symmetry energy. Following similar steps as before, we obtain
\begin{align}
 L_{1} & = 3 \rhoz\!\left(\frac{d S_{1}}{d\rho}\right)_{\!\!0}
             = 3 \rhoz\!\left[
                   \left(\frac{\partial S_{1}}{\partial\rho}\right) +
                   \left(\frac{\partial S_{1}}{\partial W_{0}}\right)          
                   \left(\frac{\partial W_{0}}{\partial\rho}\right) \right]_{0}   \nonumber \\  	
         &  = 3J_{1} 
                   \left[1-32\left(\frac{g_{\rm v}^{2}}{m_{\rm v}^{\ast 2}}\right)\!W_{0}
                   \Lambda_{\rm v} J_{1}\right]_{0} = \Big(L\!-\!L_{0}\Big)\,.
 \label{L1b}
\end {align}
This is all that is needed to achieve the desired goal of expressing $g_{\rho}^{2}/m_{\rho}^{2}$ 
and $\Lambda_{\rm v}$ in terms of $J$ and $L$. Indeed, given that $L$ is provided, and $J_{1}$ 
and $L_{0}$ have been determined from Eqs.\,(\ref{J1}) and\,(\ref{L0}), respectively, the only
unknown in the previous equation is $\Lambda_{\rm v}$. Finally, using the definition of the
effective $\rho$-meson mass given in Eq.\,(\ref{SymEnergy}), we can solve for 
$g_{\rho}^{2}/m_{\rho}^{2}$. That is,
\begin{equation}
 \frac{m_{\rho}^{2}}{g_{\rho}^{2}} =
 \frac{m_{\rho}^{\ast 2}}{g_{\rho}^{2}} - 2\Lambda_{\rm v}W_{0}^{2} =
 \frac{\rhoz}{8J_{1}} - 2\Lambda_{\rm v}W_{0}^{2} \,.
 \label{grho2}
\end {equation}

\bibliography{FSUGold2.bbl}

\begin{thebibliography}{89}%
\makeatletter
\providecommand \@ifxundefined [1]{%
 \@ifx{#1\undefined}
}%
\providecommand \@ifnum [1]{%
 \ifnum #1\expandafter \@firstoftwo
 \else \expandafter \@secondoftwo
 \fi
}%
\providecommand \@ifx [1]{%
 \ifx #1\expandafter \@firstoftwo
 \else \expandafter \@secondoftwo
 \fi
}%
\providecommand \natexlab [1]{#1}%
\providecommand \enquote  [1]{``#1''}%
\providecommand \bibnamefont  [1]{#1}%
\providecommand \bibfnamefont [1]{#1}%
\providecommand \citenamefont [1]{#1}%
\providecommand \href@noop [0]{\@secondoftwo}%
\providecommand \href [0]{\begingroup \@sanitize@url \@href}%
\providecommand \@href[1]{\@@startlink{#1}\@@href}%
\providecommand \@@href[1]{\endgroup#1\@@endlink}%
\providecommand \@sanitize@url [0]{\catcode `\\12\catcode `\$12\catcode
  `\&12\catcode `\#12\catcode `\^12\catcode `\_12\catcode `\%12\relax}%
\providecommand \@@startlink[1]{}%
\providecommand \@@endlink[0]{}%
\providecommand \url  [0]{\begingroup\@sanitize@url \@url }%
\providecommand \@url [1]{\endgroup\@href {#1}{\urlprefix }}%
\providecommand \urlprefix  [0]{URL }%
\providecommand \Eprint [0]{\href }%
\providecommand \doibase [0]{http://dx.doi.org/}%
\providecommand \selectlanguage [0]{\@gobble}%
\providecommand \bibinfo  [0]{\@secondoftwo}%
\providecommand \bibfield  [0]{\@secondoftwo}%
\providecommand \translation [1]{[#1]}%
\providecommand \BibitemOpen [0]{}%
\providecommand \bibitemStop [0]{}%
\providecommand \bibitemNoStop [0]{.\EOS\space}%
\providecommand \EOS [0]{\spacefactor3000\relax}%
\providecommand \BibitemShut  [1]{\csname bibitem#1\endcsname}%
\let\auto@bib@innerbib\@empty
\bibitem [{UNE()}]{UNEDF}%
  \BibitemOpen
  \href {http://unedf.org} {\enquote {\bibinfo {title} {Building a universal
  nuclear energy density functional},}\ }\bibinfo {note} {(UNEDF
  Collaboration)}\BibitemShut {NoStop}%
\bibitem [{\citenamefont {Kortelainen}\ \emph
  {et~al.}(2010{\natexlab{a}})\citenamefont {Kortelainen}, \citenamefont
  {Lesinski}, \citenamefont {More}, \citenamefont {Nazarewicz}, \citenamefont
  {Sarich}, \citenamefont {Schunck}, \citenamefont {Stoitsov},\ and\
  \citenamefont {Wild}}]{Kortelainen:2010hv}%
  \BibitemOpen
  \bibfield  {author} {\bibinfo {author} {\bibfnamefont {M.}~\bibnamefont
  {Kortelainen}}, \bibinfo {author} {\bibfnamefont {T.}~\bibnamefont
  {Lesinski}}, \bibinfo {author} {\bibfnamefont {J.}~\bibnamefont {More}},
  \bibinfo {author} {\bibfnamefont {W.}~\bibnamefont {Nazarewicz}}, \bibinfo
  {author} {\bibfnamefont {J.}~\bibnamefont {Sarich}}, \bibinfo {author}
  {\bibfnamefont {N.}~\bibnamefont {Schunck}}, \bibinfo {author} {\bibfnamefont
  {M.~V.}\ \bibnamefont {Stoitsov}}, \ and\ \bibinfo {author} {\bibfnamefont
  {S.}~\bibnamefont {Wild}},\ }\href@noop {} {\bibfield  {journal} {\bibinfo
  {journal} {Phys. Rev. C}\ }\textbf {\bibinfo {volume} {82}},\ \bibinfo
  {pages} {024313} (\bibinfo {year} {2010}{\natexlab{a}})}\BibitemShut
  {NoStop}%
\bibitem [{\citenamefont {Kortelainen}\ \emph {et~al.}(2012)\citenamefont
  {Kortelainen}, \citenamefont {McDonnell}, \citenamefont {Nazarewicz},
  \citenamefont {Reinhard}, \citenamefont {Sarich}, \citenamefont {Schunck},
  \citenamefont {Stoitsov},\ and\ \citenamefont {Wild}}]{Kortelainen:2012}%
  \BibitemOpen
  \bibfield  {author} {\bibinfo {author} {\bibfnamefont {M.}~\bibnamefont
  {Kortelainen}}, \bibinfo {author} {\bibfnamefont {J.}~\bibnamefont
  {McDonnell}}, \bibinfo {author} {\bibfnamefont {W.}~\bibnamefont
  {Nazarewicz}}, \bibinfo {author} {\bibfnamefont {P.-G.}\ \bibnamefont
  {Reinhard}}, \bibinfo {author} {\bibfnamefont {J.}~\bibnamefont {Sarich}},
  \bibinfo {author} {\bibfnamefont {N.}~\bibnamefont {Schunck}}, \bibinfo
  {author} {\bibfnamefont {M.~V.}\ \bibnamefont {Stoitsov}}, \ and\ \bibinfo
  {author} {\bibfnamefont {S.~M.}\ \bibnamefont {Wild}},\ }\href@noop {}
  {\bibfield  {journal} {\bibinfo  {journal} {Phys. Rev. C}\ }\textbf {\bibinfo
  {volume} {85}},\ \bibinfo {pages} {024304} (\bibinfo {year}
  {2012})}\BibitemShut {NoStop}%
\bibitem [{\citenamefont {Kortelainen}\ \emph {et~al.}(2013)\citenamefont
  {Kortelainen}, \citenamefont {McDonnell}, \citenamefont {Nazarewicz},
  \citenamefont {Olsen}, \citenamefont {Reinhard}, \citenamefont {Sarich},
  \citenamefont {Schunck}, \citenamefont {Wild}, \citenamefont {Davesne},\ and\
  \citenamefont {Pastore}}]{Kortelainen:2013}%
  \BibitemOpen
  \bibfield  {author} {\bibinfo {author} {\bibfnamefont {M.}~\bibnamefont
  {Kortelainen}}, \bibinfo {author} {\bibfnamefont {J.}~\bibnamefont
  {McDonnell}}, \bibinfo {author} {\bibfnamefont {W.}~\bibnamefont
  {Nazarewicz}}, \bibinfo {author} {\bibfnamefont {E.}~\bibnamefont {Olsen}},
  \bibinfo {author} {\bibfnamefont {P.-G.}\ \bibnamefont {Reinhard}}, \bibinfo
  {author} {\bibfnamefont {J.}~\bibnamefont {Sarich}}, \bibinfo {author}
  {\bibfnamefont {N.}~\bibnamefont {Schunck}}, \bibinfo {author} {\bibfnamefont
  {S.~M.}\ \bibnamefont {Wild}}, \bibinfo {author} {\bibfnamefont
  {D.}~\bibnamefont {Davesne}}, \bibinfo {author} {\bibfnamefont
  {J.}~\bibnamefont {Erler}}, \ and\ \bibinfo {author} {\bibfnamefont
  {A.}~\bibnamefont {Pastore}},\ }\href@noop {} {\bibfield  {journal} {\bibinfo
   {journal} {Phys. Rev. C}\ }\textbf {\bibinfo {volume} {89}},\ \bibinfo
  {pages} {054314} (\bibinfo {year} {2014})}\BibitemShut {NoStop}%
\bibitem [{\citenamefont {Walecka}(1974)}]{Walecka:1974qa}%
  \BibitemOpen
  \bibfield  {author} {\bibinfo {author} {\bibfnamefont {J.~D.}\ \bibnamefont
  {Walecka}},\ }\href@noop {} {\bibfield  {journal} {\bibinfo  {journal}
  {Ann. Phys. (NY)}\ }\textbf {\bibinfo {volume} {83}},\ \bibinfo {pages} {491}
  (\bibinfo {year} {1974})}\BibitemShut {NoStop}%
\bibitem [{\citenamefont {Serot}\ and\ \citenamefont
  {Walecka}(1986)}]{Serot:1984ey}%
  \BibitemOpen
  \bibfield  {author} {\bibinfo {author} {\bibfnamefont {B.~D.}\ \bibnamefont
  {Serot}}\ and\ \bibinfo {author} {\bibfnamefont {J.~D.}\ \bibnamefont
  {Walecka}},\ }\href@noop {} {\bibfield  {journal} {\bibinfo  {journal} {Adv.
  Nucl. Phys.}\ }\textbf {\bibinfo {volume} {16}},\ \bibinfo {pages} {1}
  (\bibinfo {year} {1986})}\BibitemShut {NoStop}%
\bibitem [{\citenamefont {Horowitz}\ and\ \citenamefont
  {Serot}(1981)}]{Horowitz:1981xw}%
  \BibitemOpen
  \bibfield  {author} {\bibinfo {author} {\bibfnamefont {C.~J.}\ \bibnamefont
  {Horowitz}}\ and\ \bibinfo {author} {\bibfnamefont {B.~D.}\ \bibnamefont
  {Serot}},\ }\href@noop {} {\bibfield  {journal} {\bibinfo  {journal} {Nucl.
  Phys. A}\ }\textbf {\bibinfo {volume} {368}},\ \bibinfo {pages} {503}
  (\bibinfo {year} {1981})}\BibitemShut {NoStop}%
\bibitem [{\citenamefont {Lalazissis}\ \emph {et~al.}(1997)\citenamefont
  {Lalazissis}, \citenamefont {Konig},\ and\ \citenamefont
  {Ring}}]{Lalazissis:1996rd}%
  \BibitemOpen
  \bibfield  {author} {\bibinfo {author} {\bibfnamefont {G.~A.}\ \bibnamefont
  {Lalazissis}}, \bibinfo {author} {\bibfnamefont {J.}~\bibnamefont {Konig}}, \
  and\ \bibinfo {author} {\bibfnamefont {P.}~\bibnamefont {Ring}},\ }\href@noop
  {} {\bibfield  {journal} {\bibinfo  {journal} {Phys. Rev. C}\ }\textbf
  {\bibinfo {volume} {55}},\ \bibinfo {pages} {540} (\bibinfo {year}
  {1997})}\ 
  \BibitemShut {NoStop}%
\bibitem [{\citenamefont {Lalazissis}\ \emph {et~al.}(1999)\citenamefont
  {Lalazissis}, \citenamefont {Raman},\ and\ \citenamefont
  {Ring}}]{Lalazissis:1999}%
  \BibitemOpen
  \bibfield  {author} {\bibinfo {author} {\bibfnamefont {G.~A.}\ \bibnamefont
  {Lalazissis}}, \bibinfo {author} {\bibfnamefont {S.}~\bibnamefont {Raman}}, \
  and\ \bibinfo {author} {\bibfnamefont {P.}~\bibnamefont {Ring}},\ }\href@noop
  {} {\bibfield  {journal} {\bibinfo  {journal} {At. Data Nucl. Data Tables}\
  }\textbf {\bibinfo {volume} {71}},\ \bibinfo {pages} {1} (\bibinfo {year}
  {1999})}\BibitemShut {NoStop}%
\bibitem [{\citenamefont {Todd-Rutel}\ and\ \citenamefont
  {Piekarewicz}(2005)}]{Todd-Rutel:2005fa}%
  \BibitemOpen
  \bibfield  {author} {\bibinfo {author} {\bibfnamefont {B.~G.}\ \bibnamefont
  {Todd-Rutel}}\ and\ \bibinfo {author} {\bibfnamefont {J.}~\bibnamefont
  {Piekarewicz}},\ }\href@noop {} {\bibfield  {journal} {\bibinfo  {journal}
  {Phys. Rev. Lett.}\ }\textbf {\bibinfo {volume} {95}},\ \bibinfo {pages}
  {122501} (\bibinfo {year} {2005})}\ \BibitemShut
  {NoStop}%
\bibitem [{\citenamefont {Brown}(2000)}]{Brown:2000}%
  \BibitemOpen
  \bibfield  {author} {\bibinfo {author} {\bibfnamefont {B.~A.}\ \bibnamefont
  {Brown}},\ }\href@noop {} {\bibfield  {journal} {\bibinfo  {journal} {Phys.
  Rev. Lett.}\ }\textbf {\bibinfo {volume} {85}},\ \bibinfo {pages} {5296}
  (\bibinfo {year} {2000})}\BibitemShut {NoStop}%
\bibitem [{\citenamefont {Furnstahl}(2002)}]{Furnstahl:2001un}%
  \BibitemOpen
  \bibfield  {author} {\bibinfo {author} {\bibfnamefont {R.~J.}\ \bibnamefont
  {Furnstahl}},\ }\href@noop {} {\bibfield  {journal} {\bibinfo  {journal}
  {Nucl. Phys. A}\ }\textbf {\bibinfo {volume} {706}},\ \bibinfo {pages} {85}
  (\bibinfo {year} {2002})}\ \BibitemShut {NoStop}%
\bibitem [{\citenamefont {Centelles}\ \emph {et~al.}(2009)\citenamefont
  {Centelles}, \citenamefont {Roca-Maza}, \citenamefont {Vi\~nas},\ and\
  \citenamefont {Warda}}]{Centelles:2008vu}%
  \BibitemOpen
  \bibfield  {author} {\bibinfo {author} {\bibfnamefont {M.}~\bibnamefont
  {Centelles}}, \bibinfo {author} {\bibfnamefont {X.}~\bibnamefont
  {Roca-Maza}}, \bibinfo {author} {\bibfnamefont {X.}~\bibnamefont {Vi\~nas}},
  \ and\ \bibinfo {author} {\bibfnamefont {M.}~\bibnamefont {Warda}},\ }\href
  {\doibase 10.1103/PhysRevLett.102.122502} {\bibfield  {journal} {\bibinfo
  {journal} {Phys. Rev. Lett.}\ }\textbf {\bibinfo {volume} {102}},\ \bibinfo
  {pages} {122502} (\bibinfo {year} {2009})}\ \BibitemShut
  {NoStop}%
\bibitem [{\citenamefont {Roca-Maza}\ \emph {et~al.}(2011)\citenamefont
  {Roca-Maza}, \citenamefont {Centelles}, \citenamefont {Vinas},\ and\
  \citenamefont {Warda}}]{RocaMaza:2011pm}%
  \BibitemOpen
  \bibfield  {author} {\bibinfo {author} {\bibfnamefont {X.}~\bibnamefont
  {Roca-Maza}}, \bibinfo {author} {\bibfnamefont {M.}~\bibnamefont
  {Centelles}}, \bibinfo {author} {\bibfnamefont {X.}~\bibnamefont {Vinas}}, \
  and\ \bibinfo {author} {\bibfnamefont {M.}~\bibnamefont {Warda}},\ }\href
  {\doibase 10.1103/PhysRevLett.106.252501} {\bibfield  {journal} {\bibinfo
  {journal} {Phys. Rev. Lett.}\ }\textbf {\bibinfo {volume} {106}},\ \bibinfo
  {pages} {252501} (\bibinfo {year} {2011})}\ \BibitemShut
  {NoStop}%
\bibitem [{\citenamefont {Erler}\ \emph {et~al.}(2012)\citenamefont {Erler},
  \citenamefont {Birge}, \citenamefont {Kortelainen}, \citenamefont
  {Nazarewicz}, \citenamefont {Olsen}, \citenamefont {Perhac},\ and\
  \citenamefont {Stoitsov}}]{Erler:2012}%
  \BibitemOpen
  \bibfield  {author} {\bibinfo {author} {\bibfnamefont {J.}~\bibnamefont
  {Erler}}, \bibinfo {author} {\bibfnamefont {N.}~\bibnamefont {Birge}},
  \bibinfo {author} {\bibfnamefont {M.}~\bibnamefont {Kortelainen}}, \bibinfo
  {author} {\bibfnamefont {W.}~\bibnamefont {Nazarewicz}}, \bibinfo {author}
  {\bibfnamefont {E.}~\bibnamefont {Olsen}}, \bibinfo {author} {\bibfnamefont
  {A.~M.}\ \bibnamefont {Perhac}}, \ and\ \bibinfo {author} {\bibfnamefont
  {M.}~\bibnamefont {Stoitsov}},\ }\href@noop {} {\bibfield  {journal}
  {\bibinfo  {journal} {Nature}\ }\textbf {\bibinfo {volume} {486}},\ \bibinfo
  {pages} {509} (\bibinfo {year} {2012})}\BibitemShut {NoStop}%
\bibitem [{\citenamefont {Afanasjev}\ \emph {et~al.}(2013)\citenamefont
  {Afanasjev}, \citenamefont {Agbemava}, \citenamefont {Ray},\ and\
  \citenamefont {Ring}}]{Afanasjev:2013}%
  \BibitemOpen
  \bibfield  {author} {\bibinfo {author} {\bibfnamefont {A.~V.}\ \bibnamefont
  {Afanasjev}}, \bibinfo {author} {\bibfnamefont {S.~E.}\ \bibnamefont
  {Agbemava}}, \bibinfo {author} {\bibfnamefont {D.}~\bibnamefont {Ray}}, \
  and\ \bibinfo {author} {\bibfnamefont {P.}~\bibnamefont {Ring}},\ }\href@noop
  {} {\bibfield  {journal} {\bibinfo  {journal} {Phys. Lett. B}\ }\textbf
  {\bibinfo {volume} {726}},\ \bibinfo {pages} {680} (\bibinfo {year}
  {2013})}\BibitemShut {NoStop}%
\bibitem [{\citenamefont {Lattimer}\ and\ \citenamefont
  {Prakash}(2004)}]{Lattimer:2004pg}%
  \BibitemOpen
  \bibfield  {author} {\bibinfo {author} {\bibfnamefont {J.~M.}\ \bibnamefont
  {Lattimer}}\ and\ \bibinfo {author} {\bibfnamefont {M.}~\bibnamefont
  {Prakash}},\ }\href {\doibase 10.1126/science.1090720} {\bibfield  {journal}
  {\bibinfo  {journal} {Science}\ }\textbf {\bibinfo {volume} {304}},\ \bibinfo
  {pages} {536} (\bibinfo {year} {2004})}\ \BibitemShut
  {NoStop}%
\bibitem [{\citenamefont {Demorest}\ \emph {et~al.}(2010)\citenamefont
  {Demorest}, \citenamefont {Pennucci}, \citenamefont {Ransom}, \citenamefont
  {Roberts},\ and\ \citenamefont {Hessels}}]{Demorest:2010bx}%
  \BibitemOpen
  \bibfield  {author} {\bibinfo {author} {\bibfnamefont {P.}~\bibnamefont
  {Demorest}}, \bibinfo {author} {\bibfnamefont {T.}~\bibnamefont {Pennucci}},
  \bibinfo {author} {\bibfnamefont {S.}~\bibnamefont {Ransom}}, \bibinfo
  {author} {\bibfnamefont {M.}~\bibnamefont {Roberts}}, \ and\ \bibinfo
  {author} {\bibfnamefont {J.}~\bibnamefont {Hessels}},\ }\href {\doibase
  10.1038/nature09466} {\bibfield  {journal} {\bibinfo  {journal} {Nature}\
  }\textbf {\bibinfo {volume} {467}},\ \bibinfo {pages} {1081} (\bibinfo {year}
  {2010})}\ \BibitemShut {NoStop}%
\bibitem [{\citenamefont {Antoniadis}\ \emph {et~al.}(2013)\citenamefont
  {Antoniadis}, \citenamefont {Freire}, \citenamefont {Wex}, \citenamefont
  {Tauris}, \citenamefont {Lynch} \emph {et~al.}}]{Antoniadis:2013pzd}%
  \BibitemOpen
  \bibfield  {author} {\bibinfo {author} {\bibfnamefont {J.}~\bibnamefont
  {Antoniadis}}, \bibinfo {author} {\bibfnamefont {P.~C.}\ \bibnamefont
  {Freire}}, \bibinfo {author} {\bibfnamefont {N.}~\bibnamefont {Wex}},
  \bibinfo {author} {\bibfnamefont {T.~M.}\ \bibnamefont {Tauris}}, \bibinfo
  {author} {\bibfnamefont {R.~S.}\ \bibnamefont {Lynch}},  \emph {et~al.},\
  }\href {\doibase 10.1126/science.1233232} {\bibfield  {journal} {\bibinfo
  {journal} {Science}\ }\textbf {\bibinfo {volume} {340}},\ \bibinfo {pages}
  {6131} (\bibinfo {year} {2013})}\ \BibitemShut {NoStop}%
\bibitem [{\citenamefont {Editors}(2011)}]{PhysRevA.83.040001}%
  \BibitemOpen
  \bibfield  {author} {\bibinfo {author} {\bibnamefont {The Editors}},\ }\href
  {\doibase 10.1103/PhysRevA.83.040001} {\bibfield  {journal} {\bibinfo
  {journal} {Phys. Rev. A}\ }\textbf {\bibinfo {volume} {83}},\ \bibinfo
  {pages} {040001} (\bibinfo {year} {2011})}\BibitemShut {NoStop}%
\bibitem [{\citenamefont {Reinhard}\ and\ \citenamefont
  {Nazarewicz}(2010)}]{Reinhard:2010wz}%
  \BibitemOpen
  \bibfield  {author} {\bibinfo {author} {\bibfnamefont {P.-G.}\ \bibnamefont
  {Reinhard}}\ and\ \bibinfo {author} {\bibfnamefont {W.}~\bibnamefont
  {Nazarewicz}},\ }\href {\doibase 10.1103/PhysRevC.81.051303} {\bibfield
  {journal} {\bibinfo  {journal} {Phys. Rev. C}\ }\textbf {\bibinfo {volume}
  {81}},\ \bibinfo {pages} {051303} (\bibinfo {year} {2010})}\ \BibitemShut
  {NoStop}%
\bibitem [{\citenamefont {Fattoyev}\ and\ \citenamefont
  {Piekarewicz}(2011)}]{Fattoyev:2011ns}%
  \BibitemOpen
  \bibfield  {author} {\bibinfo {author} {\bibfnamefont {F.~J.}~\bibnamefont
  {Fattoyev}}\ and\ \bibinfo {author} {\bibfnamefont {J.}~\bibnamefont
  {Piekarewicz}},\ }\href@noop {} {\bibfield  {journal} {\bibinfo  {journal}
  {Phys. Rev. C}\ }\textbf {\bibinfo {volume} {84}},\ \bibinfo {pages} {064302}
  (\bibinfo {year} {2011})}\ \BibitemShut {NoStop}%
\bibitem [{\citenamefont {Fattoyev}\ and\ \citenamefont
  {Piekarewicz}(2012)}]{Fattoyev:2012rm}%
  \BibitemOpen
  \bibfield  {author} {\bibinfo {author} {\bibfnamefont {F.~J.}~\bibnamefont
  {Fattoyev}}\ and\ \bibinfo {author} {\bibfnamefont {J.}~\bibnamefont
  {Piekarewicz}},\ }\href@noop {} {\bibfield  {journal} {\bibinfo  {journal}
  {Phys. Rev. C}\ }\textbf {\bibinfo {volume} {86}},\ \bibinfo {pages} {015802}
  (\bibinfo {year} {2012})}\ \BibitemShut {NoStop}%
\bibitem [{\citenamefont {Reinhard}\ and\ \citenamefont
  {Nazarewicz}(2013)}]{Reinhard:2012vw}%
  \BibitemOpen
  \bibfield  {author} {\bibinfo {author} {\bibfnamefont {P.-G.}~\bibnamefont
  {Reinhard}}\ and\ \bibinfo {author} {\bibfnamefont {W.}~\bibnamefont
  {Nazarewicz}},\ }\href {\doibase 10.1103/PhysRevC.87.014324} {\bibfield
  {journal} {\bibinfo  {journal} {Phys. Rev. C}\ }\textbf {\bibinfo {volume}
  {87}},\ \bibinfo {pages} {014324} (\bibinfo {year} {2013})}\ \BibitemShut
  {NoStop}%
\bibitem [{\citenamefont {Reinhard}\ \emph {et~al.}(2013)\citenamefont
  {Reinhard}, \citenamefont {Piekarewicz}, \citenamefont {Nazarewicz},
  \citenamefont {Agrawal}, \citenamefont {Paar} \emph
  {et~al.}}]{Reinhard:2013fpa}%
  \BibitemOpen
  \bibfield  {author} {\bibinfo {author} {\bibfnamefont {P.-G.}\ \bibnamefont
  {Reinhard}}, \bibinfo {author} {\bibfnamefont {J.}~\bibnamefont
  {Piekarewicz}}, \bibinfo {author} {\bibfnamefont {W.}~\bibnamefont
  {Nazarewicz}}, \bibinfo {author} {\bibfnamefont {B.}~\bibnamefont {Agrawal}},
  \bibinfo {author} {\bibfnamefont {N.}~\bibnamefont {Paar}},  \emph {et~al.},\
  }\href@noop {} {\bibfield  {journal} {\bibinfo  {journal}
      {Phys. Rev. C}\
  }\textbf {\bibinfo {volume} {88}},\ \bibinfo {pages} {034325} (\bibinfo
  {year} {2013})}\ \BibitemShut {NoStop}%
\bibitem [{\citenamefont {Dobaczewski}\ \emph {et~al.}(2014)\citenamefont
  {Dobaczewski}, \citenamefont {Nazarewicz},\ and\ \citenamefont
  {Reinhard}}]{Dobaczewski:2014jga}%
  \BibitemOpen
  \bibfield  {author} {\bibinfo {author} {\bibfnamefont {J.}~\bibnamefont
  {Dobaczewski}}, \bibinfo {author} {\bibfnamefont {W.}~\bibnamefont
  {Nazarewicz}}, \ and\ \bibinfo {author} {\bibfnamefont {P.-G.}\ \bibnamefont
  {Reinhard}},\ }\href {\doibase 10.1088/0954-3899/41/7/074001} {\bibfield
  {journal} {\bibinfo  {journal} {J. Phys. G}\ }\textbf {\bibinfo {volume}
  {41}},\ \bibinfo {pages} {074001} (\bibinfo {year} {2014})}\ \BibitemShut
  {NoStop}%
\bibitem [{\citenamefont {Piekarewicz}\ \emph {et~al.}(2014)\citenamefont
  {Piekarewicz}, \citenamefont {Chen},\ and\ \citenamefont
  {Fattoyev}}]{Piekarewicz:2014kza}%
  \BibitemOpen
  \bibfield  {author} {\bibinfo {author} {\bibfnamefont {J.}~\bibnamefont
  {Piekarewicz}}, \bibinfo {author} {\bibfnamefont {W.-C.}\ \bibnamefont
  {Chen}}, \ and\ \bibinfo {author} {\bibfnamefont {F.~J.}~\bibnamefont
  {Fattoyev}},\ }\href@noop {} {\bibinfo  {journal} {arXiv:1407.0911}\
  (\bibinfo {year} {2014})} \BibitemShut
  {NoStop}%
\bibitem [{\citenamefont {Ozel}\ \emph {et~al.}(2010)\citenamefont {Ozel},
  \citenamefont {Baym},\ and\ \citenamefont {Guver}}]{Ozel:2010fw}%
  \BibitemOpen
  \bibfield  {author} {\bibinfo {author} {\bibfnamefont {F.}~\bibnamefont
  {Ozel}}, \bibinfo {author} {\bibfnamefont {G.}~\bibnamefont {Baym}}, \ and\
  \bibinfo {author} {\bibfnamefont {T.}~\bibnamefont {Guver}},\ }\href
  {\doibase 10.1103/PhysRevD.82.101301} {\bibfield  {journal} {\bibinfo
  {journal} {Phys. Rev. D}\ }\textbf {\bibinfo {volume} {82}},\ \bibinfo {pages}
  {101301} (\bibinfo {year} {2010})}\ \BibitemShut {NoStop}%
\bibitem [{\citenamefont {Steiner}\ \emph {et~al.}(2010)\citenamefont
  {Steiner}, \citenamefont {Lattimer},\ and\ \citenamefont
  {Brown}}]{Steiner:2010fz}%
  \BibitemOpen
  \bibfield  {author} {\bibinfo {author} {\bibfnamefont {A.~W.}\ \bibnamefont
  {Steiner}}, \bibinfo {author} {\bibfnamefont {J.~M.}\ \bibnamefont
  {Lattimer}}, \ and\ \bibinfo {author} {\bibfnamefont {E.~F.}\ \bibnamefont
  {Brown}},\ }\href {\doibase 10.1088/0004-637X/722/1/33} {\bibfield  {journal}
  {\bibinfo  {journal} {Astrophys. J.}\ }\textbf {\bibinfo {volume} {722}},\
  \bibinfo {pages} {33} (\bibinfo {year} {2010})}\ \BibitemShut
  {NoStop}%
\bibitem [{\citenamefont {Suleimanov}\ \emph {et~al.}(2011)\citenamefont
  {Suleimanov}, \citenamefont {Poutanen}, \citenamefont {Revnivtsev},\ and\
  \citenamefont {Werner}}]{Suleimanov:2010th}%
  \BibitemOpen
  \bibfield  {author} {\bibinfo {author} {\bibfnamefont {V.}~\bibnamefont
  {Suleimanov}}, \bibinfo {author} {\bibfnamefont {J.}~\bibnamefont
  {Poutanen}}, \bibinfo {author} {\bibfnamefont {M.}~\bibnamefont
  {Revnivtsev}}, \ and\ \bibinfo {author} {\bibfnamefont {K.}~\bibnamefont
  {Werner}},\ }\href {\doibase 10.1088/0004-637X/742/2/122} {\bibfield
  {journal} {\bibinfo  {journal} {Astrophys. J.}\ }\textbf {\bibinfo {volume}
  {742}},\ \bibinfo {pages} {122} (\bibinfo {year} {2011})}\ \BibitemShut
  {NoStop}%
\bibitem [{\citenamefont {Guillot}\ \emph {et~al.}(2013)\citenamefont
  {Guillot}, \citenamefont {Servillat}, \citenamefont {Webb},\ and\
  \citenamefont {Rutledge}}]{Guillot:2013wu}%
  \BibitemOpen
  \bibfield  {author} {\bibinfo {author} {\bibfnamefont {S.}~\bibnamefont
  {Guillot}}, \bibinfo {author} {\bibfnamefont {M.}~\bibnamefont {Servillat}},
  \bibinfo {author} {\bibfnamefont {N.~A.}\ \bibnamefont {Webb}}, \ and\
  \bibinfo {author} {\bibfnamefont {R.~E.}\ \bibnamefont {Rutledge}},\
  }\href@noop {} {\bibfield  {journal} {\bibinfo  {journal} {Astrophys. J.}\
  }\textbf {\bibinfo {volume} {772}},\ \bibinfo {pages} {7} (\bibinfo {year}
  {2013})}\ \BibitemShut {NoStop}%
\bibitem [{\citenamefont {Schwenk}\ and\ \citenamefont
  {Pethick}(2005)}]{Schwenk:2005ka}%
  \BibitemOpen
  \bibfield  {author} {\bibinfo {author} {\bibfnamefont {A.}~\bibnamefont
  {Schwenk}}\ and\ \bibinfo {author} {\bibfnamefont {C.~J.}\ \bibnamefont
  {Pethick}},\ }\href@noop {} {\bibfield  {journal} {\bibinfo  {journal} {Phys.
  Rev. Lett.}\ }\textbf {\bibinfo {volume} {95}},\ \bibinfo {pages} {160401}
  (\bibinfo {year} {2005})}\ \BibitemShut {NoStop}%
\bibitem [{\citenamefont {Gezerlis}\ and\ \citenamefont
  {Carlson}(2010)}]{Gezerlis:2009iw}%
  \BibitemOpen
  \bibfield  {author} {\bibinfo {author} {\bibfnamefont {A.}~\bibnamefont
  {Gezerlis}}\ and\ \bibinfo {author} {\bibfnamefont {J.}~\bibnamefont
  {Carlson}},\ }\href {\doibase 10.1103/PhysRevC.81.025803} {\bibfield
  {journal} {\bibinfo  {journal} {Phys. Rev. C}\ }\textbf {\bibinfo {volume}
  {81}},\ \bibinfo {pages} {025803} (\bibinfo {year} {2010})}\ \BibitemShut
  {NoStop}%
\bibitem [{\citenamefont {Vidana}\ \emph {et~al.}(2009)\citenamefont {Vidana},
  \citenamefont {Providencia}, \citenamefont {Polls},\ and\ \citenamefont
  {Rios}}]{Vidana:2009is}%
  \BibitemOpen
  \bibfield  {author} {\bibinfo {author} {\bibfnamefont {I.}~\bibnamefont
  {Vidana}}, \bibinfo {author} {\bibfnamefont {C.}~\bibnamefont {Providencia}},
  \bibinfo {author} {\bibfnamefont {A.}~\bibnamefont {Polls}}, \ and\ \bibinfo
  {author} {\bibfnamefont {A.}~\bibnamefont {Rios}},\ }\href {\doibase
  10.1103/PhysRevC.80.045806} {\bibfield  {journal} {\bibinfo  {journal} {Phys.
  Rev. C}\ }\textbf {\bibinfo {volume} {80}},\ \bibinfo {pages} {045806}
  (\bibinfo {year} {2009})}\ \BibitemShut {NoStop}%
\bibitem [{\citenamefont {Hebeler}\ \emph {et~al.}(2010)\citenamefont
  {Hebeler}, \citenamefont {Lattimer}, \citenamefont {Pethick},\ and\
  \citenamefont {Schwenk}}]{Hebeler:2010jx}%
  \BibitemOpen
  \bibfield  {author} {\bibinfo {author} {\bibfnamefont {K.}~\bibnamefont
  {Hebeler}}, \bibinfo {author} {\bibfnamefont {J.~M.}~\bibnamefont {Lattimer}},
  \bibinfo {author} {\bibfnamefont {C.~J.}~\bibnamefont {Pethick}}, \ and\
  \bibinfo {author} {\bibfnamefont {A.}~\bibnamefont {Schwenk}},\ }\href
  {\doibase 10.1103/PhysRevLett.105.161102} {\bibfield  {journal} {\bibinfo
  {journal} {Phys. Rev. Lett.}\ }\textbf {\bibinfo {volume} {105}},\ \bibinfo
  {pages} {161102} (\bibinfo {year} {2010})}\ \BibitemShut
  {NoStop}%
\bibitem [{\citenamefont {Hebeler}\ \emph {et~al.}(2013)\citenamefont
  {Hebeler}, \citenamefont {Lattimer}, \citenamefont {Pethick},\ and\
  \citenamefont {Schwenk}}]{Hebeler:2013nza}%
  \BibitemOpen
  \bibfield  {author} {\bibinfo {author} {\bibfnamefont {K.}~\bibnamefont
  {Hebeler}}, \bibinfo {author} {\bibfnamefont {J.}~\bibnamefont {Lattimer}},
  \bibinfo {author} {\bibfnamefont {C.}~\bibnamefont {Pethick}}, \ and\
  \bibinfo {author} {\bibfnamefont {A.}~\bibnamefont {Schwenk}},\ }\href
  {\doibase 10.1088/0004-637X/773/1/11} {\bibfield  {journal} {\bibinfo
  {journal} {Astrophys. J.}\ }\textbf {\bibinfo {volume} {773}},\ \bibinfo
  {pages} {11} (\bibinfo {year} {2013})}\ \BibitemShut
  {NoStop}%
\bibitem [{\citenamefont {Mueller}\ and\ \citenamefont
  {Serot}(1996)}]{Mueller:1996pm}%
  \BibitemOpen
  \bibfield  {author} {\bibinfo {author} {\bibfnamefont {H.}~\bibnamefont
  {Mueller}}\ and\ \bibinfo {author} {\bibfnamefont {B.~D.}\ \bibnamefont
  {Serot}},\ }\href@noop {} {\bibfield  {journal} {\bibinfo  {journal} {Nucl.
  Phys. A}\ }\textbf {\bibinfo {volume} {606}},\ \bibinfo {pages} {508}
  (\bibinfo {year} {1996})}\ \BibitemShut {NoStop}%
\bibitem [{\citenamefont {Serot}\ and\ \citenamefont
  {Walecka}(1997)}]{Serot:1997xg}%
  \BibitemOpen
  \bibfield  {author} {\bibinfo {author} {\bibfnamefont {B.~D.}\ \bibnamefont
  {Serot}}\ and\ \bibinfo {author} {\bibfnamefont {J.~D.}\ \bibnamefont
  {Walecka}},\ }\href@noop {} {\bibfield  {journal} {\bibinfo  {journal} {Int.
  J. Mod. Phys. E}\ }\textbf {\bibinfo {volume} {6}},\ \bibinfo {pages} {515}
  (\bibinfo {year} {1997})}\ \BibitemShut {NoStop}%
\bibitem [{\citenamefont {Horowitz}\ and\ \citenamefont
  {Piekarewicz}(2001{\natexlab{a}})}]{Horowitz:2000xj}%
  \BibitemOpen
  \bibfield  {author} {\bibinfo {author} {\bibfnamefont {C.~J.}\ \bibnamefont
  {Horowitz}}\ and\ \bibinfo {author} {\bibfnamefont {J.}~\bibnamefont
  {Piekarewicz}},\ }\href@noop {} {\bibfield  {journal} {\bibinfo  {journal}
  {Phys. Rev. Lett.}\ }\textbf {\bibinfo {volume} {86}},\ \bibinfo {pages}
  {5647} (\bibinfo {year} {2001}{\natexlab{a}})}\ \BibitemShut
  {NoStop}%
\bibitem [{\citenamefont {Furnstahl}\ \emph
  {et~al.}(1997{\natexlab{a}})\citenamefont {Furnstahl}, \citenamefont
  {Serot},\ and\ \citenamefont {Tang}}]{Furnstahl:1996wv}%
  \BibitemOpen
  \bibfield  {author} {\bibinfo {author} {\bibfnamefont {R.~J.}~\bibnamefont
  {Furnstahl}}, \bibinfo {author} {\bibfnamefont {B.~D.}\ \bibnamefont
  {Serot}}, \ and\ \bibinfo {author} {\bibfnamefont {H.-B.}\ \bibnamefont
  {Tang}},\ }\href {\doibase 10.1016/S0375-9474(96)00472-1} {\bibfield
  {journal} {\bibinfo  {journal} {Nucl. Phys. A}\ }\textbf {\bibinfo {volume}
  {615}},\ \bibinfo {pages} {441} (\bibinfo {year} {1997}{\natexlab{a}})}\
  \BibitemShut {NoStop}%
\bibitem [{\citenamefont {Furnstahl}\ \emph
  {et~al.}(1997{\natexlab{b}})\citenamefont {Furnstahl}, \citenamefont
  {Serot},\ and\ \citenamefont {Tang}}]{Furnstahl:1996zm}%
  \BibitemOpen
  \bibfield  {author} {\bibinfo {author} {\bibfnamefont {R.~J.}\ \bibnamefont
  {Furnstahl}}, \bibinfo {author} {\bibfnamefont {B.~D.}\ \bibnamefont
  {Serot}}, \ and\ \bibinfo {author} {\bibfnamefont {H.-B.}\ \bibnamefont
  {Tang}},\ }\href@noop {} {\bibfield  {journal} {\bibinfo  {journal} {Nucl.
  Phys. A}\ }\textbf {\bibinfo {volume} {618}},\ \bibinfo {pages} {446}
  (\bibinfo {year} {1997}{\natexlab{b}})}\ \BibitemShut
  {NoStop}%
\bibitem [{\citenamefont {Rusnak}\ and\ \citenamefont
  {Furnstahl}(1997)}]{Rusnak:1997dj}%
  \BibitemOpen
  \bibfield  {author} {\bibinfo {author} {\bibfnamefont {J.~J.}\ \bibnamefont
  {Rusnak}}\ and\ \bibinfo {author} {\bibfnamefont {R.~J.}~\bibnamefont
  {Furnstahl}},\ }\href {\doibase 10.1016/S0375-9474(97)00598-8} {\bibfield
  {journal} {\bibinfo  {journal} {Nucl. Phys. A}\ }\textbf {\bibinfo {volume}
  {627}},\ \bibinfo {pages} {495} (\bibinfo {year} {1997})}\ 
  \BibitemShut {NoStop}%
\bibitem [{\citenamefont {Furnstahl}\ and\ \citenamefont
  {Hackworth}(1997)}]{Furnstahl:1997hq}%
  \BibitemOpen
  \bibfield  {author} {\bibinfo {author} {\bibfnamefont {R.~J.}~\bibnamefont
  {Furnstahl}}\ and\ \bibinfo {author} {\bibfnamefont {J.~C.}\ \bibnamefont
  {Hackworth}},\ }\href {\doibase 10.1103/PhysRevC.56.2875} {\bibfield
  {journal} {\bibinfo  {journal} {Phys. Rev. C}\ }\textbf {\bibinfo {volume}
  {56}},\ \bibinfo {pages} {2875} (\bibinfo {year} {1997})}\ 
  \BibitemShut {NoStop}%
\bibitem [{\citenamefont {Kortelainen}\ \emph
  {et~al.}(2010{\natexlab{b}})\citenamefont {Kortelainen}, \citenamefont
  {Furnstahl}, \citenamefont {Nazarewicz},\ and\ \citenamefont
  {Stoitsov}}]{Kortelainen:2010dt}%
  \BibitemOpen
  \bibfield  {author} {\bibinfo {author} {\bibfnamefont {M.}~\bibnamefont
  {Kortelainen}}, \bibinfo {author} {\bibfnamefont {R.~J.}~\bibnamefont
  {Furnstahl}}, \bibinfo {author} {\bibfnamefont {W.}~\bibnamefont
  {Nazarewicz}}, \ and\ \bibinfo {author} {\bibfnamefont {M.~V.}~\bibnamefont
  {Stoitsov}},\ }\href {\doibase 10.1103/PhysRevC.82.011304} {\bibfield
  {journal} {\bibinfo  {journal} {Phys. Rev. C}\ }\textbf {\bibinfo {volume}
  {82}},\ \bibinfo {pages} {011304} (\bibinfo {year} {2010}{\natexlab{b}})}\
  \BibitemShut {NoStop}%
\bibitem [{\citenamefont {Boguta}\ and\ \citenamefont
  {Bodmer}(1977)}]{Boguta:1977xi}%
  \BibitemOpen
  \bibfield  {author} {\bibinfo {author} {\bibfnamefont {J.}~\bibnamefont
  {Boguta}}\ and\ \bibinfo {author} {\bibfnamefont {A.~R.}\ \bibnamefont
  {Bodmer}},\ }\href@noop {} {\bibfield  {journal} {\bibinfo  {journal} {Nucl.
  Phys. A}\ }\textbf {\bibinfo {volume} {292}},\ \bibinfo {pages} {413}
  (\bibinfo {year} {1977})}\BibitemShut {NoStop}%
\bibitem [{\citenamefont {Horowitz}\ and\ \citenamefont
  {Piekarewicz}(2001{\natexlab{b}})}]{Horowitz:2001ya}%
  \BibitemOpen
  \bibfield  {author} {\bibinfo {author} {\bibfnamefont {C.~J.}\ \bibnamefont
  {Horowitz}}\ and\ \bibinfo {author} {\bibfnamefont {J.}~\bibnamefont
  {Piekarewicz}},\ }\href@noop {} {\bibfield  {journal} {\bibinfo  {journal}
  {Phys. Rev. C}\ }\textbf {\bibinfo {volume} {64}},\ \bibinfo {pages} {062802}
  (\bibinfo {year} {2001}{\natexlab{b}})}\ \BibitemShut
  {NoStop}%
\bibitem [{\citenamefont {Carriere}\ \emph {et~al.}(2003)\citenamefont
  {Carriere}, \citenamefont {Horowitz},\ and\ \citenamefont
  {Piekarewicz}}]{Carriere:2002bx}%
  \BibitemOpen
  \bibfield  {author} {\bibinfo {author} {\bibfnamefont {J.}~\bibnamefont
  {Carriere}}, \bibinfo {author} {\bibfnamefont {C.~J.}\ \bibnamefont
  {Horowitz}}, \ and\ \bibinfo {author} {\bibfnamefont {J.}~\bibnamefont
  {Piekarewicz}},\ }\href@noop {} {\bibfield  {journal} {\bibinfo  {journal}
  {Astrophys. J.}\ }\textbf {\bibinfo {volume} {593}},\ \bibinfo {pages} {463}
  (\bibinfo {year} {2003})}\ \BibitemShut {NoStop}%
\bibitem [{\citenamefont {Horowitz}\ \emph {et~al.}(2004)\citenamefont
  {Horowitz}, \citenamefont {Perez-Garcia},\ and\ \citenamefont
  {Piekarewicz}}]{Horowitz:2004yf}%
  \BibitemOpen
  \bibfield  {author} {\bibinfo {author} {\bibfnamefont {C.~J.}\ \bibnamefont
  {Horowitz}}, \bibinfo {author} {\bibfnamefont {M.~A.}\ \bibnamefont
  {Perez-Garcia}}, \ and\ \bibinfo {author} {\bibfnamefont {J.}~\bibnamefont
  {Piekarewicz}},\ }\href@noop {} {\bibfield  {journal} {\bibinfo  {journal}
  {Phys. Rev. C}\ }\textbf {\bibinfo {volume} {69}},\ \bibinfo {pages} {045804}
  (\bibinfo {year} {2004})}\ \BibitemShut {NoStop}%
\bibitem [{\citenamefont {Todd}\ and\ \citenamefont
  {Piekarewicz}(2003)}]{Todd:2003xs}%
  \BibitemOpen
  \bibfield  {author} {\bibinfo {author} {\bibfnamefont {B.~G.}\ \bibnamefont
  {Todd}}\ and\ \bibinfo {author} {\bibfnamefont {J.}~\bibnamefont
  {Piekarewicz}},\ }\href@noop {} {\bibfield  {journal} {\bibinfo  {journal}
  {Phys. Rev. C}\ }\textbf {\bibinfo {volume} {67}},\ \bibinfo {pages} {044317}
  (\bibinfo {year} {2003})}\ \BibitemShut {NoStop}%
\bibitem [{\citenamefont {Piekarewicz}\ and\ \citenamefont
  {Centelles}(2009)}]{Piekarewicz:2008nh}%
  \BibitemOpen
  \bibfield  {author} {\bibinfo {author} {\bibfnamefont {J.}~\bibnamefont
  {Piekarewicz}}\ and\ \bibinfo {author} {\bibfnamefont {M.}~\bibnamefont
  {Centelles}},\ }\href {\doibase 10.1103/PhysRevC.79.054311} {\bibfield
  {journal} {\bibinfo  {journal} {Phys. Rev. C}\ }\textbf {\bibinfo {volume}
  {79}},\ \bibinfo {pages} {054311} (\bibinfo {year} {2009})}\ \BibitemShut
  {NoStop}%
\bibitem [{\citenamefont {Baym}\ \emph {et~al.}(1971)\citenamefont {Baym},
  \citenamefont {Pethick},\ and\ \citenamefont {Sutherland}}]{Baym:1971pw}%
  \BibitemOpen
  \bibfield  {author} {\bibinfo {author} {\bibfnamefont {G.}~\bibnamefont
  {Baym}}, \bibinfo {author} {\bibfnamefont {C.}~\bibnamefont {Pethick}}, \
  and\ \bibinfo {author} {\bibfnamefont {P.}~\bibnamefont {Sutherland}},\
  }\href@noop {} {\bibfield  {journal} {\bibinfo  {journal} {Astrophys. J.}\
  }\textbf {\bibinfo {volume} {170}},\ \bibinfo {pages} {299} (\bibinfo {year}
  {1971})}\BibitemShut {NoStop}%
\bibitem [{\citenamefont {Link}\ \emph {et~al.}(1999)\citenamefont {Link},
  \citenamefont {Epstein},\ and\ \citenamefont {Lattimer}}]{Link:1999ca}%
  \BibitemOpen
  \bibfield  {author} {\bibinfo {author} {\bibfnamefont {B.}~\bibnamefont
  {Link}}, \bibinfo {author} {\bibfnamefont {R.~I.}\ \bibnamefont {Epstein}}, \
  and\ \bibinfo {author} {\bibfnamefont {J.~M.}\ \bibnamefont {Lattimer}},\
  }\href {\doibase 10.1103/PhysRevLett.83.3362} {\bibfield  {journal} {\bibinfo
   {journal} {Phys. Rev. Lett.}\ }\textbf {\bibinfo {volume} {83}},\ \bibinfo
  {pages} {3362} (\bibinfo {year} {1999})}\ \BibitemShut
  {NoStop}%
\bibitem [{\citenamefont {Glendenning}(2000)}]{Glendenning:2000}%
  \BibitemOpen
  \bibfield  {author} {\bibinfo {author} {\bibfnamefont {N.~K.}\ \bibnamefont
  {Glendenning}},\ }\enquote {\bibinfo {title} {Compact stars},}\ \ (\bibinfo
  {publisher} {Springer-Verlag New York},\ \bibinfo {year} {2000})\BibitemShut
  {NoStop}%
\bibitem [{\citenamefont {Agrawal}\ \emph {et~al.}(2005)\citenamefont
  {Agrawal}, \citenamefont {Shlomo},\ and\ \citenamefont
  {Au}}]{Agrawal:2005ix}%
  \BibitemOpen
  \bibfield  {author} {\bibinfo {author} {\bibfnamefont {B.~K.}\ \bibnamefont
  {Agrawal}}, \bibinfo {author} {\bibfnamefont {S.}~\bibnamefont {Shlomo}}, \
  and\ \bibinfo {author} {\bibfnamefont {V.~K.}\ \bibnamefont {Au}},\
  }\href@noop {} {\bibfield  {journal} {\bibinfo  {journal}
      {Phys. Rev. C}\
  }\textbf {\bibinfo {volume} {72}},\ \bibinfo {pages} {014310} (\bibinfo
  {year} {2005})}\ \BibitemShut {NoStop}%
\bibitem [{\citenamefont {Chen}\ \emph {et~al.}(2010)\citenamefont {Chen},
  \citenamefont {Ko}, \citenamefont {Li},\ and\ \citenamefont
  {Xu}}]{Chen:2010qx}%
  \BibitemOpen
  \bibfield  {author} {\bibinfo {author} {\bibfnamefont {L.-W.}\ \bibnamefont
  {Chen}}, \bibinfo {author} {\bibfnamefont {C.~M.}\ \bibnamefont {Ko}},
  \bibinfo {author} {\bibfnamefont {B.-A.}\ \bibnamefont {Li}}, \ and\ \bibinfo
  {author} {\bibfnamefont {J.}~\bibnamefont {Xu}},\ }\href {\doibase
  10.1103/PhysRevC.82.024321} {\bibfield  {journal} {\bibinfo  {journal}
  {Phys. Rev. C}\ }\textbf {\bibinfo {volume} {82}},\ \bibinfo {pages} {024321}
  (\bibinfo {year} {2010})}\ \BibitemShut {NoStop}%
\bibitem [{\citenamefont {Brandt}(1999)}]{Brandt:1999}%
  \BibitemOpen
  \bibfield  {author} {\bibinfo {author} {\bibfnamefont {S.}~\bibnamefont
  {Brandt}},\ }\enquote {\bibinfo {title} {Data analysis: Statistical and
  computational methods for scientists and engineers},}\ \ (\bibinfo
  {publisher} {Springer},\ \bibinfo {address} {New York},\ \bibinfo {year}
  {1999})\ \bibinfo {edition} {3rd}\ ed.\BibitemShut {Stop}%
\bibitem [{\citenamefont {Bevington}\ and\ \citenamefont
  {Robinson}(2003)}]{Bevington2003}%
  \BibitemOpen
  \bibfield  {author} {\bibinfo {author} {\bibfnamefont {P.}~\bibnamefont
  {Bevington}}\ and\ \bibinfo {author} {\bibfnamefont {D.}~\bibnamefont
  {Robinson}},\ }\enquote {\bibinfo {title} {Data reduction and error
  analysis},}\ \ (\bibinfo  {publisher} {McGraw Hill},\ \bibinfo {address} {New
  York},\ \bibinfo {year} {2003})\ \bibinfo {edition} {3rd}\ ed.\BibitemShut
  {Stop}%
\bibitem [{\citenamefont {Press}\ \emph {et~al.}(1989)\citenamefont {Press},
  \citenamefont {Flannery}, \citenamefont {Teukolsky},\ and\ \citenamefont
  {Vetterling}}]{NumericalRecipes}%
  \BibitemOpen
  \bibfield  {author} {\bibinfo {author} {\bibfnamefont {W.~H.}\ \bibnamefont
  {Press}}, \bibinfo {author} {\bibfnamefont {B.~P.}\ \bibnamefont {Flannery}},
  \bibinfo {author} {\bibfnamefont {S.~A.}\ \bibnamefont {Teukolsky}}, \ and\
  \bibinfo {author} {\bibfnamefont {W.~T.}\ \bibnamefont {Vetterling}},\
  }\enquote {\bibinfo {title} {Numerical recipes: The art of scientific
  computing},}\ \ (\bibinfo  {publisher} {Cambridge University Press},\
  \bibinfo {year} {1989})\BibitemShut {NoStop}%
\bibitem [{\citenamefont {Wang}\ \emph {et~al.}(2012)\citenamefont {Wang},
  \citenamefont {Audi}, \citenamefont {Wapstra}, \citenamefont {Kondev},
  \citenamefont {MacCormick}, \citenamefont {Xu},\ and\ \citenamefont
  {Pfeiffer}}]{Wang:2012}%
  \BibitemOpen
  \bibfield  {author} {\bibinfo {author} {\bibfnamefont {M.}~\bibnamefont
  {Wang}}, \bibinfo {author} {\bibfnamefont {G.}~\bibnamefont {Audi}}, \bibinfo
  {author} {\bibfnamefont {A.~H.}\ \bibnamefont {Wapstra}}, \bibinfo {author}
  {\bibfnamefont {F.~G.}\ \bibnamefont {Kondev}}, \bibinfo {author}
  {\bibfnamefont {M.}~\bibnamefont {MacCormick}}, \bibinfo {author}
  {\bibfnamefont {X.}~\bibnamefont {Xu}}, \ and\ \bibinfo {author}
  {\bibfnamefont {B.}~\bibnamefont {Pfeiffer}},\ }\href@noop {} {\bibfield
  {journal} {\bibinfo  {journal} {Chinese Phys. C}\ }\textbf {\bibinfo {volume}
  {36}},\ \bibinfo {pages} {1603} (\bibinfo {year} {2012})}\BibitemShut
  {NoStop}%
\bibitem [{\citenamefont {Angeli}\ and\ \citenamefont
  {Marinova}(2013)}]{Angeli:2013}%
  \BibitemOpen
  \bibfield  {author} {\bibinfo {author} {\bibfnamefont {I.}~\bibnamefont
  {Angeli}}\ and\ \bibinfo {author} {\bibfnamefont {K.}~\bibnamefont
  {Marinova}},\ }\href@noop {} {\bibfield  {journal} {\bibinfo  {journal} {At.
  Data Nucl. Data Tables}\ }\textbf {\bibinfo {volume} {99}},\ \bibinfo {pages}
  {69 } (\bibinfo {year} {2013})}\BibitemShut {NoStop}%
\bibitem [{\citenamefont {Youngblood}\ \emph {et~al.}(1999)\citenamefont
  {Youngblood}, \citenamefont {Clark},\ and\ \citenamefont
  {Lui}}]{Youngblood:1999}%
  \BibitemOpen
  \bibfield  {author} {\bibinfo {author} {\bibfnamefont {D.~H.}\ \bibnamefont
  {Youngblood}}, \bibinfo {author} {\bibfnamefont {H.~L.}\ \bibnamefont
  {Clark}}, \ and\ \bibinfo {author} {\bibfnamefont {Y.-W.}\ \bibnamefont
  {Lui}},\ }\href {\doibase 10.1103/PhysRevLett.82.691} {\bibfield  {journal}
  {\bibinfo  {journal} {Phys. Rev. Lett.}\ }\textbf {\bibinfo {volume} {82}},\
  \bibinfo {pages} {691} (\bibinfo {year} {1999})}\BibitemShut {NoStop}%
\bibitem [{\citenamefont {Uchida}\ \emph {et~al.}(2004)\citenamefont {Uchida},
  \citenamefont {Sakaguchi}, \citenamefont {Itoh}, \citenamefont {Yosoi},
  \citenamefont {Kawabata} \emph {et~al.}}]{Uchida:2004bs}%
  \BibitemOpen
  \bibfield  {author} {\bibinfo {author} {\bibfnamefont {M.}~\bibnamefont
  {Uchida}}, \bibinfo {author} {\bibfnamefont {H.}~\bibnamefont {Sakaguchi}},
  \bibinfo {author} {\bibfnamefont {M.}~\bibnamefont {Itoh}}, \bibinfo {author}
  {\bibfnamefont {M.}~\bibnamefont {Yosoi}}, \bibinfo {author} {\bibfnamefont
  {T.}~\bibnamefont {Kawabata}},  \emph {et~al.},\ }\href {\doibase
  10.1103/PhysRevC.69.051301} {\bibfield  {journal} {\bibinfo  {journal}
  {Phys. Rev. C}\ }\textbf {\bibinfo {volume} {69}},\ \bibinfo {pages} {051301}
  (\bibinfo {year} {2004})}\BibitemShut {NoStop}%
\bibitem [{\citenamefont {Li}\ \emph {et~al.}(2007)\citenamefont {Li} \emph
  {et~al.}}]{Li:2007bp}%
  \BibitemOpen
  \bibfield  {author} {\bibinfo {author} {\bibfnamefont {T.}~\bibnamefont {Li}}
  \emph {et~al.},\ }\href {\doibase 10.1103/PhysRevLett.99.162503} {\bibfield
  {journal} {\bibinfo  {journal} {Phys. Rev. Lett.}\ }\textbf {\bibinfo
  {volume} {99}},\ \bibinfo {pages} {162503} (\bibinfo {year} {2007})}\
  \BibitemShut {NoStop}%
\bibitem [{\citenamefont {Li}\ \emph {et~al.}(2010)\citenamefont {Li} \emph
  {et~al.}}]{Li:2010kfa}%
  \BibitemOpen
  \bibfield  {author} {\bibinfo {author} {\bibfnamefont {T.}~\bibnamefont {Li}}
  \emph {et~al.},\ }\href {\doibase 10.1103/PhysRevC.81.034309} {\bibfield
  {journal} {\bibinfo  {journal} {Phys. Rev. C}\ }\textbf {\bibinfo {volume}
  {81}},\ \bibinfo {pages} {034309} (\bibinfo {year} {2010})}\ \BibitemShut
  {NoStop}%
\bibitem [{\citenamefont {Patel}\ \emph {et~al.}(2013)\citenamefont {Patel},
  \citenamefont {Garg}, \citenamefont {Fujiwara}, \citenamefont {Adachi},
  \citenamefont {Akimune} \emph {et~al.}}]{Patel:2013uyt}%
  \BibitemOpen
  \bibfield  {author} {\bibinfo {author} {\bibfnamefont {D.}~\bibnamefont
  {Patel}}, \bibinfo {author} {\bibfnamefont {U.}~\bibnamefont {Garg}},
  \bibinfo {author} {\bibfnamefont {M.}~\bibnamefont {Fujiwara}}, \bibinfo
  {author} {\bibfnamefont {T.}~\bibnamefont {Adachi}}, \bibinfo {author}
  {\bibfnamefont {H.}~\bibnamefont {Akimune}},  \emph {et~al.},\ }\href
  {\doibase 10.1016/j.physletb.2013.08.027} {\bibfield  {journal} {\bibinfo
  {journal} {Phys. Lett. B}\ }\textbf {\bibinfo {volume} {726}},\ \bibinfo
  {pages} {178} (\bibinfo {year} {2013})}\ \BibitemShut
  {NoStop}%
\bibitem [{\citenamefont {Patel}\ and\ \citenamefont {Garg}()}]{PatelPC:2013}%
  \BibitemOpen
  \bibfield  {author} {\bibinfo {author} {\bibfnamefont {D.}~\bibnamefont
  {Patel}}\ and\ \bibinfo {author} {\bibfnamefont {U.}~\bibnamefont {Garg}},\
  }\href@noop {} {}\bibinfo {howpublished} {private communication}\BibitemShut
  {NoStop}%
\bibitem [{\citenamefont {Chen}\ \emph {et~al.}(2013)\citenamefont {Chen},
  \citenamefont {Piekarewicz},\ and\ \citenamefont {Centelles}}]{Chen:2013tca}%
  \BibitemOpen
  \bibfield  {author} {\bibinfo {author} {\bibfnamefont {W.-C.}\ \bibnamefont
  {Chen}}, \bibinfo {author} {\bibfnamefont {J.}~\bibnamefont {Piekarewicz}}, \
  and\ \bibinfo {author} {\bibfnamefont {M.}~\bibnamefont {Centelles}},\
  }\href@noop {} {\bibfield  {journal} {\bibinfo  {journal} {Phys. Rev. C}\
  }\textbf {\bibinfo {volume} {88}},\ \bibinfo
  {pages} {024319} (\bibinfo {year} {2013})}\ \BibitemShut
  {NoStop}%
\bibitem [{\citenamefont {Ring}\ and\ \citenamefont
  {Schuck}(2004)}]{Ring:2004}%
  \BibitemOpen
  \bibfield  {author} {\bibinfo {author} {\bibfnamefont {P.}~\bibnamefont
  {Ring}}\ and\ \bibinfo {author} {\bibfnamefont {P.}~\bibnamefont {Schuck}},\
  }\enquote {\bibinfo {title} {The nuclear many-body problem},}\ \ (\bibinfo
  {publisher} {Springer, New York},\ \bibinfo {year} {2004})\BibitemShut
  {NoStop}%
\bibitem [{\citenamefont {Harakeh}\ and\ \citenamefont {van~der
  Woude}(2001)}]{Harakeh:2001}%
  \BibitemOpen
  \bibfield  {author} {\bibinfo {author} {\bibfnamefont {M.~N.}\ \bibnamefont
  {Harakeh}}\ and\ \bibinfo {author} {\bibfnamefont {A.}~\bibnamefont {van~der
  Woude}},\ }\enquote {\bibinfo {title} {Giant resonances-fundamental
  high-frequency modes of nuclear excitation},}\ \ (\bibinfo  {publisher}
  {Clarendon, Oxford},\ \bibinfo {year} {2001})\BibitemShut {NoStop}%
\bibitem [{\citenamefont {Piekarewicz}(2007)}]{Piekarewicz:2007us}%
  \BibitemOpen
  \bibfield  {author} {\bibinfo {author} {\bibfnamefont {J.}~\bibnamefont
  {Piekarewicz}},\ }\href {\doibase 10.1103/PhysRevC.76.031301} {\bibfield
  {journal} {\bibinfo  {journal} {Phys. Rev. C}\ }\textbf {\bibinfo {volume}
  {76}},\ \bibinfo {pages} {031301} (\bibinfo {year} {2007})}\ \BibitemShut
  {NoStop}%
\bibitem [{\citenamefont {Sagawa}\ \emph {et~al.}(2007)\citenamefont {Sagawa},
  \citenamefont {Yoshida}, \citenamefont {Zeng}, \citenamefont {Gu},\ and\
  \citenamefont {Zhang}}]{Sagawa:2007sp}%
  \BibitemOpen
  \bibfield  {author} {\bibinfo {author} {\bibfnamefont {H.}~\bibnamefont
  {Sagawa}}, \bibinfo {author} {\bibfnamefont {S.}~\bibnamefont {Yoshida}},
  \bibinfo {author} {\bibfnamefont {G.-M.}\ \bibnamefont {Zeng}}, \bibinfo
  {author} {\bibfnamefont {J.-Z.}\ \bibnamefont {Gu}}, \ and\ \bibinfo {author}
  {\bibfnamefont {X.-Z.}\ \bibnamefont {Zhang}},\ }\href {\doibase
  10.1103/PhysRevC.76.034327} {\bibfield  {journal} {\bibinfo  {journal} {Phys.
  Rev. C}\ }\textbf {\bibinfo {volume} {76}},\ \bibinfo {pages} {034327}
  (\bibinfo {year} {2007})}\ \BibitemShut {NoStop}%
\bibitem [{\citenamefont {Avdeenkov}\ \emph {et~al.}(2009)\citenamefont
  {Avdeenkov}, \citenamefont {Grummer}, \citenamefont {Kamerdzhiev},
  \citenamefont {Krewald}, \citenamefont {Litvinova} \emph
  {et~al.}}]{Avdeenkov:2008bi}%
  \BibitemOpen
  \bibfield  {author} {\bibinfo {author} {\bibfnamefont {V.}~\bibnamefont
  {Tselyaev}}, \bibinfo {author} {\bibfnamefont {J.}~\bibnamefont {Speth}},
  \bibinfo {author} {\bibfnamefont {S.}~\bibnamefont {Krewald}}, \bibinfo
  {author} {\bibfnamefont {E.}~\bibnamefont {Litvinova}}, \bibinfo {author}
  {\bibfnamefont {S.}~\bibnamefont {Kamerdzhiev}}, \bibinfo {author}
  {\bibfnamefont {N.}~\bibnamefont {Lyutorovich}}, \bibinfo {author}
  {\bibfnamefont {A.}~\bibnamefont {Avdeenkov}}, \ and\ \bibinfo {author}
  {\bibfnamefont {F.}~\bibnamefont {Grummer}},\ }\href
  {\doibase 10.1103/PhysRevC.79.034309} {\bibfield  {journal} {\bibinfo
  {journal} {Phys. Rev. C}\ }\textbf {\bibinfo {volume} {79}},\ \bibinfo {pages}
  {034309} (\bibinfo {year} {2009})}\ \BibitemShut {NoStop}%
\bibitem [{\citenamefont {Piekarewicz}(2010)}]{Piekarewicz:2009gb}%
  \BibitemOpen
  \bibfield  {author} {\bibinfo {author} {\bibfnamefont {J.}~\bibnamefont
  {Piekarewicz}},\ }\href {\doibase 10.1088/0954-3899/37/6/064038} {\bibfield
  {journal} {\bibinfo  {journal} {J. Phys. G}\ }\textbf {\bibinfo {volume}
  {37}},\ \bibinfo {pages} {064038} (\bibinfo {year} {2010})}\ \BibitemShut
  {NoStop}%
\bibitem [{\citenamefont {Cao}\ \emph {et~al.}(2012)\citenamefont {Cao},
  \citenamefont {Sagawa},\ and\ \citenamefont {Colo}}]{Cao:2012dt}%
  \BibitemOpen
  \bibfield  {author} {\bibinfo {author} {\bibfnamefont {L.-G.}\ \bibnamefont
  {Cao}}, \bibinfo {author} {\bibfnamefont {H.}~\bibnamefont {Sagawa}}, \ and\
  \bibinfo {author} {\bibfnamefont {G.}~\bibnamefont {Colo}},\ }\href@noop {}
  {\bibfield {journal} {\bibinfo  {journal} {arXiv:1206.6552}\ } 
  (\bibinfo {year} {2012})}\ \BibitemShut {NoStop}%
\bibitem [{\citenamefont {Vesely}\ \emph {et~al.}(2012)\citenamefont {Vesely},
  \citenamefont {Toivanen}, \citenamefont {Carlsson}, \citenamefont
  {Dobaczewski}, \citenamefont {Michel} \emph {et~al.}}]{Vesely:2012dw}%
  \BibitemOpen
  \bibfield  {author} {\bibinfo {author} {\bibfnamefont {P.}~\bibnamefont
  {Vesely}}, \bibinfo {author} {\bibfnamefont {J.}~\bibnamefont {Toivanen}},
  \bibinfo {author} {\bibfnamefont {B.}~\bibnamefont {Carlsson}}, \bibinfo
  {author} {\bibfnamefont {J.}~\bibnamefont {Dobaczewski}}, \bibinfo {author}
  {\bibfnamefont {N.}~\bibnamefont {Michel}},  \emph {et~al.},\ }\href
  {\doibase 10.1103/PhysRevC.86.024303} {\bibfield  {journal} {\bibinfo
  {journal} {Phys. Rev. C}\ }\textbf {\bibinfo {volume} {86}},\ \bibinfo {pages}
  {024303} (\bibinfo {year} {2012})}\ \BibitemShut {NoStop}%
\bibitem [{\citenamefont {Piekarewicz}(2013)}]{Piekarewicz:2013bea}%
  \BibitemOpen
  \bibfield  {author} {\bibinfo {author} {\bibfnamefont {J.}~\bibnamefont
  {Piekarewicz}},\ }\href {\doibase 10.1140/epja/i2014-14025-x} {\bibfield
  {journal} {\bibinfo  {journal} {Eur. Phys. J. A}\ }\textbf {\bibinfo {volume}
  {50}},\ \bibinfo {pages} {25} (\bibinfo {year} {2013})}\ \BibitemShut
  {NoStop}%
\bibitem [{\citenamefont {Chen}\ \emph {et~al.}(2014)\citenamefont {Chen},
  \citenamefont {Piekarewicz},\ and\ \citenamefont {Volya}}]{Chen:2013jsa}%
  \BibitemOpen
  \bibfield  {author} {\bibinfo {author} {\bibfnamefont {W.-C.}\ \bibnamefont
  {Chen}}, \bibinfo {author} {\bibfnamefont {J.}~\bibnamefont {Piekarewicz}}, \
  and\ \bibinfo {author} {\bibfnamefont {A.}~\bibnamefont {Volya}},\
  }\href@noop {} {\bibfield  {journal} {\bibinfo  {journal}
      {Phys. Rev. C}\
  }\textbf {\bibinfo {volume} {89}},\ \bibinfo {pages} {014321} (\bibinfo
  {year} {2014})}\ \BibitemShut {NoStop}%
\bibitem [{\citenamefont {Piekarewicz}(2004)}]{Piekarewicz:2003br}%
  \BibitemOpen
  \bibfield  {author} {\bibinfo {author} {\bibfnamefont {J.}~\bibnamefont
  {Piekarewicz}},\ }\href@noop {} {\bibfield  {journal} {\bibinfo  {journal}
  {Phys. Rev. C}\ }\textbf {\bibinfo {volume} {69}},\ \bibinfo {pages} {041301}
  (\bibinfo {year} {2004})}\ \BibitemShut {NoStop}%
\bibitem [{\citenamefont {Lattimer}\ and\ \citenamefont
  {Prakash}(2007)}]{Lattimer:2006xb}%
  \BibitemOpen
  \bibfield  {author} {\bibinfo {author} {\bibfnamefont {J.~M.}\ \bibnamefont
  {Lattimer}}\ and\ \bibinfo {author} {\bibfnamefont {M.}~\bibnamefont
  {Prakash}},\ }\href {\doibase 10.1016/j.physrep.2007.02.003} {\bibfield
  {journal} {\bibinfo  {journal} {Phys. Rept.}\ }\textbf {\bibinfo {volume}
  {442}},\ \bibinfo {pages} {109} (\bibinfo {year} {2007})}\ \BibitemShut
  {NoStop}%
\bibitem [{\citenamefont {Abrahamyan}\ \emph {et~al.}(2012)\citenamefont
  {Abrahamyan}, \citenamefont {Ahmed}, \citenamefont {Albataineh},
  \citenamefont {Aniol}, \citenamefont {Armstrong} \emph
  {et~al.}}]{Abrahamyan:2012gp}%
  \BibitemOpen
  \bibfield  {author} {\bibinfo {author} {\bibfnamefont {S.}~\bibnamefont
  {Abrahamyan}}, \bibinfo {author} {\bibfnamefont {Z.}~\bibnamefont {Ahmed}},
  \bibinfo {author} {\bibfnamefont {H.}~\bibnamefont {Albataineh}}, \bibinfo
  {author} {\bibfnamefont {K.}~\bibnamefont {Aniol}}, \bibinfo {author}
  {\bibfnamefont {D.}~\bibnamefont {Armstrong}},  \emph {et~al.},\ }\href
  {\doibase 10.1103/PhysRevLett.108.112502} {\bibfield  {journal} {\bibinfo
  {journal} {Phys. Rev. Lett.}\ }\textbf {\bibinfo {volume} {108}},\ \bibinfo
  {pages} {112502} (\bibinfo {year} {2012})}\ \BibitemShut
  {NoStop}%
\bibitem [{\citenamefont {Horowitz}\ \emph {et~al.}(2012)\citenamefont
  {Horowitz}, \citenamefont {Ahmed}, \citenamefont {Jen}, \citenamefont
  {Rakhman}, \citenamefont {Souder} \emph {et~al.}}]{Horowitz:2012tj}%
  \BibitemOpen
  \bibfield  {author} {\bibinfo {author} {\bibfnamefont {C.~J.}~\bibnamefont
  {Horowitz}}, \bibinfo {author} {\bibfnamefont {Z.}~\bibnamefont {Ahmed}},
  \bibinfo {author} {\bibfnamefont {C.}~\bibnamefont {Jen}}, \bibinfo {author}
  {\bibfnamefont {A.}~\bibnamefont {Rakhman}}, \bibinfo {author} {\bibfnamefont
  {P.}~\bibnamefont {Souder}},  \emph {et~al.},\ }\href {\doibase
  10.1103/PhysRevC.85.032501} {\bibfield  {journal} {\bibinfo  {journal}
  {Phys. Rev. C}\ }\textbf {\bibinfo {volume} {85}},\ \bibinfo {pages} {032501}
  (\bibinfo {year} {2012})}\ \BibitemShut {NoStop}%
\bibitem [{\citenamefont {Lattimer}\ and\ \citenamefont
  {Steiner}(2013)}]{Lattimer:2013hma}%
  \BibitemOpen
  \bibfield  {author} {\bibinfo {author} {\bibfnamefont {J.~M.}\ \bibnamefont
  {Lattimer}}\ and\ \bibinfo {author} {\bibfnamefont {A.~W.}\ \bibnamefont
  {Steiner}},\ }\href@noop {} 
  {\bibfield {journal} {\bibinfo  {journal} {arXiv:1305.3242}\ } 
  (\bibinfo {year} {2013})}\ \BibitemShut {NoStop}%
\bibitem [{\citenamefont {Fattoyev}\ and\ \citenamefont
  {Piekarewicz}(2013)}]{Fattoyev:2013yaa}%
  \BibitemOpen
  \bibfield  {author} {\bibinfo {author} {\bibfnamefont {F.~J.}~\bibnamefont
  {Fattoyev}}\ and\ \bibinfo {author} {\bibfnamefont {J.}~\bibnamefont
  {Piekarewicz}},\ }\href {\doibase 10.1103/PhysRevLett.111.162501} {\bibfield
  {journal} {\bibinfo  {journal} {Phys. Rev. Lett.}\ }\textbf {\bibinfo
  {volume} {111}},\ \bibinfo {pages} {162501} (\bibinfo {year}
  {2013})}\ \BibitemShut {NoStop}%
\bibitem [{\citenamefont {Farine}\ \emph {et~al.}(1978)\citenamefont {Farine},
  \citenamefont {Pearson},\ and\ \citenamefont {Rouben}}]{Farine:1978}%
  \BibitemOpen
  \bibfield  {author} {\bibinfo {author} {\bibfnamefont {M.}~\bibnamefont
  {Farine}}, \bibinfo {author} {\bibfnamefont {J.}~\bibnamefont {Pearson}}, \
  and\ \bibinfo {author} {\bibfnamefont {B.}~\bibnamefont {Rouben}},\ }\href
  {\doibase http://dx.doi.org/10.1016/0375-9474(78)90241-5} {\bibfield
  {journal} {\bibinfo  {journal} {Nucl. Phys. A}\ }\textbf {\bibinfo {volume}
  {304}},\ \bibinfo {pages} {317} (\bibinfo {year} {1978})}\BibitemShut
  {NoStop}%
\bibitem [{\citenamefont {Zhang}\ and\ \citenamefont
  {Chen}(2013)}]{Zhang:2013wna}%
  \BibitemOpen
  \bibfield  {author} {\bibinfo {author} {\bibfnamefont {Z.}~\bibnamefont
  {Zhang}}\ and\ \bibinfo {author} {\bibfnamefont {L.-W.}\ \bibnamefont
  {Chen}},\ }\href {\doibase 10.1016/j.physletb.2013.08.002} {\bibfield
  {journal} {\bibinfo  {journal} {Phys. Lett. B}\ }\textbf {\bibinfo {volume}
  {726}},\ \bibinfo {pages} {234} (\bibinfo {year} {2013})}\ \BibitemShut
  {NoStop}%
\bibitem [{\citenamefont {Brown}(2013)}]{Brown:2013mga}%
  \BibitemOpen
  \bibfield  {author} {\bibinfo {author} {\bibfnamefont {B.~A.}\ \bibnamefont
  {Brown}},\ }\href {\doibase 10.1103/PhysRevLett.111.232502} {\bibfield
  {journal} {\bibinfo  {journal} {Phys. Rev. Lett.}\ }\textbf {\bibinfo {volume}
  {111}},\ \bibinfo {pages} {232502} (\bibinfo {year} {2013})}\ \BibitemShut
  {NoStop}%
\bibitem [{\citenamefont {Horowitz}\ \emph {et~al.}(2014)\citenamefont
  {Horowitz}, \citenamefont {Brown}, \citenamefont {Kim}, \citenamefont
  {Lynch}, \citenamefont {Michaels} \emph {et~al.}}]{Horowitz:2014bja}%
  \BibitemOpen
  \bibfield  {author} {\bibinfo {author} {\bibfnamefont {C.~J.}~\bibnamefont
  {Horowitz}}, \bibinfo {author} {\bibfnamefont {E.}~\bibnamefont {Brown}},
  \bibinfo {author} {\bibfnamefont {Y.}~\bibnamefont {Kim}}, \bibinfo {author}
  {\bibfnamefont {W.}~\bibnamefont {Lynch}}, \bibinfo {author} {\bibfnamefont
  {R.}~\bibnamefont {Michaels}},  \emph {et~al.},\ }\href {\doibase
  10.1088/0954-3899/41/9/093001} {\bibfield  {journal} {\bibinfo  {journal} {J.
  Phys. G}\ }\textbf {\bibinfo {volume} {41}},\ \bibinfo {pages} {093001}
  (\bibinfo {year} {2014})}\ \BibitemShut {NoStop}%
\bibitem [{\citenamefont {Fattoyev}\ \emph {et~al.}(2010)\citenamefont
  {Fattoyev}, \citenamefont {Horowitz}, \citenamefont {Piekarewicz},\ and\
  \citenamefont {Shen}}]{Fattoyev:2010mx}%
  \BibitemOpen
  \bibfield  {author} {\bibinfo {author} {\bibfnamefont {F.~J.}\ \bibnamefont
  {Fattoyev}}, \bibinfo {author} {\bibfnamefont {C.~J.}\ \bibnamefont
  {Horowitz}}, \bibinfo {author} {\bibfnamefont {J.}~\bibnamefont
  {Piekarewicz}}, \ and\ \bibinfo {author} {\bibfnamefont {G.}~\bibnamefont
  {Shen}},\ }\href {\doibase 10.1103/PhysRevC.82.055803} {\bibfield  {journal}
  {\bibinfo  {journal} {Phys. Rev. C}\ }\textbf {\bibinfo {volume} {82}},\
  \bibinfo {pages} {055803} (\bibinfo {year} {2010})}\ \BibitemShut
  {NoStop}%
\bibitem [{\citenamefont {Gezerlis}\ and\ \citenamefont
  {Carlson}(2008)}]{Gezerlis:2007fs}%
  \BibitemOpen
  \bibfield  {author} {\bibinfo {author} {\bibfnamefont {A.}~\bibnamefont
  {Gezerlis}}\ and\ \bibinfo {author} {\bibfnamefont {J.}~\bibnamefont
  {Carlson}},\ }\href {\doibase 10.1103/PhysRevC.77.032801} {\bibfield
  {journal} {\bibinfo  {journal} {Phys. Rev. C}\ }\textbf {\bibinfo {volume}
  {77}},\ \bibinfo {pages} {032801} (\bibinfo {year} {2008})}\ \BibitemShut
  {NoStop}%
\end{thebibliography}%

\end{document}